\documentclass[letterpaper,11pt]{article}
\usepackage{jheppub}
\usepackage{amsmath,latexsym}
\usepackage{graphicx}      
\usepackage{subcaption}    
\usepackage{tikz-feynman}  
\tikzfeynmanset{compat=1.1.0}  
\usepackage{booktabs}
\usepackage{comment}
\usepackage{siunitx}
\usepackage{setspace}
\usepackage{amssymb}
\usepackage{amsmath}
\usepackage{amstext}
\usepackage{amsfonts}
\usepackage{bbold}
\usepackage{slashed}
\usepackage{tikz-feynman}
\usepackage{xcolor}
\usepackage{subcaption}
\usepackage{physics}
\usepackage{slashed}
\usepackage{setspace}
\usepackage{amssymb}
\usepackage{amsmath}
\usepackage{amstext}
\usepackage{amsfonts}
\usepackage{bbold}
\usepackage{slashed}
\usepackage{tikz-feynman}
\usepackage{xcolor}
\usepackage{subcaption}
\usepackage{physics}
\usepackage{slashed}
\newcommand{\beq}{\begin{equation}}
\newcommand{\eeq}{\end{equation}}
\renewcommand{\l}{\left}
\renewcommand{\r}{\right}

\newcommand{\bea}{\begin{eqnarray}}
\newcommand{\eea}{\end{eqnarray}}

\newcommand{\nn}{\nonumber}

\newcommand{\be}{\begin{eqnarray}}
\newcommand{\ee}{\end{eqnarray}}
\newcommand{\cP}{{\cal P}}

\definecolor{SGreen}{rgb}{0.0547,0.613,0.328}
\definecolor{NodeBlue}{rgb}{0.0547,0.148,0.578}

\newcommand{\dbar}{d\hspace*{-0.08em}\bar{}\hspace*{0.1em}}
\newcommand{\deltabar}{\delta\hspace*{-0.2em}\bar{}\hspace*{0.15em}}
\begin{document}

\title{A Systematic Lagrangian Formulation for Quantum and Classical Gravity at High Energies} 

\author{Ira Z. Rothstein and}
\affiliation{Department of Physics, Carnegie Mellon University,\\
Pittsburgh, PA 15213, USA}
\author{Michael Saavedra}
\emailAdd{izr@andrew.cmu.edu}
\emailAdd{msaavedr@andrew.cmu.edu}

\abstract{

We derive a systematic Lagrangian approach for quantum gravity in the super-Planckian limit where $s\gg M_{pl}^2\gg t$.
The action can be used to calculate to arbitrary accuracy in the quantum and classical expansion parameters $\alpha_Q= \frac{t}{M_{pl}^2}$ and $\alpha_C= \frac{st}{M_{pl}^4}$, respectively,  for the scattering of massless particles.
The perturbative series  contains powers of $\log(s/t)$ which can be resummed
using a rapidity renormalization group equation (RRGE) that follows from a factorization theorem which
allows us to write the amplitude as a convolution of a soft and collinear jet functions.  We prove that the soft function is composed of a tower of operators, enumerated by the number of t-channel graviton exchanges.  The running of the one graviton exchange  soft
operator leads to the graviton Regge trajectory while the two graviton operator running corresponds to
the gravitational BFKL equation.  For the former, we find agreement with one of the two results previously presented
in the literature, while for the latter our result agrees (up to regulator dependent pieces) with those
of Lipatov. We find  that the convolutive piece of the gravitational BFKL kernel is the square
 of that of QCD at leading order.
The power counting simplifies considerably in the classical limit where we can use our formalism
to extract logs at any order in the PM expansion. The leading log  at $(2N+1)$ order in the Post-Minkowskian expansion follows from calculating
the one loop anomalous dimension for the $N+1$th order piece of the soft function
and perturbatively solving the RRGE $N-1$ times. The factorization theorem implies that
the leading classical logs which arise alternate between being real and imaginary in nature as $N$ increases.

}

\maketitle

\section{Introduction}\label{sec: 1}

Quantum gravity as an effective field theory, at least formally, is  well understood\cite{Donoghue:1995cz}
as long as all invariants are sufficiently small compared to the fundamental scale $M_{pl}$.
 In this regime the non-renormalizability of gravity is tamed,  save for the fact that as  we aspire to higher accuracy
we introduce more unknown UV parameters that must be fixed from experiment, or matched from some UV completion.
The renormalization group flow into the IR is not terribly interesting since all logs are power suppressed and there is no 
limit in which a resummation can be done systematically. 

  However,  we know this can not be the only kinematic regime for which we can maintain calculational control as, after all,  we certainly can predict astronomical orbits with high accuracy.
This super-Planckian scattering, corresponding to the limit  $s\gg t$, is the so-called ``Regge" regime, must be within our calculational reach even though the graviton coupling scales as $s/M_{pl}^2$ and $t/M_{pl}^2$ when the emission occurs off of energetic and soft partons respectively. 
Note that even 
if we work in the regime $s\gg M_{pl}^2\gg t$,  we are immediately faced with a severe power counting challenge given the growth of  effecitve coupling ($Gs\gg 1)$ in the super-Planckian limit. In fact, matters are made worse by the existence of large (``Regge'') logs of
the ratio $s/t$, and, more importantly, this regime of  forward scattering is enhanced due to the t-channel graviton exchange by a factor of $1/t$. 

The super-Planckian limit is a double edged sword. On the one hand, the growth of the cross section in $s$, at fixed $t$, leads to, at least naively,  a violation of  unitarity, but it also pushes the process into the semi-classical (eikonal) regime (for an extensive review of progress using the eikonal approximation see \cite{DiVecchia:2023frv})  over which we would expect to have calculational control. In the case of massless particle scattering, the classical picture of the initial state consists of two Aichelburg-Sexl shock wave metrics, and for impact parameter $b\sim 1/t$ large compared to the effective Schwarzchild radius $R_S= 2 G \sqrt{s}$, is tractable by classical GR\cite{Eardley:2002re}.  As the impact parameter diminishes we reach the regime of black hole production and a thermal  final state, as per Hawkings' result.  Thus it seems that, at least for $t<R_S^{-1}$,   super Planckian scattering
is dominated by IR physics.  We might glibly conclude that we may maintain calculational control by simply working at large impact parameter such that only-non local interactions will contribute, since contact interactions will be suppressed for localized incoming wave packets. However, this is premature as it is possible for local operators
to mix with non-local operators via soft exchanges of gauge bosons. In fact this occurs in NRQCD \cite{Caswell:1985ui,Luke:1999kz,Brambilla:1999xf},  the theory of non-relativistic bound states.


 The goal of this paper is to build a Lagrangian formalism which allows one to calculate systematically in a double expansion in $\alpha_Q \equiv t/M_{pl}^2$ and $\alpha_C\equiv st/M_{pl}^4$.  These ratios control quantum and classical corrections respectively. In addition, we will be working to leading
 order in $\lambda= \sqrt{t/s}$.
Our motivations are:  Formally, we would like define, in a gauge invariant operator formalism, the notion of a Regge trajectory and a BFKL equation for gravity, and to search for commonalities between QCD and gravity that go beyond what is known in the double copy relations \cite{Bern:2019crd,Bern:2022wqg}. Practically, we wish to show that this a Lagrangian effective field theory formalism can greatly simplify calcuations  of the Regge trajectory, as well as  higher order corrections in the PM expansion which are of interest to gravitational wave experiments.

Significant effort has been put into the calculation of  higher order  PM corrections to the classical scattering angle for
the purposes of increasing the accuracy of parameter extractions for binary inspirals.
There are various ways of approaching
these corrections, including using the classical world line approch \cite{Goldberger:2004jt} in the PM expasion \cite{Dlapa:2021npj}, the QFT world line approach \cite{Mogull:2020sak,Driesse:2024xad}, and the S-matrix approach
\cite{Neill:2013wsa,Bjerrum-Bohr:2013bxa,Bern:2021dqo}. All of these calculations utilize the physical limit, $s \sim m$,
and since we will be considering light-like scattering,  our results will only overlap  with a subset of the contributions\footnote{The soft loops, are insensitive to the existence of a mass.}. 

As concrete calculations in this paper we will show a simple way to extract the, mass independent,  classical log at 3PM, as well as the 
leading order Regge trajectory for which the two calculations in the literature  \cite{Bartels:2012ra,Melville:2013qca}  seems to disagree\footnote{The Regge trajectory is IR divergent and thus the local pieces are regulator dependent. However, the results in  \cite{Bartels:2012ra,Melville:2013qca}  also disagree in their $\log(t)$ dependence of the trajectory which should be independent of the regulator.}.
 It is our hope that by illuminating the all orders structure of the series we may be able to perform suitable resummations.

The technical details of our calculations will be couched in terms of EFT language. However, in an effort to make the physics 
accesible to a more general audience, we have relegated most of the EFT details to appendices. The EFT is the scaffolding that allows for all orders proofs
of factorization of the leading order, in $t/s$ amplitude for near forward scattering which can be written as
a convolution of soft and collinear functions
\beq
\langle O \rangle =  \langle O_n \rangle  \otimes \langle O_s \rangle  \otimes  \langle O_{\bar n} \rangle.
\eeq
Readers interested in generating fixed order results 
can do so using the full theory in conjunction with the method of regions \cite{Smirnov:1990rz,Beneke:1997zp,Smirnov:2002pj} to find the appropriate integrands dictated by the EFT.  However, the resummations are based upon operator rapidity anomalous dimensions which are defined within the EFT.

There are several existing approaches in the literature to studying the super-Planckian limit.  Early work tended to focus on obtaining the leading classical Eikonal phase through a variety of approaches \cite{tHooft:1987vrq, Verlinde:1991iu, Muzinich:1987in}.
Amati, Ciafaloni, and Veneziano (ACV)  expanded string amplitudes in the semi-classical limit\cite{Amati:1987wq,Amati:1987uf,Amati:1990xe, Amati:1992zb}, which allowed them to extract the two loop contribution to the classical phase.  More recent approaches involve Wilson lines \cite{Melville:2013qca, Luna:2016idw} and double copy considerations \cite{Schnitzer:2007kh, Schnitzer:2007rn, Akhoury:2011kq, SabioVera:2011wy, Naculich:2020clm, Raj:2023irr, Raj:2023iqn, Raj:2024xsi}.  
Lipatov \cite{Lipatov:1991nf} introduced ``effective actions" for high energy scattering which involved "Reggeon fields", which is quite distinct
from our approach; several other effective actions approaches closely related to Reggeon fields have also appeared in the literature \cite{Kirschner:1994xk, Amati:1993tb, Amati:2007ak, Lipatov:2011ab}. Recently the authors of \cite{Gao:2024qsg}, have given a nice explanation of the differences between Reggeon fields, in the context of QCD, and the theory formulated in \cite{Rothstein:2016bsq}, upon which our theory is based.

\subsection{Conventions and Definitions}
In our conventions we use  the mostly minus metric and define $\kappa^2= 16 \pi G= \frac{1}{2M_{pl}^2}$. We will go back and 
forth between using $\kappa$ and $M_{pl}$,  when discussing power counting.
The light cone momenta are defined as follows
\beq
p^\mu= n \cdot p \frac{\bar n^\mu}{2}+\bar n \cdot p \frac{n^\mu}{2}+p_\perp^\mu\equiv (n \cdot p, \bar n \cdot p,p_\perp^\mu).
\eeq
To reduce equation clutter we will sometimes use $p_+=\bar n \cdot p$ and $p_-= n \cdot p$.
 The measures are defined with
$ [d^Dk]=\frac{d^{D}k}{(2\pi)^{D}}$, $d=4-2 \epsilon$ and $d^\prime=2-2\epsilon$, $\deltabar^n(x)={(2\pi)^n}{\delta^n(x)}.$
Transverse inner products with vectors ($\vec k$) are Euclidean, while those without are Minkowskian.

It will be useful to define $B(a,b)$ which is the coefficient of the one-loop bubble integral in $2-2\epsilon$ dimensions:
\beq
(4\pi)^{-\epsilon}\int\frac{[d^{2-\epsilon}k_{1\perp}]}{[\vec{k}_\perp^2]^a [(\vec{k}_\perp + \vec{q}_\perp)^2]^b} = \frac{B(a,b)}{4\pi}(\vec{q}_\perp^2)^{1-\epsilon-a-b},
\label{Bubble Def}
\eeq
with 
\beq
B(a,b) = \frac{\Gamma(1-a-\epsilon)\Gamma(1-b-\epsilon)\Gamma(-1+a+b+\epsilon)}{\Gamma(a)\Gamma(b)\Gamma(2-a-b-2\epsilon)}.
\label{Bubble Coef}
\eeq

We will be considering only a complex scalar fields coupling to gravity and is otherwise free, with action
\beq
S=S_{EH}+ \int d^4x \sqrt{-g} (\partial_\mu \phi ) (\partial_\nu \phi^\star) g^{\mu \nu},
\eeq
and our convention for the linearized field is $g_{\mu \nu}= \eta_{\mu \nu}+\frac{\kappa}{\sqrt{2}} h_{\mu \nu}.$
where $\kappa= \frac{1}{\sqrt{2}M_{pl}}$.
 \section{Lessons from YM Theory}

To gain insight into gravity in the Regge limit it behooves us to consider  the case of YM theory
which is, in some ways, simpler than gravity, since the coupling is dimensionless so power counting is almost trivial, but in other ways
more complicated due to the color structures which arise at higher orders.
However, as we shall  see, the structure of the gravitational theory is considerably simpler than QCD once
we know how to tame the seemingly non-perturbative coupling behavior, as will be discussed in the next section.

YM theory has the nice property that hard processes are power suppressed \footnote{For on-shell operators. i.e. those which only contain physical polarizations.}, as a consequence of the fact that it is
classically conformal.  Let us first consider the case of  a generic hard scattering process away from the forward limit, where  we integrate out the hard modes and match onto
a theory of light-like scatterers.
 The systematics of this theory are based on a double expansion in $\alpha_s$ and $\Lambda/Q$, where $Q\sim s\sim t $ is the hard scattering scale and $\Lambda$ is the appropriate IR scale for the observable of interest.
The amplitude will contain large (double) logs of the ratio $Q/\Lambda^2$ whose resummation 
can be achieved by working in the EFT called  SCET, (Soft-Collinear Effective theory) \cite{Bauer:2001cu,Bauer:2000ew,Bauer:2001yt}.
Some of these logs are due to loops of large virtuality (``hard loops") which can be resummed using
renormalization group techniques, while other logs, of the ratio $s/t$ are actually due to the large ratio of rapidities which can
be resummed using rapidity renormalization group (RRG) methods \cite{Chiu:2011qc,Chiu2012ATheory}.

 Once we consider the Regge limit the power counting changes drastically, as the higher dimensional
 near forward scattering operator which arises from the exchange of a so-called  (off-shell) ``Glauber gluon", becomes order one. This is not to say that there are no near forward
 interactions in a hard scattering event (all invariants being large), however, it can be proven \cite{Collins:1984kg}\footnote{ Spectator-spectator interactions are the main bottleneck. All factorization proofs for hard scattering observables in SCET, to date, assume that Glaubers do not contribute.}, that for sufficiently 
 inclusive observables these interactions cancel  up to corrections which are suppressed by 
 the hard scattering scale. Thus in the Regge regime (no hard scattering) the forward scattering interaction
 dominates and amplitude. The interactions are characterized by the t-channel exchange of  a Glauber gluon with light-cone momentum 
 scalings $p^\mu \sim \sqrt{s} (\lambda^2,\lambda^2,\lambda)$, which  are off-shell ($p_+ p_- \ll p_\perp^2$) and can be integrated out, leading to an interaction which is non-local
 in the transverse direction. 
 These  Glauber modes have an analagous mode in QED  which is relevant for  quantum optics.
 The canonical definition of SCET  does not include these modes, which lead to a generalized version of SCET, GSCET,  \cite{Rothstein:2016bsq}.  The resummation of these Glauber exchanges generates the eikonal phase characteristic of the semi-classical
 nature of the near forward scattering process.
 
Let us consider the form of the  one loop amplitude
\beq
\label{avatar}
M \sim\frac{\alpha_s}{t} (1+i \pi C_1  \alpha_s \Gamma[\epsilon]\l(\frac{t}{\mu^2}\r)+C_2 \alpha_s \log(\frac{s}{-t})+...
\eeq 
Here we have ignored color which  leads to complex structure at higher orders.
It is important to note that there are {\it no hard loop} correction to any order  in $\alpha$
for forward scattering kinematics in QCD, as such contributions are all power suppressed by factors of $t/s$ \cite{Rothstein:2016bsq}.
This is again a consequence of the conformal nature of classical YM theory.
The imaginary term in Eq.(\ref{avatar}) is the avatar of the classical phase and the large $\log(\frac{s}{-t})$, which needs to be resummed
to regain calculational control in the asymptotic limit, leads to the ``Regge trajectory"\footnote{The Regge trajectory is defined as
the running of the octet operator.}. There is a storied history of the resummation of these logs  that goes under the name of ``Reggeization". 
Gribov's original approach \cite{Gribov:1967vfb} to  the problem has  led to a  number of perspectives including the classic work of Balitsky, Fadin, Kuraev and Lipatov \cite{Fadin:1975cb}, Lipatov’s effective action \cite{Lipatov:1995pn}, and more modern approaches in terms of Wilson lines\cite{Balitsky:1995ub,Caron-Huot:2013fea,Caron-Huot:2017fxr, Vernazza:2018gyb, Falcioni:2021buo}.
For an historical review
see \cite{DelDuca:2018nsu}.
 In some instances, e.g. in the anti-symmetric octet color channel, the resummations of these logs leads to so-called Regge form  of the amplitude
where  the amplitude can be written as \footnote{In momentum space the eikonal phase is not manifest, but instead the series $C$ includes both classical (eikonal) and quantum contributions.}
\beq
\label{reggie}
M \sim C(\alpha_s) \l( \frac{s}{t}\r)^{\alpha(t)}.
\eeq
$\alpha(t)$ is the, infrared divergent, Regge trajectory. 
This form of the amplitude holds up to next-to leading log in general  \cite{Duca_2001} and to all orders in the planar limit \cite{Kuraev:1976ge}.
Amplitudes of this form have ``Regge pole'' behavior since they arise when there is a pole in complex angular momentum plane. This is as opposed to the case where cuts arise and the amplitude takes on a more complicated form.
Recent progress has shown that there are relations between $\alpha(t)$ and the series in $\alpha_s$ that defines $C$ \cite{Del_Duca_2018,Moult:2022lfy}.  In addition, it has been shown that, by considering ampltiudes with definate crossing symmetry,   unitarity implies that there are relations between 
the Regge trajectory and the eikonal phase \cite{Rothstein:2024fpx}, as well as between various anomalous dimensions.

\section{The Gravitational Case}

Now let us return to the gravitational case. 
 We would expect this amplitude  to take
a form identical to (\ref{reggie}). However, as discussed in the introduction the
hard scattering S-channel operators are enhanced by powers of $s/M_{pl}^2$.
For instance, for scalar scattering the tree level s-channel graviton exchange will
generate a local operator with a Wilson coefficient that scales as $s/M_{pl}^2$.
\[
\begin{tikzpicture}
\begin{feynman}
	\vertex (i1)  at (-2,1);
	\vertex (f1) at (2, 1);
	\vertex (i2)  at (-2,-1);
	\vertex (f2) at (2, -1);
	\node[dot] (g1) at (-1,0);
	\node[dot] (g2) at (1,0);
\diagram*{
	(i1)--[ line width = 0.3mm](g1)--[ line width = 0.3mm](i2),
	(f1)--[ line width = 0.3mm](g2)--[line width = 0.3mm](f2),
	(g1)--[gluon, line width = 0.3mm](g2),
};
\end{feynman}
\end{tikzpicture}
\]
Any observable sensitive to this operator will not be under calculational control. In fact, we could insert higher dimensional
operators  with unknown Wilson coefficients at  the vertices, and they too would be super-leading.
Notice that simply specifying the kinematics as being Regge does not eliminate the contributions from such operators.
However, if  we consider a set of observables  ($ O$)   for which the incoming wave packets have a compact region of support and are separated
in the transverse direction by an amount greater than Schwarzchild radius, then operators which interpolate  for a fixed number of partial waves
wont contribute. It is interesting to note that this is NOT the end of story, as under some circumstatnces soft emissions can mix local and non-local
operators. In fact, this is exactly what happens in the case of non-relativistic bound states \cite{Titard:1993nn,Manohar:2000hj}, such a  quarkonium where
the annihilation diagram generates a local color octet  $T^a \otimes T^a \delta(x)$ potential that gets corrected by a soft exchance as in this diagram \footnote{In the EFT
NRQCD \cite{Caswell:1985ui,Luke:1999kz,Brambilla:1999xf}, this gluon is called ``Ultra-Soft" because all of its momentum components scale as $mv^2$, where $v$ is the relative
velocity in the bound state. }
$$
\begin{tikzpicture}
\tikzfeynmanset{ every vertex/.style={red, dot}, every particle/.style={blue}, every blob/.style={draw=green!40!black, pattern color=green!40!black},
}
\begin{feynman}
	\vertex (p1) at (-1.5,1);
	\vertex (p2) at (-1.5,-1);
	\vertex (p3) at (1.5,-1);
	\vertex (p4) at (1.5,1);
    \node[dot] (s1) at (-1, .6666);
    \node[dot] (s2) at (1,.6666);
    \node[crossed dot](a) at (0,0);
\diagram*{
	(p1)--[  line width = 0.3mm](a)--[  line width = 0.3mm](p3),
	(p2)--[  line width = 0.3mm](a)--[  line width = 0.3mm](p4),
    (s2)--[line width = 0.3mm, gluon, quarter right](s1)
};
\end{feynman}
\end{tikzpicture}
$$
that  generates a counter term for a non-local potential $V(x) \sim \frac{1}{r^3}$ which would contribute to
the set of observables $O$.  Physically we can imagine two widely separated partons one of which emits a
soft quanta which shifts its momenta, leading to a head on annihilation after which the quantum is reabsorbed and
the final state is again well separated. In NRQCD this poses no challenge to the power counting since the annihilation
graph is down by $\alpha_s\sim v$\footnote{In the bound state the power counting is such that $\alpha \sim v$.}.

We may worry that something similar can happen in the gravitational case and indeed it would, however only if the the matter
propagator assumes the dispersion relation $E=\frac{p^2}{2m}$, that is the soft exchange would cause the source line to
recoil and thus  the matter lines are not eikonal, and must behave quantum mechanically which, in turn, implies $mv \sim 1/r$ or
$L \sim 1$, which is outside the set $O$. 
In the Regge limit, as long as the collinear lines eikonalize, the soft corrections to local vertices can not generate
the  non-analyticity in $t$ required to contribute to our collection of observables.

Finally, one may worry that the exclusion of the s-channel operators will pose a challenge to the Ward identity once
we put gravitons on external states. In the EFT the Ward identity must be satisfied order by order in each of the
expansion parameters. As we shall see, by building operators using explicitly gauge invariant building blocks we are assured that
that the Ward identities will be satisfied and the contribution from local interactions (s-channel processes) will automatically
be included \footnote{This may seem strange from the point of view that local interactions are suppressed. However,
unphysical polarizations on external lines can lead to leading order local contributions which will automatically be accounted for in the EFT operators.}.


\section{Glauber Gravitational SCET}

\subsection{ Power Counting}

In QCD there are two countings, one based on the coupling ($\alpha_s$), and the other upon the kinematic variables $t/s$.
This will no longer be true in gravity, since the charges are stress-energy,  which will complicate matters. On the other hand in both theories
the field scalings, to be discussed once we define the relevant modes,  will be  indepedent of the couplings since they will be determined by the scaling in the free theory.

There are multiple kinematic scenarios of interest depending upon whether or not the scatterers are massive or not.
In this paper we will consider massless case. 
As mentioned in the introduction the EFT will be valid when the following hierarchy is satisfied
\beq
s\gg M_{pl} \gg t.
\eeq 
As in the case of QCD we will be working to leading order in the parameter $\lambda= \sqrt{t/s}$, but  to
all order in the couplings 
\beq
\label{choice}
\alpha_Q \equiv \frac{t}{ M_{pl}^2}< 1~~~~~\alpha_C \equiv \frac{st}{M_{pl}^4}< 1,
\eeq
which control the quantum and classical loop corrections, respectively. 
 $\alpha_C \equiv \frac{st}{M_{pl}^4}< 1$ implies that classical non-linearities are sub-leading such that we are not in the regime where black hole formation occurs\footnote{Here $t$ is the typical momentum in one graviton (Glauber) exchange, which is called $t_{individual}$ in \cite{Amati:1990xe, Amati:1992zb}, and should not be confused
    with the physical (impulse) $t \sim Gs/b$ which results from a coherent field of Glaubers which constitutes the shock wave.}. However, we can study the approach to black hole formation
as a function of $\alpha_C$.

As opposed to QCD, the gravitational Glauber interaction is power {\it enhanced}, i.e. $1/\lambda$. Such a state of affairs is usually a death knell for any EFT since  power counting forces us to rescale the action such that
the superleading interaction scales as unity, which would make the kinetic pieces sub-leading, and the theory would have
no propagating degrees of freedom. 
 However, since the Regge limit is a semi-classical in nature the amplitude has sufficient structure that calculational
control can be maintained.  To see this note that 
the semi-classical nature of the process  ensures that the amplitude can be written in 
impact parameter space  \cite{DiVecchia:2023frv} in the form
\beq
\label{amp}
M(b,s) \sim  \l((1+[\sum_{i=0} \alpha_Q^iC_i(bs) ]) e^{ i\delta_{\text{Cl}}^{(0)} \sum_{j=0} (\alpha_C^j D_j(bs) )} -1\r),
\eeq
where $ \delta_{\text{Cl}}^{(0)}$ is the Fourier transform of the leading order Glauber result
\begin{equation}
   \delta_{\text{Cl}}^{(0)} = Gs \pi^\epsilon (\bar{\mu}^2b^2)^\epsilon \Gamma(-\epsilon).
\end{equation}
The function $D( bs)$  is a series of logs.

Given this form of the amplitude we may  treat the kinetic term as being as the same order
as the Glauber interaction. Furthermore, this form of the amplitude allows us to cleanly separate the classical
from the quantum \footnote{In the massive case there is an alternative path one can take by working in an EFT
of potentials. Then the super classical terms that show up in iterations are cancelled when matching onto the
EFT \cite{Neill:2013wsa,Cheung:2018wkq}.}
. This need not have been the case
given that we have three dimensionful parameters  ($s,t,M_{pl}^2)$, the dimensionless couplings $\lambda,\alpha_Q$ 
and $\alpha_C$ are not independent ($\frac{\alpha_Q^2}{\lambda^2}=\alpha_C$). 
This would not be a problem save for the fact that the  theory contains (Glauber) operators which scale as inverse powers of $\lambda$
 which complicates the power counting.  At the diagrammatic level we may distinguish classical and quantum corrections when considering soft gravitons as any soft loop that does not involve an eikonal line will be quantum mechanical.
 In the massless case, as we consider here,  \cite{DiVecchia:2019kta} finds that eikonlization fails   at  order $O(G^4)$, but
  remains in tact in the soft sector \cite{Du:2024rkf}, and thus will not affect our analysis.

Notice that a direct calculation of terms which scale as powers of  $ \alpha_C$ would not suffice to extract the classical 
piece since, as we can see from the form of the amplitude, the expansion of the exponent  will yield powers of $\alpha_C$  (from the leading term) that will hit  quantum terms
in the prefactor and generate  classical scaling  contributions ($\frac{s}{t} \alpha_Q^2=\alpha_C$).
Figure one shows the general structure of the series. We see that the classical contribution skips orders in the PM expansion since we need an extra Glauber exchange to get a factor of $s/M_{pl}^2$ to accompany a quantum suppression
of $t/M_{pl}^2$. The RRG sums  all the logs along the green lines, as each step to the right generates another log, whereas vertical motion does not. The bottom green line generates the leading order  Regge trajectory.  These logs can arise from 
either soft or collinear emissions. As we will discuss below in the EFT all diagrams get contributions from soft and collinear partons. In the soft sector it is easy to determine which diagrams are classical and which are quantum, as any loop which does  not involve an eikonal line is necessarily quantum. 
The RRG running of the soft function, which  sums diagrams which involve adding rungs  between Glauber lines, will include both quantum as well
a classical piece. It also sums  the soft eye graphs which are purely quantum mechanical.

As previously mentioned, in the collinear sector 
the massless parton can split \footnote{When working in the limit where $s \sim m^2$, collinear emissions are no longer relevant and the source always behaves classically, and can be treated  as in NRGR \cite{Goldberger:2004jt}.} and, 
as such, the amplitude will become sensitive to the gravitational UV completion even if at large impact parameter.
The physical picture is analgous to the case discussed above for NRQCD.  The collinear loop generator a $\log(q^2)$ generating
the required non-analyticity to generate a long range force from the contact interaction.

 Nevertheless, if one is only interested
in logs one can calculate solely in the soft sector since the logs are all fixed by the anomalous dimensions which
can be calculated by choosing to work either in the soft sector or the collinear sectors \footnote{ In the EFT the full amplitude has no rapidity divergences which cancel between the collinear and soft sectors.}, as will be discussed below.
As one goes to higher orders in the quantum expansion one must include power corrected  Glauber operators which
can be lifted up by subsequent Glauber exchange. In this paper we will not be working to sufficiently high
order for this to be an issue.

 \begin{figure}
\label{fd}
	\centering
	\includegraphics[width=.7\linewidth]{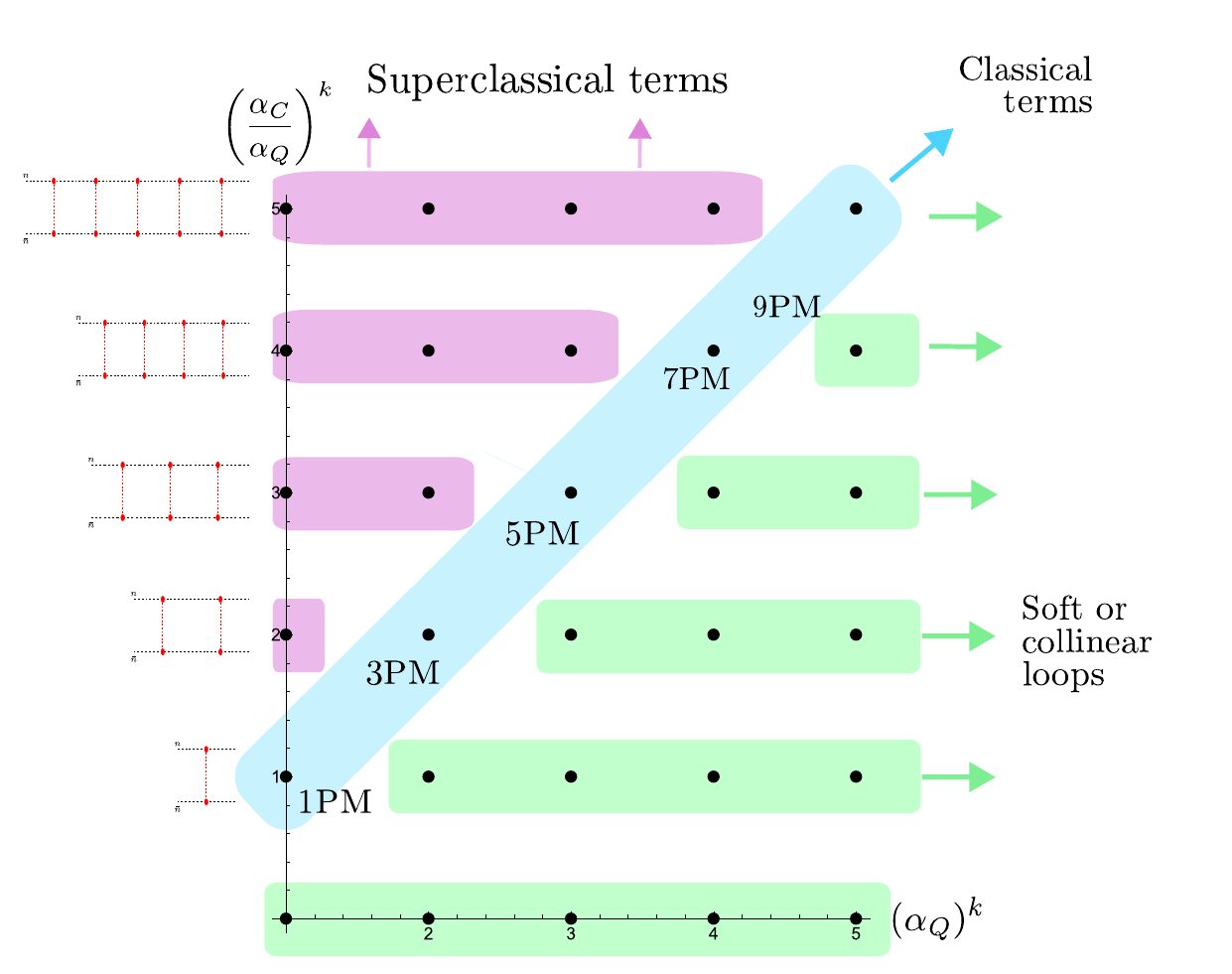}
	\caption{The structure of the perturbative series.  The blue  circles correspond to classical contributions in the 
	Post-Minkowskian expansion. The pink circles are super classical (box diagrams) while the greens lines indicate quantum corrections from soft and collinear loops. The classical contributions occur at odd orders in the PM expansion. 
	Each soft/collinear loop generates a log while Glauber loops generate $i\pi$.}\label{Perturbative Structure}
\end{figure}

If one wishes to power count by diagrams instead of operators, 
it is simple to read off the scaling of a given diagram. Each matter vertex gives as factor of $s$, while each matter line gives
a factor of $1/\sqrt{s}$.  All vertices give a power of $1/M_{pl}$ and given that the amplitude for scalars is scaleless, the
remaining units are made by powers of $t$ with a minimum exponent of $-1$.  Note that   each operator
scales homogeneously with $\lambda$ but not in $\alpha_Q$, or $\alpha_C$, thus even though those couplings are ratios of scales we should think of them in the same as we would $\alpha_s$ in YM theory. Also all amplitudes are analytic in $s$ since classically  there are no hard momenta flowing through the loops. In the EFT the apparent non-analyticity in $s$ due to Regge logs should be thought of as being a matching (rapidity) scale (like the way $Q$ shows up
in a canonical Wilsonian RG).
This also explains why in the massless case we only have odd PM powers contributing to the classical result, since the amplitude can only
have integer powers of $s$. In the massive theory we can have odd powers of the mass.

\subsection{The Action}

As mentioned in the introduction, the scaffolding of our calculations will be  Glauber SCET, and here will
will quickly review this topic in preparation for the introduction of the crucial factorization theorem (\ref{fact})
upon which our analysis hinges. 
For the case of hard scattering  a version of SCET for gravity was developed in \cite{Beneke:2012xa,Okui:2017all}.
Here we will be considering the complementary case which describes the Regge region and has a very different power
counting which substantially changes the nature of the theory.
Some of  atomistic gauge invariant objects upon which we build our theory can be ported over from 
the EFT for gravity for the case of hard scattering in \cite{Beneke:2012xa,Okui:2017all}. 
Also it is important to emphasize that  the forward case considered here will involve different challenges than the foundational
theory discussed in \cite{Beneke:2012xa,Okui:2017all}, as we will be working in what is known as SCETII versus 
those works which discussed SCETI. The distinction between these theories lies in scaling of the invariant mass of the
soft modes. In SCETI, the soft modes have invariant masses which are parameterically smaller than  the SCETII soft modes.
In SCETI therefore, the soft modes can be treated like a background field which is not true true in our case.

The 
 EFT for near forward scattering in gravity is structurally very similar to the case  of YM theory \cite{Rothstein:2016bsq}, thought it will significantly differ in important ways.
The starting point for building the EFT is to determine the modes necessary to reproduce the IR physics of the full theory.
The relevant modes are fixed by determining, given the regime of validity of the theory, the kinematic regions for which IR singularities arise.  
There is no distinction between the modal analysis in gravity and in YM theory, though the power counting
of the fields components are different (see below).
For high energy scattering the relevant  modes
are soft ($s$), collinear ($n$) and anti-colinear ($\bar n$) where the light cone momentum scale as
$p_s^\mu \sim (\lambda,\lambda,\lambda)$, $p_n^\mu\sim (1, \lambda^2,\lambda)$ and $p_{\bar n}^\mu\sim ( \lambda^2,1,\lambda)$, respectively. Here $\lambda\sim \sqrt{t}/\sqrt{s}$ is the power counting parameter. The overall scaling dimension will be $\sqrt{s}$.
Any prediction will be made in the context of a triple expansion in $\lambda$, $\alpha_C$  and $\alpha_Q$, though we will only work to leading
order in $\lambda$.  In this paper our matter fields will be complex scalars. Both soft and collinear scalar modes exist in the theory and both fields scale as $\lambda$.
The scalar soft mode loops will not generate rapidity logs and wont play a role at the order we will be working.
In the de Donder gauge the polarization of the  collinear graviton field ($h_{\mu \nu}^n) $
will scale as (in the $(+,-,\perp))$ basis
\beq
\label{pc}
h_{\mu \nu}^n \sim \frac{p^n_\mu p^n_\nu}{\lambda}
\eeq
and all of the components of the soft graviton scale as $\lambda$.  We can see that for the collinear modes
the large components $n^\mu n^\nu h^n_{\mu \nu}$ scale as $1/\lambda$, which poses a challenge to the power
counting that will be resolved when we write down the diff. invariant building blocks (see the appendix).

The total Lagrangian is written as
\beq
{\cal L}= {\cal L}_n+{\cal L}_{\bar n}+{\cal L}_s+ {\cal L_G}.
\eeq
where  ${\cal L}_n+{\cal L}_{\bar n}+{\cal L}_s$ correspond to the Lagrangian  for soft and collinear modes 
while ${\cal L_G}$ accounts for the factorization violating interactions, i.e Glauber contribution which connects modes in different sectors.  The theory has three distinct gauge symmetries (diffeomorphism invariances), collinear, anti-collinear and soft.
The actions for the purely collinear and soft fields are simply the full theory actions \footnote{We will not be including soft scalars in
our theory as they will not play a role for what we will be intertested in.}. The challenge is to build $ {\cal L_G}$ which is invariant under all three gauge symmetries constructed from gauge invariant operator building blocks, the details of which can be found in the appendix.  It is because the leading order 
action can be factored in this way  \eqref{eq:Onsnb} that is it relatively simple to write down factorization theorems when the Glauber mode is
included, as it is the only mode which has the ability to connect various sectors.
 Glauber exchanges will generate the following set of non-local (in the transverse plane) 
gauge invariant operators
\begin{align}  \label{eq:Onsnb}
     O^{\phi \phi}_{ns\bar{n}} &=  {\cal O}_n^{\phi }   
     \frac{1}{\cP_\perp^2} {\cal O}_s 
     \frac{1}{\cP_\perp^2} 
     {\cal O}_{\bar n}^{\phi } \,
    &  O^{ h \phi}_{ns\bar{n}} &=  {\cal O}_n^{\phi }   \frac{1}{\cP_\perp^2}  {\cal O}_s \frac{1}{\cP_\perp^2}{\cal O}_{\bar n}^{h } \,,
    \nn\\
        O^{\phi h}_{ns\bar{n}} &=  {\cal O}_n^{\phi } \frac{1}{\cP_\perp^2} {\cal O}_s \frac{1}{\cP_\perp^2} {\cal O}_{\bar n}^{h } \,,
    &  O^{hh}_{ns\bar{n}} &=  {\cal O}_n^{h } \frac{1}{\cP_\perp^2} {\cal O}_s
    \frac{1}{\cP_\perp^2} {\cal O}_{\bar n}^{h } \,.
\end{align}
${\cP_\perp^2}$ is the transverse derivative operator in coordinate space, or the ``label
 operator in momentum space.
On the left-hand side the subscripts indicate that these operators involve three sectors $\{n,s,\bar{n}\}$, while the first and second superscript determine whether we take a (scalar) quark or graviton operator in the $n$-collinear or $\bar{n}$-collinear sectors. 
There are other operators involving only two rapidity sectors (soft and (anti)collinear) that are discussed in the appendix.

\subsection{The Need for Power Suppressed non-Local Operators}
\label{power}
If we are interested in higher order corrections we will need to include operators suppressed by powers of $\lambda$ that
are non-local, i.e. sub-leading Glauber operators.  This is due to the superleading $\lambda$-scaling of the leading power Glauber Lagrangian, which scales as $\mathcal{O}_{ns\bar{n}}^{ij}\sim \frac{s}{M_{pl}^2}\sim \frac{1}{\lambda^2}\alpha_Q $.
  As we can see in Fig. \ref{Perturbative Structure}, we may add leading power Glaubers to obtain  power enhancements such that an operator $O_k$ which scales as $O_k\sim\lambda^{2k}$, $k>0$ may then be inserted into a diagram with $k + 1$ Glaubers and lead to a leading order $\lambda$ scaling.  Sub-leading operators represent quantum corrections and thus if we are interested in classical pieces one would
think that they can be ignored at the outset. In general this is true, as the interference like terms between super classical and quantum will
not contribute to the classical phase in Eq. (\ref{amp}). However, there are exceptions, as there are power corrections to the
Glauber operators that lead to classical corrections that must be included in the classical phase.
The need for power-suppressed operators is not unique to the SCET approach to forward scattering in gravity presented here.  In the Heavy Particle Effective Theory (HEFT) formalism for example, power suppressed, or quantum, operators are known to be necessary for higher order classical results \cite{Damgaard:2019lfh, Brandhuber:2021kpo, Brandhuber:2021eyq, Herderschee:2023fxh}.  
To the order we will be working in this paper, we wont need any such power suppressed operators as their effects
will contribute to the target of interest namely logarithmics classical pieces  at 5PM.


\subsection{The Rapidity Regulator}
In QCD there is no hard matching at leading power, therefore there  are no hard logs that need to
be resummed. For hard scattering gravity, hard logs are again  suppressed, this time by the Planck mass,  however, for super-Planckian forward scattering
this is no longer true.  These hard logs are quantum in nature and, despite the breakdown of eikonalization for the massless case, will not play a roll at the order we are working.
On the other hand, we will encounter power counting relevant rapidity logs which are generated by  loop integrals which will need to be regulated even in the presence
of dimensional regularization. These divergences arise as  consequence of the eikonal propagators generated by  the Wilson lines
as well as the fact that the Glauber gluon propagator is proportional to $1/p_\perp^2$, i.e. one drops the sub-leading light 
cone momentum, which is a step that is necessary to maintain manifest power counting.
Rapidity divergences arise in the EFT, since collinear and soft modes have the same invariant mass and one must define 
a rapidity factorization scale to distinguish them in integrals (for a discussion see \cite{Chiu:2011qc,Chiu2012ATheory}). A typical rapidity divergent integral takes the form of the box diagram
\beq
I= \int [d^dk]\frac{1}{(n \cdot k-\Delta_1)}\frac{1}{(\bar n\cdot k -\Delta_2)} \frac{1}{k_{1\perp}^2}\frac{1}{(k_{1\perp}-q_\perp)^2}
\eeq
This divergence is due to the transverse nature of the Glauber momentum. There are other rapidity divergences which are due to
the eikonal lines resulting from the Wilson lines. Such divergences take the form 
\beq
I= \int [d^dk]\frac{1}{(n \cdot k-\Delta_1+i\epsilon)}\frac{1}{(\bar n\cdot k -\Delta_2+i\epsilon)} \frac{1}{k^2}\frac{1}{(k-q_\perp)^2}.
\eeq
To regulate these divergences we introduce a rapidity regulator  following \cite{Chiu:2011qc,Chiu2012ATheory} in both the Glauber operator as well as both soft and collinear Wilson line couplings. As emphasized in \cite{Moult:2022lfy},
these two regulators must be kept independent of each starting at the two loops level where
 one must use two distinct parameters $\eta$ and $\eta^\prime$ for the collinear/soft loops and Glauber loops respectively and that $\eta^\prime$ must be taken to zero prior to $\eta$.
These regulators are implemented at the level of the action by modifying the Wilson lines and Glauber operators such that
Glauber loops between collinear generates a factor of $w\mid 2 k_z /\nu \mid^{-\eta}$ while collinear loops 
aquire a factor of $w\mid n\cdot k/\nu \mid^{-\eta}$ and soft loops $2w\mid 2k_z/\nu \mid^{-\eta/2}$. 
Here $w$ is a book keeping coupling used to calculate anomalous dimensions by endowing it with the following $\nu$ dependence
\beq
\frac{\partial w^2(\nu)}{\partial \nu}=-\eta w^2~~~~~~~~~~\lim_{\eta \rightarrow 0}w=1.
\eeq
More details about the use of the regulator at two loops can be found in \cite{Moult:2022lfy}.  For readers not interested in performing resummations these issues are irrelevant.

\subsection{Factorization of the Amplitude from Glauber SCET for YM}

Using this action we can write down a factorized form for the amplitude that looks effectively two dimensional.
To include effects of the Glaubers within the EFT following \cite{Rothstein:2016bsq, Moult:2022lfy,Gao:2024qsg} we start with the time evolution operator 
\begin{align}
U(a,b;T ) & = \lim_{T\to \infty (1-i0)} \int \big[{\cal D}\phi\big] \exp\bigg[ i \int_{-T}^{T}\!\! d^4x\: \big( {\cal L}_{n\bar{n} s}^{(0)}(x) + {\cal L}_G^{\rm II(0)}(x)\big) \bigg] \,,
\end{align}
one then expands in the number of Glauber potential insertions attaching to the $n$ and $\bar{n}$ projectiles, given by $i$ and $j$ respectively, so that
\begin{align}
\label{fact}
\exp\bigg[ i \int_{-T}^{T}\!\! d^4x\: \big( {\cal L}_G^{\rm II(0)}(x)\big) \bigg]= 1 +\sum\limits_{i=1}^\infty \sum\limits_{j=1}^\infty U_{(i,j)}.
\end{align}
For any number of Glauber potential insertions, one can then factorize the soft and collinear operators to give a factorized expression for the amplitude for scattering of projectile $\kappa$ with $\kappa^\prime$ is
\bea
\label{ampFactorized}
iM^{\kappa \kappa^\prime}=-\sum_{MN} \int \int_{\perp(N,M)} J_{\kappa N}(\{l_{\perp i}\},\epsilon,\eta) 
S_{ (N,M)}(\{l_{\perp i}\};\{l^\prime_{\perp i}\},\epsilon,\eta)
\bar J_{\kappa^\prime M}(\{l^\prime_{\perp i}\},\epsilon,\eta) \nn \\
\eea
where, following the notation in \cite{Gao:2024qsg}, we defined
\bea
\label{conv}
\int \int_{\perp(N,M)}= \frac{(-i)^{N+M}}{N! M!} \int \prod_{i=1}^N\prod_{j=1}^M
 \frac{[d^{d^\prime}l_{i\perp}]}{l^2_{i\perp}}\frac{[d^{d^\prime}l^\prime_{j\perp}]}{l^{\prime 2}_{j\perp}}\deltabar^{d^\prime}(\sum l_{i\perp}-q_\perp )\deltabar^{d^\prime}(\sum {l^\prime}_{j\perp}-q_\perp),
\eea
$\kappa$ and $\kappa^\prime$ label the external states, i.e. scalars or gravitons.
As we will see in gravity $N=M$ as a consequence of symmetries at the order we are working.

Note that in Eq.(\ref{ampFactorized}) all of the Glauber light cone momentum integrals have been performed, as have all of the soft and
collinear loops, that is why $J,S$ depend upon the regulators $\epsilon$ and $\eta$.
 All of the Glauber loops  correspond to box integrals \footnote{Cross box integrals vanish with this regulator.} which are rapidity finite and give a result
independent of the perp momenta,  
since the Glauber light-cone momenta are dropped in the soft function.
After performing the Glauber energy integral by contours we then use the result for the rapidity regulated $k_z$ integration
\beq
\label{gint}
\int \frac{[dk_z]}{-2k_z+A +i \epsilon}\mid \frac{2 k_z}{\nu} \mid^{-\eta}=-\frac{1}{4}.
\eeq
More generally, it was shown in \cite{Rothstein:2016bsq} that  the n-Glauber box diagram generates a factor of $\frac{i^n}{n!}$ 
which is necessary to form the semi-classical phase. This explains why the amplitude is defined with 
the factorial prefactors in Eq.(\ref{conv}).

The jet function are defined as time ordered products, e.g. at the one and two Glauber gluon level
\bea
J(k_{1\perp})&=&\int dx^\pm_1  \langle p \mid T( (O^{\phi}_n+O^{h}_n)(k_{1\perp},x_1^\pm) \mid p^\prime \rangle \nn \\
J(k_{1\perp},k^\prime_\perp)&=&\int dx^\pm_1 dx^\pm_2 \langle p \mid T( (O^{\phi}_n+O^h_n)(k_{1\perp},x_2^\pm)  (O^{\phi}_n+O^h_n)(k_{1\perp},x_1^\pm)  \mid p^\prime \rangle,
\eea
The jets are written in this way because the combination $(O^{\phi}_n+O^h_n)$, see equation (\ref{Collinear Ops}) for the definition,  is an eigen-vector of $\nu\frac{d}{d\nu}$.
At tree level the individual jet functions for $n$ Glauber exchange
\bea
J^{(0)}_{\phi n}&=& (n \cdot p)^{n+1} \left(\frac{\kappa}{2}\right)^n  \nn \\
J^{(0)}_{h n} &=& (b_{\mu \nu} \epsilon^\mu \epsilon^\nu)^2 \left(n \cdot p\right)^{n+1} (\frac{\kappa}{2})^n
\label{Jn}
\eea
where
 \beq b_{\mu \nu}= \bar n \cdot p_1 g_{\perp}^{\mu \nu}-\bar n^\mu p_{1\perp}^\nu-\bar n^\nu p_{4\perp}^\nu+\frac{p_{1 \perp}\cdot p_{4\perp} \bar n^\mu \bar n^\nu}{\bar n \cdot p_2}.
 \eeq
$p_1/p_4$ are the incoming and out going momenta respectively.  The tree level soft function for $(i,j)$ 
 is given by
\beq
S_{(i,j)}^{(0)}(l_{i \perp};l_{i\perp}^\prime)=2 \delta_{ij}i^j j!  \prod_{a=1}^j l_{i \perp}^{\prime 2}\prod_{n=1}^{j-1} \deltabar^{d^\prime}(l_{n \perp}-l_{n \perp}^\prime)\label{treeS}
\eeq
Note that $S^{(0)}_{(1,1)}=2i l_\perp^2$ and $J^{(0)}_{1}=(n \cdot p)^2\frac{\kappa}{2}$, such that the leading order, one Glauber, tree level exchange gives
\beq
iM_0=i s^2\frac{\kappa^2}{2\vec q_\perp^2}. \label{Tree Amplitude}
\eeq
We will need the form of the tree level results for the purposes of renormalization.

\subsection{The All Orders Glauber Integrals}

One advantage of the EFT is that we see immediately that all diagrams at any order can immediately be reduced to
a set of integrals over transverse momenta. The reason or this is that Glauber light cone momenta only appear
in the collinear propagators. Moreover, the soft light cone momenta only show up in soft lines since they are
sub-dominant when they run through collinear lines \footnote{It is crucial that the plus/minus soft momentum flow the appropriate collinear side.} and thus multi-pole expanded away.  Thus we may perform all of the 
Glauber light cone momentum integrals with relative ease.

In this paper we will not be concerned with collinear corrections since we can calculate all of the terms of interest (Logs) purely from the soft function \footnote{Since the full amplitude is independent of $\nu$ the $log(\nu)$ in the collinear functions sum to minus the the $log(\nu)$ in the soft function.}, so a generic diagram will take the following form:
\begin{figure}[h!]
\centering
\begin{tikzpicture}
  \begin{feynman}
    \vertex (i1) at (0,2);
    \vertex (f1) at (6,2);
    \vertex (i2) at (0,-2);
    \vertex (f2) at (6,-2);

    \vertex (a1) at (1.5,2);
    \vertex (a2) at (3.0,2);
    \vertex (a3) at (4.5,2);

    \vertex (b1) at (1.5,-2);
    \vertex (b2) at (3.0,-2);
    \vertex (b3) at (4.5,-2);

    \diagram*{
      (i1) -- [plain] (a1) -- [plain] (a2) -- [plain] (a3) -- [plain] (f1),
      (i2) -- [plain] (b1) -- [plain] (b2) -- [plain] (b3) -- [plain] (f2),
      (a1) -- [boson] (b1),
      (a2) -- [boson] (b2),
      (a3) -- [boson] (b3),
    };
  \end{feynman}

  \begin{scope}
    \filldraw[fill=lightgray, draw=black, thick]
      (3,0) ellipse [x radius=2.25, y radius=0.75];
  \end{scope}

  \foreach \x in {0.8, 1.1, 1.9, 2.2, 2.7, 3.3, 3.8, 4.2, 4.8, 5.1} {
    \node at (\x, 0) {\large\(\cdot\)};
  }

\end{tikzpicture}
\caption{The vertical lines are Glaubers while the blob represents all possible soft loops.}
\end{figure}
Note that all cross boxes vanish when we use the regulator where each Glauber rung with momentum $k_i$ is weighted by
a factor of $\mid \frac{2 k_i^z}{\nu}\mid^{-\eta}$. In the absense of any soft corrections we can write down the result
for the integral with $n$ Glaubers
\beq
iM_n= \frac{(-i)^{n}\kappa^{2n}}{2^{2n-1}}\frac{s^{n+1}}{n!} \prod_{i=1}^{n-1} \int [d^{d^\prime}k_i]\frac{(i^\epsilon \mu^{2\epsilon})^{n-1}}{(k_{1\perp}+q_\perp)^2 (k_{2\perp}-k_{1\perp})^2....(k_{n-1\perp}-k_{n-2\perp})^2k_{n -1\perp}^2}
\eeq
The momenta here are Minkowskian.

The sum over n can be performed by going to impact parameter space leaving
\beq
\tilde {\mathcal M}_n(b_\perp)=\frac{(-i)^{n+1}\kappa^{2n}}{2^{2n}}\frac{s^{n}}{n!}  \left[-
  \frac{\Gamma(-\epsilon)}{4\pi}
  \left(
     \frac{\mu \,\lvert \vec{b}_{\perp} \rvert\, e^{\gamma_{E}}}{2}
  \right)^{2\epsilon}
\right]^{n}.
\eeq
where
\begin{equation}
\label{M}
    \tilde{\mathcal{M}}(s,b) = \int\dbar^{d-2}q_\perp\frac{\mathcal{M}(s,\vec q_\perp^2)}{2s}e^{i\vec q_\perp\cdotp \vec b}.
\end{equation}

The series resums to get the usual eikonal phase (see eq, (\ref{M}) for norm)
\beq
\label{phase}
{\mathcal M}(b)=i (1-e^{i \delta})
\eeq
where 
\beq
\delta= \frac{\kappa^2s}{4}\left[
  \frac{\Gamma(-\epsilon)}{4\pi}
  \left(
     \frac{\mu \,\lvert \vec{b}_{\perp} \rvert\, e^{\gamma_{E}}}{2}
  \right)^{2\epsilon}
\right]
\eeq

In the presence of soft corrections (the blob in the figure) the Glauber light-cone integrals goes through unscathed as 
a consequence of factorization. The remaining integrals are purely in transverse momenta.

\subsection{Summing the Logs using the Rapidity Renormalization Group (RRG)}
While the amplitude is free of rapidity divergences, the individual components are not, and they obey the RRG equations
\begin{align}
\label{RRG}
    \nu\frac{\partial}{\partial\nu}J_{\kappa(i)} &= \sum_{j=1}^\infty J_{\kappa(j)}\otimes\gamma^J_{(j,i)},\nonumber\\
    \nu\frac{\partial}{\partial\nu}S_{(i,j)}& =- \sum_{k = 1}^\infty \gamma^S_{(i,k)}\otimes S_{(k,j)} - \sum_{k = 1}^\infty S_{(i,k)}\otimes \gamma^S_{(k,j)}, \\
    \nu\frac{\partial}{\partial\nu}\bar{J}_{\kappa'(i)} &= \sum_{j=1}^\infty \gamma^{\bar J}_{(i,j)}\otimes \bar{J}_{\kappa'(j)}.\nonumber
\end{align}
$\gamma_{(i,j)}$ are the rapidity anomalous dimensions, which will be defined below.

In Yang-Mills theory Each $J_i$ and $S_{(i,j)}$ is decomposed  into irreducible representations of the  symmetry group.
Operators with different numbers of Glaubers, but in the same irrep, can mix (for a discussion of the general structure see \cite{Gao:2024qsg}).
This is one complication that will obviously not arise in the case of gravity which will present a different set of challenges.
Another significant simplification that arises in the gravitational case is that $S_{M,N}\propto \delta_{MN}$ due to
RPI invariance which is the invariance of the physics under small deformations of the choice of light cone directions
for the partons \cite{Manohar:2002fd}.  Which is to say that, in the EFT we must choose a large light cone momentum around which to expand and there is arbitrariness in that choice.  Technically this correponds to invariance under a shifts of the light
cone directions $n$ and $\bar n$ that leave the inner products $n \cdot n=\bar n \cdot \bar n=0$ and $n \cdot \bar n=2$ invariant. In the case at hand we will utilize the fact that RPI implies that every  amplitude scales as $n^a\bar n^b$ with $a=b$\footnote{This is called RPIII in the language of \cite{Manohar:2002fd}.}.
 Any amplitude can only depend upon
the product of the two large incoming (conserved) light cone momenta $n \cdot p \bar n \cdot p$. 
While the jet functions are not RPI invariant, the product of the two is. Morever, the soft function is independent of $n$ and $\bar n $, see Eqs. (\ref{Jn},\ref{treeS}).
This invariance implies that diagrams with a different numbers of Glaubers connecting and top and the bottom must vanish.
 This is a significant
 simplification from QCD where diagrams such as the ``tennis court" diagram arise at three loops which vanish
 in gravity. This result holds independent of the type of collinear parton being considered.
 QCD allows for such diagrams because the Jet functions are linear in $n,\bar n$ independent of the number of insertions of Glaubers.
  Operationally, the vanishing of diagrams with a different number of  Glauber connections on the top and bottom of the diagram
 arises due to the vanishing  of the tensor integrals. 
 It is important to note however, that once power suppressed operators are included, the RPI argument will fail because the
 jet funtion $J^i$ will no longer scale as $n^i$, and thus diagrams with differing number of Glaubers connecting the top and the bottom 
 line will be non-vanishing. The contribution  of these non-diagonal soft contributions will no contribute to the 5PM leading log which
 we will calculate in a later section.
 
 
   Now that we know that $S$ is diagonal, to the order we are working, this simplifies the RRG equations considerably.
 In addition it allows us to write down the following simple constraint
 \beq
J_{ (i)}\otimes\gamma^J_{(i)}+\gamma^{\bar J}_{(i)}\otimes \bar{J}_{(i)}-  \gamma^S_{(i)}\otimes S_{(i)}
 -  S_{(i)}\otimes \gamma^S_{(i)}=0,
\eeq
which follows from the fact that the full result must be independent of the $\nu$.
Note that since $S$ is diagonal we have simplified its index structure.

With this simplification we have 
\bea
J_{i}(\{l_{\perp i}\},\epsilon,\eta,\nu)&=& \int_{\perp(i)} J_i({k_{\perp i}},\epsilon,\nu) Z^J_i(\{k_{\perp i}\};\{l_{\perp i}\},\epsilon,\eta,\nu) \nn \\
S_{i}(\{l_{\perp i}\};\{l_{\perp i}^\prime\},\eta,\nu,\epsilon) &=& \int_{\perp(i)}  \int_{\perp(i)}  Z^S_i[\{l_{\perp i}\};\{k_{\perp i}\},\epsilon,\eta,\nu] S_{i}[\{k_{\perp i}\};\{k_{\perp i}^\prime\},\nu]Z^S_i[\{k^\prime_{\perp i}\};\{l^\prime_{\perp i}\})\epsilon,\eta,\nu]. \nn \\
\eea
where the left hand sides are bare quantities which have poles in $\eta$. Note that there is $\epsilon$ dependence
in the renormalized quantities because these objects are not IR safe.
The integrations are defined by
\beq
  \int_{\perp(A)} \equiv \frac{(-i)^A}{A!}\int \prod_{a=1,A} \frac{[d^{d^\prime}k_{1\perp}^a]}{(k_{1\perp}^{a})^2}
  \deltabar^{d^\prime}(\sum_{a=1,A} k_{\perp }^a-q_\perp).
\eeq

 The anomalous dimensions 
are  defined by imposing
\beq
\nu \frac{d}{d\nu} J_{i}(\{l_{\perp i}\},\epsilon,\eta,\nu)=0,
\eeq
and  are then given by
\beq
\gamma_J^{(i)}= -(\nu \frac{d}{d\nu} { Z}^J_i)\otimes({{Z}}_i^{J})^{-1},
\eeq
Due to our choice of normalization in the convolution eq.(\ref{conv}), the $Z$ factor has units of plus two.

\section{The Rapidity Renormalization Group and the Regge Trajectory}\label{S11}

Let us calculate the leading order running of the $S_{(1,1)}$, which will yield the Regge trajectory.
This correction is down by a factor of $\alpha_Q$ relative to the Glauber contribution.  There is only one
diagram to calculate in the EFT, the so-called ``eye-graph", which if opened up into the full theory would
correspond to the soft graph topologies corresponding to  vacuum bubble, box and cross box graphs. 
The  flower graph  also contributes at this order but does not contain any rapidity divergences.
 The same can be said for the scalar vacuum bubble.
 To calculate
the anomalous dimensions we are only interested in the rapidity divergent term which is given by
\bea
\label{eye}
\begin{gathered}
\scalebox{.7}{
\begin{tikzpicture}
\begin{feynman}
\vertex (p3) at (1.5, 1.5);
\vertex (p2) at (-1.5, 1.5);
\vertex (p1) at (-1.5, -1.5);
\vertex (p4) at (1.5, -1.5);
\vertex [label= \(n\)] at (-1.5, 1);
\vertex [label= \(\bar{n}\)] at (-1.5, -1.5);
\vertex (g1) at (-0, 1.5);
\vertex (g2) at (-0, -1.5);
\vertex (s1) at (-0,.622);
\vertex (s2) at (-0, -.622);
\diagram*{
(p2)--[scalar,  line width = 0.3mm](g1)--[ scalar,  line width = 0.3mm](p3),
(p4)--[scalar,  line width = 0.3mm](g2)--[ scalar,  line width = 0.3mm](p1),
(g1)--[scalar, red, line width = 0.3mm](s1),
(g2)--[scalar, red, line width = 0.3mm](s2),
(s2)--[SGreen, gluon,looseness = 1.75, line width=0.3mm, half right](s1)--[SGreen, gluon,looseness = 1.75, line width=0.3mm, half right](s2),
};
\end{feynman}
\filldraw[red] (-0,1.5) ellipse (0.6mm and 1.2mm);
\filldraw[red] (-0,-1.5) ellipse (0.6mm and 1.2mm);
\filldraw[red] (-0,.622) ellipse (0.6mm and 1.2mm);
\filldraw[red] (-0,-.622) ellipse (0.6mm and 1.2mm);
\end{tikzpicture}
}
\end{gathered}
&=& -i\frac{\kappa^4s^2\,w^2 }{8 \pi \eta }(3-2\epsilon)q_\perp^2\int \frac{[d^{d^\prime }k_{1\perp}]}{k_{1\perp}^2(k_{1\perp}-q_\perp)^2}\nn \\
&=&i\frac{\kappa^4s^2\,w^2}{32 \pi^2 \eta }(3-2\epsilon)B[1,1]
\l( \frac{-t}{\bar\mu^2}\r)^{-\epsilon}.
\label{Soft Eye}
\eea

Since there is only one Glauber exchanged the renormalization is multiplicative, as opposed to convolutive.    We then can write
\beq
S^B_{(1,1)}= \tilde Z^S_{(1,1)} S_{1,1}^R,
\eeq
Here we have introduced $\tilde Z$ as the (dimensionless) multiplicative renormaliation factor.  The anomalous dimension in this
case will also be written  as $\tilde \gamma_{(1,1)}$, since the RRG is multiplicative.


Recalling that that at leading order $S_{(1,1)}=2 it$, and that two factors of $\kappa s/2$ get absorbed into the $J$'s
we find 
\beq
\tilde{Z}^S_{(1,1)}=\frac{\kappa^2 t\,w^2}{16 \pi^2 \eta} (3-2\epsilon)B[1,1]\l( \frac{-t}{\bar\mu^2}\r)^{-\epsilon}.
\eeq
which leads to the RRG equation
\beq
\nu \frac{d S^R_{(1,1)}}{d\nu}= -\tilde{\gamma}^S_{(1,1)}(t) S_R,
\eeq
with $\tilde{\gamma}^S_{(1,1)}$ being given by
\begin{equation}
\tilde{\gamma}_{(1,1)}^S =- \frac{\kappa^2 t}{16\pi^2}(3-2\epsilon)B[1,1]\l( \frac{-t}{\bar\mu^2}\r)^{-\epsilon}.
\end{equation}

We way then identify $\omega_G(t)=-\frac{1}{2} \tilde{\gamma}^S_{(1,1)}(t)$ as the graviton Regge trajectory \footnote{Note that there is an additional factor of  $1/2$   because the trajectory is defined as $M \sim (s/-t)^\omega$.}
\begin{equation}
    \omega_G(t) = \frac{\kappa^2 t}{32\pi^2}(3-2\epsilon)B[1,1]\l( \frac{-t}{\bar\mu^2}\r)^{-\epsilon}=\frac{\kappa^2 t}{16\pi^2}\left(-\frac{3}{\epsilon}  + 3\log\frac{-t}{\mu^2} + 2 + O(\epsilon)\right). \label{Regge Trajectory}
\end{equation}
The divergence is purely IR in nature.

We can compare is this leading order Regge trajectory found in the literature.
Our results agree with those given in \cite{Bartels:2012ra}, for the physical non-local piece. The only other result that we are aware of for the Regge trajectory was given
in \cite{Melville:2013qca}. The $\log(t)$ coefficient seems to disagree with ours result, but  the result in \cite{Melville:2013qca}
has dependence on both $t$ as well as an impact parameter  $z$ (the transverse separation between the Wilson lines), so it's not clear how to  compare.
\subsection{The Systematics of the Regge Trajectory}

The Regge trajectory is defined as the IR divergent anomalous dimensions of $S_{(1,1)}$ which  is  not physical.
Nonetheless, it is of considerable theoretical interest. In this paper we have calculated the leading order trajectory
which sums terms of the form $\frac{t}{M_{pl}^2} \log(\frac{s}{-t})$. 
For this to be a sysematic resummation we would need  $\frac{t}{M_{pl}^2} \log(\frac{s}{-t})\sim 1$, at least for  Einstein Gravity,  since the existence of counter-terms starting at order $(\frac{t}{M_{pl}^2})^6$ 
will dominate higher order terms in the re-summation if the criteria above is not met.  
Thus in principle one could sum sub-leading logs until one reaches the accuracy of $\alpha_Q^^3$ at which point the counter-term will contribute some
in a model dependent fashion. 
The same conclusion applies to the running of higher dimension
soft operators (or collinear for that matter), and their subsequent BFKL type of equations. Note that for there to be any systematic controld whatsoever
we are still restricting our observable to have a large impact parameter, as discussed previously. 


\section{The Gravitational BFKL Equation}

In this section we derive the gravitational BFKL equation, which was first given for the total cross section in \cite{Lipatov:1982it}.  As shown in \cite{Gao:2024qsg}, the BFKL equation is derived in the EFT through the renormalization of $S_{(2,2)}$.  We perform this renormalization, and then we generalize this result and renormalize $S_{(N,N)}$ for arbitrary $N$. It is worth emphasizing that there is nothing special about $N=2$ other than the fact that this is the first soft operator which obeys convolutional
running. 
There are also BFKL like equations for higher $N$.
The caveats about the systematics in the previous discussion of the Regge trajectory apply here as well.

\subsection{Renormalizing $S_{(2,2)}$}\label{BFKL}

\begin{figure}
\begin{subfigure}[b]{0.31\textwidth}
\centering
\begin{tikzpicture}
\begin{feynman}
	\vertex (f1) at (1,1.5);
	\vertex (f2) at (1,-1.5);
	\vertex (p3) at (1.75, 1.5);
	\vertex (p2) at (-1.75, 1.5);
	\vertex (p1) at (-1.75, -1.5);
	\vertex (p4) at (1.75, -1.5);
	\vertex [label= \(n\)] at (-1.75, 1);
	\vertex [label= \(\bar{n}\)] at (-1.75, -1.5);
	\vertex (g1) at (-1, 1.5);
	\vertex (g2) at (-1, -1.5);
	\vertex (c1) at (0, 2.0);
	\vertex (c2) at (0, -2.0);
 	\vertex (c3) at (0, 1.5);
	\vertex (c4) at (0, -1.5);
	\vertex (s1) at (-1,.622);
	\vertex (s2) at (-1, -.622);
\diagram*{
	(f1)--[scalar, red, line width = 0.3mm](f2),
	(p2)--[scalar,  line width = 0.3mm](g1)--[scalar,  line width = 0.3mm](f1)--[scalar,  line width = 0.3mm](p3),
	(p4)--[scalar,  line width = 0.3mm](f2)--[scalar,  line width = 0.3mm](g2)--[scalar,  line width = 0.3mm](p1),
	(g1)--[scalar, red, line width = 0.3mm](s1),
	(g2)--[scalar, red, line width = 0.3mm](s2),
	(s2)--[SGreen, gluon,looseness = 1.75, line width=0.3mm, half right](s1)--[SGreen, gluon,looseness = 1.75, line width=0.3mm, half right](s2),
};
\end{feynman}
	\filldraw[red] (-1,1.5) ellipse (0.6mm and 1.2mm);
	\filldraw[red] (-1,-1.5) ellipse (0.6mm and 1.2mm);
	\filldraw[red] (1,1.5) ellipse (0.6mm and 1.2mm);
	\filldraw[red] (1, -1.5) ellipse (0.6mm and 1.2mm);
	\filldraw[red] (-1,.622) ellipse (0.6mm and 1.2mm);
	\filldraw[red] (-1,-.622) ellipse (0.6mm and 1.2mm);
\end{tikzpicture}
\end{subfigure}
\begin{subfigure}[b]{0.31\textwidth}
\centering
\begin{tikzpicture}
\begin{feynman}
	\vertex (f1) at (1,1.5);
	\vertex (f2) at (1,-1.5);
	\vertex (p3) at (1.75, 1.5);
	\vertex (p2) at (-1.75, 1.5);
	\vertex (p1) at (-1.75, -1.5);
	\vertex (p4) at (1.75, -1.5);
	\vertex [label= \(n\)] at (-1.75, 1);
	\vertex [label= \(\bar{n}\)] at (-1.75, -1.5);
	\vertex (g1) at (-1, 1.5);
	\vertex (g2) at (-1, -1.5);
	\vertex (c1) at (0, 2.0);
	\vertex (c2) at (0, -2.0);
 	\vertex (c3) at (0, 1.5);
	\vertex (c4) at (0, -1.5);
	\vertex (s1) at (1,.622);
	\vertex (s2) at (1, -.622);
\diagram*{
	(g1)--[scalar, red, line width = 0.3mm](g2),
	(p2)--[scalar,  line width = 0.3mm](g1)--[scalar,  line width = 0.3mm](f1)--[scalar,  line width = 0.3mm](p3),
	(p4)--[scalar,  line width = 0.3mm](f2)--[scalar,  line width = 0.3mm](g2)--[scalar,  line width = 0.3mm](p1),
	(f1)--[scalar, red, line width = 0.3mm](s1),
	(f2)--[scalar, red, line width = 0.3mm](s2),
	(s2)--[SGreen, gluon,looseness = 1.75, line width=0.3mm, half right](s1)--[SGreen, gluon,looseness = 1.75, line width=0.3mm, half right](s2),
};
\end{feynman}
	\filldraw[red] (-1,1.5) ellipse (0.6mm and 1.2mm);
	\filldraw[red] (-1,-1.5) ellipse (0.6mm and 1.2mm);
	\filldraw[red] (1,1.5) ellipse (0.6mm and 1.2mm);
	\filldraw[red] (1, -1.5) ellipse (0.6mm and 1.2mm);
	\filldraw[red] (1,.622) ellipse (0.6mm and 1.2mm);
	\filldraw[red] (1,-.622) ellipse (0.6mm and 1.2mm);
\end{tikzpicture}
\end{subfigure}
\begin{subfigure}[b]{0.31\textwidth}
\centering
\begin{tikzpicture}
\begin{feynman}
	\vertex (f1) at (1,1.5);
	\vertex (f2) at (1,-1.5);
	\vertex (p3) at (1.75, 1.5);
	\vertex (p2) at (-1.75, 1.5);
	\vertex (p1) at (-1.75, -1.5);
	\vertex (p4) at (1.75, -1.5);
	\vertex [label= \(n\)] at (-1.75, 1.);
	\vertex [label= \(\bar{n}\)] at (-1.75, -1.5);
	\vertex (g1) at (-1, 1.5);
	\vertex (g2) at (-1, -1.5);
	\vertex (c1) at (0, 2.0);
	\vertex (c2) at (0, -2.0);
	\vertex (s1) at (1,0);
	\vertex (s2) at (-1, 0);
    \vertex (c3) at (0, 1.5);
	\vertex (c4) at (0, -1.5);
\diagram*{
	(f1)--[scalar, red, line width = 0.3mm](s1),
	(f2)--[scalar, red, line width = 0.3mm](s1),
	(p2)--[scalar,  line width = 0.3mm](g1)--[scalar,  line width = 0.3mm](f1)--[scalar,  line width = 0.3mm](p3),
	(p4)--[scalar,  line width = 0.3mm](f2)--[scalar,  line width = 0.3mm](g2)--[scalar,  line width = 0.3mm](p1),
	(g1)--[scalar, red, line width = 0.3mm](g2),
	(s2)--[SGreen, gluon,line width=0.3mm](s1),
};
\end{feynman}
	\filldraw[red] (-1,1.5) ellipse (0.6mm and 1.2mm);
	\filldraw[red] (-1,-1.5) ellipse (0.6mm and 1.2mm);
	\filldraw[red] (1,1.5) ellipse (0.6mm and 1.2mm);
	\filldraw[red] (1, -1.5) ellipse (0.6mm and 1.2mm);
	\filldraw[red] (1,0) ellipse (0.6mm and 1.2mm);
	\filldraw[red] (-1,0) ellipse (0.6mm and 1.2mm);
\end{tikzpicture}
\end{subfigure}
\caption{Diagrams needed for the renormalization of $S_{(2,2)}$. The first two diagrams are soft eye insertions into a Glauber rung, while the third diagrams is H graph.}\label{fig3}
\end{figure}

There are only two  loop topologies which renormalize  $S_{(2,2)}$ corresponding the H graphs and eye graphs as shown in figure (\ref{fig3}). Graphs such as those involving scalar contributions to
the Glauber polarization have no rapidity divergences.  The H-graph, shown on the right hand side of figure (\ref{fig3}) with no additional Glauber rungs,   is calculated using the Feynman rule for the Lipatov vertex in Fig. (\ref{Lipatov Vertex}), and  is given by \footnote{See figure 14 of \cite{Rothstein:2016bsq} for a discussion of the momentum routing choices and the associated regularization.}
\begin{equation}
   i\mathcal{M}_H =  \frac{i\kappa^6 s^3}{8}\int\frac{[d^dk_1][d^d k_2]\,{w'}^4\left|\frac{k_1^{-} - k_1^{+}}{\nu'}\right|^{-2\eta'}\mathcal{N}(k_1,k_2)}{((k_1 - k_2)^2 + i\epsilon)\left(p_1^+ +k_2^{+} + \frac{(k_{2\perp} +  q_\perp/ 2)^2}{p_1^-} + i\epsilon \right)\left(p_2^- -k_1^{-} + \frac{(k_2 + q_\perp/ 2)^2}{p_2^+} + i\epsilon \right)D_G},
\end{equation}
where we have defined
\begin{align}
    D_G &= d_1\,d_2\,d_3\,d_4,\nonumber\\
    \mathcal{N}(k_1,k_2) &=\bigg(q_\perp^2 ((k_1 - k_2)^2 + q_\perp^2) -( d_1 + d_4)(d_2 + d_3)
    - \frac{1}{k_1^+ k^-_2}(d_1 d_4 + d_2 d_3)(d_1 + d_2 + d_3 + d_4 -2q_\perp^2)\nonumber\\
    &-\frac{1}{(k_1^+ k^-_2)^2}\big[(d_1-d_3)(d_2-d_4)(d_1 d_4 + d_2 d_3) + d_5(d_1 d_4 -d_2 d_3)(d_1-d_2-d_3+d_4)\nonumber\\
    &-d_5^2(d_1 d_4 + d_2 d_3)\big]\bigg) w^2\left|\frac{k_1^+ + k^-_2}{\nu}\right|^{-\eta},\\
    d_1 &= k_{1\perp^2},\qquad d_2 = (k_{1\perp}+q_\perp)^2,\qquad d_3 = k_{2\perp}^2,\qquad d_4 = (k_{2\perp} + q_\perp)^2,\qquad d_5 = (k_{1\perp}-k_{2\perp})^2.\nonumber
\end{align}
Here $k_1$ and $k_2$ are whats known as $n-s$ and $\bar n -s$ Glauber respectily whose momenta  scale
as $(\lambda,\lambda^2,\lambda)$ and $(\lambda^2,\lambda,\lambda)$ respectively.
The Glauber momentum $O(\lambda^2)$ and soft momenta $O(\lambda)$ regulated by $\eta^\prime$ and $\eta$ respectively.
As discussed in \cite{Gao:2024qsg}, one  integrates over the  Glauber $k_1^-$ and $k_2^+$ components of momenta,  and first expands in $\eta'$ 
prior to taking $\eta$ to zero\footnote{This throws away terms of the form $\eta^\prime/\eta$ which would lead to violations of the collapse rule (see \cite{Rothstein:2016bsq}).}.

In principle we must make a choice of $\pm i\epsilon$ in the eikonal factors of $( k^+_1\pm i\epsilon)$ and $(k^-_2\pm i\epsilon)$, but as discussed in \cite{Rothstein:2016bsq, Gao:2024qsg}, any additional contributions are removed by zero-bin subtractions  \cite{Manohar:2006nz}(which vanish), and so the result for the integral is independent of the choice. 
 Changing variables to $k_1^+ -k_2^- = \bar k^0$ and $k_1^+ +k_2^- = \bar k^3$, we can perform the $\bar k^0$ integral by contours and integrate over $\bar k^3$ to obtain
the rapidity divergent piece of the graph
\begin{align}
    i\mathcal{M}_H = -\frac{\kappa^6 s^3\,w^2}{2^6\pi\,\eta}\int \frac{[d^{d^\prime}k_{1\perp}][d^{d^\prime}k_{2\perp}]}{d_1\,d_2\,d_3\,d_4}\left( q_\perp^4 -2\frac{(d_1 d_4 + d_2 d_3)q_\perp^2}{d_5}+ \frac{d_1^2 d_4^2 + d^2_2 d^2_3}{d_5^2}\right),
    \label{H Graph Integrand}
\end{align}

Note that there is no corresponding ghost graph. The reason is that collinear soft operator is composed of gauge invariant
building blocks. Something similar happens in NRQCD \cite{Rothstein:2018dzq} as well as Glauber SCET \cite{ Rothstein:2016bsq}. This is not to say that
ghosts dont show up at higher orders.  Indeed, if we were to include a vacuum polarization in the soft or collinear
sectors themselves, we would require ghost loops to keep the theory unitary.

The other relevant topolgy is the double box  with an soft eye subgraph.
These topologies will not contribute to any classical observable as they corresponds to cross terms between the classical exponent and the quantum corrections in (\ref{amp}).
  These diagrams are simply the one loop soft eye diagram convoluted with the Glauber box diagram in $\perp$-momenta.  This is because the soft loop is insensitive to the Glauber $k^\pm\sim\lambda^2$, while soft $l^\pm\sim\lambda$.  Given the soft eye has been computed already in Eq. (\ref{Soft Eye}), we may  write down the divergent result for the sum of the two soft eye boxes as
\begin{align}
    i\mathcal{M}_{SEB} = -\frac{\kappa^6 s^3 w^2}{32\pi \eta}
    \int [d^{d^\prime}k_{1\perp}] [d^{d^\prime}k_{2\perp}]
     \frac{(k_{1\perp} + q_\perp)^2(3-2\epsilon)}
     {(k_{1\perp})^2(k_{2\perp})^2
 (k_{1\perp} + k_{2\perp} + q_\perp)^2}
    \left(\frac{\nu^2}{-(q_\perp + k^1_\perp)^2}\right)^{\eta/2} 
    + O(\eta^0).
    \label{SEB Integrand}
\end{align}
In the above, we have already performed the small Glauber $k^\pm\sim\lambda^2$ integrals.

 The factorized $O(\alpha_Q)$ matrix elements is then written as
\begin{equation}
    i\mathcal{M}_H + i\mathcal{M}_{SEB} =  J_{(2)}^{(0)}\otimes S_{(2,2)}^{(1)}\otimes\bar{J}_{(2)}^{(0)}.
\end{equation}
Expanding out the convolutions, the amplitude becomes
\begin{equation}
 J_{(2)}^{(0)}\otimes S_{(2,2)}^{(1)}\otimes\bar{J}_{(2)}^{(0)} = \frac{1}{4}\int \frac{[d^{d^\prime}k_{1\perp}] [d^{d^\prime}k_{2\perp}]}{(k_{1\perp})^{2} (k_{1 \perp} +q_\perp)^2 (k_{2 \perp})^{2} (k_{2\perp}+q_\perp)^2} J_{(2)}^{(0)}(k_{1\perp}) S_{(2,2)}^{(1)}(k_{1\perp},k_{2\perp})\bar{J}_{(2)}^{(0)}(k_{2 \perp}).
 \end{equation}
 Since $J^{(0)}_{(2)}$ and  $J^{(0)}_{(2)}$ are independent of the transverse momentum (from their definition in Eq.( \ref{Jn}), we can extract $S_{(2,2)}^{(1)}$ from the amplitudes in Eqs. (\ref{H Graph Integrand}) and (\ref{SEB Integrand}).  The bare \footnote{We will drop the $B$ superscript from here on.  The bare objects dependence  upon $\epsilon$ and $\eta$  will be made explicit.} one loop soft function then given by 
 \begin{equation}
     S_{(2,2)}^{(1)}(k_{1\perp},k_{2\perp},\eta) = \frac{-8w^2}{\eta}\left[\frac{\kappa^2}{8\pi}K_{\text{GR}}(k_{1\perp},k_{2\perp})+\deltabar^{d-2}(k_{1\perp}-k_{2\perp})k_{2\perp}^2(k_{2\perp}+q_\perp)^2(\omega_G(k_{1\perp}) +\omega_G(k_{1\perp} +q_\perp))\right] ,
 \end{equation} 
  where $\omega_G$ is the Regge trajectory given in Eq. (\ref{Regge Trajectory}),
 and $K_{GR}$ is the convolutional kernel
 \begin{equation}
     K_{\text{GR}}(k_{1\perp}, k_{2\perp}) = \left( q_\perp^4 -2q_\perp^2\frac{(k_{2\perp}^2 (k_{2\perp}-q_\perp)^2 + (k_{1\perp}-q_\perp)^2 k_{2\perp}^2)}{(k_{1\perp}-k_{2\perp})^2}+ \frac{(k_{1\perp}^4 (k_{2\perp}-q_\perp)^4 + (k_{1\perp}-q_\perp)^4 k_{2\perp}^4}{(k_{1\perp}-k_{2\perp})^4}\right).
 \end{equation}
 The leading RRGE is
 \begin{equation}
     \nu\frac{\partial}{\partial \nu}S_{(2,2)}(k_{1\perp},k_{2\perp}) = \frac{1}{2}\int\frac{[d^{d^\prime}p_\perp]}{p_\perp^2(p_\perp -q_\perp)^2}\left(\gamma_{(2,2)}(k_{1\perp}, p_\perp)S_{(2,2)}(p_\perp,k_{2\perp}) + S_{(2,2)}(k_{1\perp},p_\perp)\gamma_{(2,2)}( p_\perp,k_{2\perp})\right).
 \end{equation}
 Using the result of Eq. (\ref{treeS}) for $S_{(2,2)}^{(0)}$, we can then extract the anomalous dimension $\gamma_{(2,2)}$ at leading order:
 \begin{equation}
     \gamma_{(2,2)}^{(1)}(k_{1\perp},p_\perp) = \frac{\kappa^2}{4\pi}K_{\text{GR}}(k_{1\perp}, p_\perp) + 2\deltabar^{d-2}(k_{1\perp}-p_\perp)p_\perp^2(p_\perp-q_\perp)^2(\omega_G(p_\perp) + \omega_G(p_\perp -q_\perp)).
     \label{gamma22}
 \end{equation}
 This rapidity RGE reproduces the gravitational analogue of the BFKL equation, given by Lipatov  \cite{Lipatov:1982it}, 
 in his Eq. (80).
 It is interesting to compare this anomalous dimension to the one computed in QCD.  There, one has\cite{Gao:2024qsg}
\begin{align}
    \gamma_{(2,2),\text{YM}}^{A_1A_2;B_1B_2} = &4\alpha_s f^{A_1 B_1C}f^{A_2B_2C} K_{\text{YM}}(k_{1\perp}, k_{2\perp})\\ 
    &+ 2\delta^{A_1B_1}\delta^{A_2 B_2}\deltabar^{d-2}(k_{1\perp}-k_{2\perp})k_{2\perp}^2(k_{2\perp}-q_\perp)^2(\alpha_R(k_{2\perp}) + \alpha_R(k_{2\perp} -q_\perp)),\nonumber
\end{align}
where $\alpha_R$ is the gluon Regge trajectory, and the QCD kernel is given by
\begin{equation}
K_{\text{YM}}(k_{1\perp}, k_{2\perp}) = q_\perp^2 + \frac{k_{2\perp}^2(q_\perp-k_{1\perp})^2 + k_{1\perp}^2(k_{2\perp}-q_\perp)^2}{(k_{2\perp}-k_{1\perp})^2}
\end{equation}
The QCD and gravity anomalous dimensions have obvious structural similarities, in that they are both the sums of a kernel representing a gluon/graviton exchange and a Reggeization term on each Glauber exchange.  Quite remarkably, there is also a relation between the convolutional kernels.  Specifically, one has
\begin{equation}
    K_{\text{GR}}(k_{1\perp},k_{2\perp}) = \left(K_{\text{YM}}(k_{1\perp}, k_{2\perp})\right)^2 + \text{scaleless},
\end{equation}
where the ``$+\, \text{scaleless}$" means terms which lead to scaleless integrals in the convolutions and thus vanish.  It has long been appreciated that there exists a double copy relation between the QCD and gravitational Lipatov vertices \cite{Lipatov:1982it, Lipatov:1982vv, SabioVera:2011wy}, so it is perhaps not too surprising that this extends to the emission piece of the anomalous dimension.  The authors are unaware of any previous mentions of this squaring relation in the literature, although it could have been noticed as early as \cite{Lipatov:1982it}. At higher orders the terms that are scaleless
become scaleful \footnote{We thank Vittorio del Duca, Riccardo Gonzo, Emanuele Rossi and Franscesco Alessio for pointing this out to us.}  thus breaking the direct relations between the GR and YM  BFKL kernels.

\subsection{The BFKL Equation for all Soft Functions}

\begin{figure}
\begin{subfigure}[b]{0.48\textwidth}
\centering
\begin{tikzpicture}
\begin{feynman}
	\vertex (f1) at (1,1.5);
	\vertex (f2) at (1,-1.5);
	\vertex (p3) at (3.75, 1.5);
	\vertex (p2) at (-1.75, 1.5);
	\vertex (p1) at (-1.75, -1.5);
	\vertex (p4) at (3.75, -1.5);
	\vertex [label= \(n\)] at (-1.75, 1);
	\vertex [label= \(\bar{n}\)] at (-1.75, -1.5);
	\vertex (g1) at (-1, 1.5);
	\vertex (g2) at (-1, -1.5);
	\vertex (c1) at (0, 2.0);
	\vertex (c2) at (0, -2.0);
 	\vertex (c3) at (0, 1.5);
	\vertex (c4) at (0, -1.5);
	\vertex (s1) at (-1,.622);
	\vertex (s2) at (-1, -.622);
    \vertex (gn1) at (3,1.5);
    \vertex (gn2) at (3,-1.5);
    \node[dot, color = red] at (2,0);
    \node[dot, color = red] at (1.75,0);
    \node[dot, color = red] at (2.25,0);    
\diagram*{
	(f1)--[scalar, red, line width = 0.3mm](f2),
	(p2)--[scalar,  line width = 0.3mm](g1)--[scalar,  line width = 0.3mm](f1)--[scalar,  line width = 0.3mm](p3),
	(p4)--[scalar,  line width = 0.3mm](f2)--[scalar,  line width = 0.3mm](g2)--[scalar,  line width = 0.3mm](p1),
	(g1)--[scalar, red, line width = 0.3mm](s1),
	(g2)--[scalar, red, line width = 0.3mm](s2),
	(s2)--[SGreen, gluon,looseness = 1.75, line width=0.3mm, half right](s1)--[SGreen, gluon,looseness = 1.75, line width=0.3mm, half right](s2),
    (gn1)--[scalar, red, line width = 0.3mm](gn2),
};
\end{feynman}
	\filldraw[red] (-1,1.5) ellipse (0.6mm and 1.2mm);
	\filldraw[red] (-1,-1.5) ellipse (0.6mm and 1.2mm);
	\filldraw[red] (1,1.5) ellipse (0.6mm and 1.2mm);
	\filldraw[red] (1, -1.5) ellipse (0.6mm and 1.2mm);
	\filldraw[red] (-1,.622) ellipse (0.6mm and 1.2mm);
	\filldraw[red] (-1,-.622) ellipse (0.6mm and 1.2mm);
    \filldraw[red] (3,1.5) ellipse (0.6mm and 1.2mm);
    \filldraw[red] (3,-1.5) ellipse (0.6mm and 1.2mm);
\end{tikzpicture}
\end{subfigure}
\begin{subfigure}[b]{0.48\textwidth}

\centering
\begin{tikzpicture}
\begin{feynman}
	\vertex (f1) at (1,1.5);
	\vertex (f2) at (1,-1.5);
	\vertex (p3) at (3.75, 1.5);
	\vertex (p2) at (-1.75, 1.5);
	\vertex (p1) at (-1.75, -1.5);
	\vertex (p4) at (3.75, -1.5);
	\vertex [label= \(n\)] at (-1.75, 1.);
	\vertex [label= \(\bar{n}\)] at (-1.75, -1.5);
	\vertex (g1) at (-1, 1.5);
	\vertex (g2) at (-1, -1.5);
	\vertex (c1) at (0, 2.0);
	\vertex (c2) at (0, -2.0);
	\vertex (s1) at (1,0);
	\vertex (s2) at (-1, 0);
    \vertex (c3) at (0, 1.5);
	\vertex (c4) at (0, -1.5);
    \vertex (gn1) at (3,1.5);
    \vertex (gn2) at (3,-1.5);
    \node[dot, color = red] at (2,0);
    \node[dot, color = red] at (1.75,0);
    \node[dot, color = red] at (2.25,0);
\diagram*{
	(f1)--[scalar, red, line width = 0.3mm](s1),
	(f2)--[scalar, red, line width = 0.3mm](s1),
	(p2)--[scalar,  line width = 0.3mm](g1)--[scalar,  line width = 0.3mm](f1)--[scalar,  line width = 0.3mm](p3),
	(p4)--[scalar,  line width = 0.3mm](f2)--[scalar,  line width = 0.3mm](g2)--[scalar,  line width = 0.3mm](p1),
	(g1)--[scalar, red, line width = 0.3mm](g2),
	(s2)--[SGreen, gluon,line width=0.3mm](s1),
    (gn1)--[scalar, red, line width = 0.3mm](gn2),
};
\end{feynman}
	\filldraw[red] (-1,1.5) ellipse (0.6mm and 1.2mm);
	\filldraw[red] (-1,-1.5) ellipse (0.6mm and 1.2mm);
	\filldraw[red] (1,1.5) ellipse (0.6mm and 1.2mm);
	\filldraw[red] (1, -1.5) ellipse (0.6mm and 1.2mm);
	\filldraw[red] (1,0) ellipse (0.6mm and 1.2mm);
	\filldraw[red] (-1,0) ellipse (0.6mm and 1.2mm);
    \filldraw[red] (3,1.5) ellipse (0.6mm and 1.2mm);
    \filldraw[red] (3,-1.5) ellipse (0.6mm and 1.2mm);
\end{tikzpicture}
\end{subfigure}

\caption{Prototypical diagrams needed to renormalize $S_{(N+1, N+1)}$.  The diagram on the left is the $N+1$-rung Glauber box with a soft eye insertion, and the diagram on the right is the multi-rung H diagram.  The soft graviton exchange can be between any two Glauber rungs, and the soft eye can similarly be inserted into any individual rung. The $H$ graph contribution to $S_{(2,2)}$  has no additional Glauber rungs.}
\label{fig4}
\end{figure}

We now extract the one loop anomalous dimensions of $S_{(N+1,N+1)}$ for arbitrary $N$.  There is a very limited class of diagrams which can contribute: $N$-Glauber boxes with a soft eye insertion on one rung, or $N$-Glauber boxes with a graviton exchanged between two rungs, i.e. the H diagram with additional Glauber rungs.  We may write the contribution of the amplitude then as
\begin{equation}
    \sum_{j>k}i\mathcal{M}_H^{jk} + \sum_{j}i\mathcal{M}_{SEB}^j =  J_{(N+1)}^{(0)}\otimes S_{(N+1,N+1)}^{(1)}\otimes\bar{J}_{(N+1)}^{(0)},
\end{equation}
where $\mathcal{M}_H^{jk}$ denotes a graviton exchange between Glauber rungs $j$ and $k$, and $\mathcal{M}_{SEB}^j$ denotes an insertion of the soft eye on rung $j$.  Adding additional Glauber rungs does not complicate the calculation of the diagrams, since, as discussed above, the soft loops are insensitive to the Glauber $k^\pm$.  Each additional Glauber loop, beyond the first, adds a factor of $i\kappa^2 \frac{s}{4}[d^{d^\prime}k_{i\perp}]/k_{i\perp2}^2$, as well an additional factor  of $-\frac{1}{(N+1)}$  that arises, from performing the Glauber lightcone integrals.  The result for  $\mathcal{M}_{SEB}^{j}$ is then
\begin{align}
    i\mathcal{M}^j_{SEB} =&\frac{(-i)^{N+1}\kappa^{4 + 2N}s^{2 + N} w^2}{2^{2N + 3}\pi\,\eta (N+1)!}\int\left(\prod_{m=1}^{N+1}\frac{[d^{d^\prime}k_{m\perp}]}{k_{m\perp}^2}\right)\deltabar^{d^\prime}\left(\sum_{m=1}^{N+1}k_{m\perp}-q_\perp\right)\label{Multi Eye}\\ &\times\int\frac{d^{d^\prime}l_{\perp}\,k_{j\perp} ^4(3-2\epsilon)}{ l_{\perp}^2\,(k_{j\perp}- l_{\perp} )^2}.\nonumber
    \end{align}
  The multi-rung H graph may similarly be computed as
\begin{align}
    i\mathcal{M}_H^{jk} &= \frac{(-i)^{N+1}\kappa^{4 + 2N}s^{2 + N} w^2}{2^{3 + 2N}\pi\eta\,(N+1)!}\int\left(\prod_{m=1}^{N+1}\frac{[d^{d^\prime}k_{m\perp}]}{k_{m\perp}^2}\right)\deltabar^{d^\prime}\left(\sum_{m=1}^{N+1}k_{m\perp}-q_\perp\right)\label{Multi H}\\
    &\times\int\frac{[d^{d^\prime}l_{\perp}]}{l_{\perp}^2(l_{\perp}-k_{j\perp}-k_{k\perp})^2} K(k_{j\perp},k_{k\perp};l_\perp,\bar  l_{\perp}\equiv l_{\perp} - k_{j\perp}-k_{k\perp}),\nonumber
\end{align}
where $K$ is given by

\begin{equation}
    K(k_j, k_k; \ell, \bar \ell) = \left( (k_j +k_k)^4 -2(k_j +k_k)^2\frac{(k_j^2 \bar \ell ^2 + k_k^2 \ell^2)}{(k_j-\ell)^2}+ \frac{(k_j^4 \bar \ell^4 + k_j^4 \ell^4)}{(k_j-\ell)^4}\right).
    \label{K}
\end{equation}

The amplitude in terms of the convolutions is given by
\begin{align}
J_{(N+1)}^{(0)}\otimes S_{(N+1,N+1)}^{(1)}\otimes\bar{J}_{(N+1)}^{(0)} &= (-1)^{N+1}\frac{\kappa^{2N+2 }s^{N + 2}}{2^{2 + 2N}(N+1)!^2}\int\left(\prod_{m=1}^{N+1}\frac{[d^{d^\prime}k_{m\perp}]}{k_{m\perp}^2}\right)\left(\prod_{n=1}^{N+1}\frac{[d^{d^\prime}\ell_{n\perp}]}{\ell_{n\perp}^2}\right)\\
&\times S_{(N+1,N+1)}^{(1)}(\{k_{m\perp}\};\{\ell_{n\perp}\})\deltabar^{d^\prime}\left(\sum_{m}k_{m\perp}-q_\perp\right)\deltabar^{d^\prime}\left(\sum_{n}l_{n\perp}-q_\perp\right),\nonumber
\end{align}

Comparing the sum of Eqs. (\ref{Multi Eye}) and (\ref{Multi H}), we can obtain $S^{(1)}_{(N+1, N+1)}$:
\begin{align}
    S^{(1)}_{(N+1, N+1)} &= \frac{4i^{N+1}(N+1)!w^2}{\eta}\bigg[\sum_{i<j}\frac{\kappa^2}{8\pi}K(k_{i\perp},k_{j\perp};\ell_{i\perp},\ell_{j\perp})\prod_{p\neq i,j} \ell_{p\perp}^2\deltabar^{d^\prime}(\ell_{p\perp}-k_{p\perp})\nonumber\\
     &+\sum_j \ell^2_{j\perp} \omega_G(\ell_j)\prod_{p\neq j} \ell_{p\perp}^2\deltabar^{d^\prime}(\ell_{p\perp}-k_{p\perp}) \bigg].
\end{align}
 The leading RRGE is then given by
 \begin{align}
     \nu\frac{\partial}{\partial \nu}S_{(N+1, N+1)}(\{k_{i\perp}\},\{\ell_{\perp_i}\}) &= -\int_{\perp(N+1)}\bigg(\gamma^{(1)}_{(N+1, N+1)}(\{k_{i\perp}\};\{\ell'_{j\perp}\})S_{(N+1, N+1)}(\{\ell'_{i\perp}\};\{\ell_{j\perp}\})\\
     &+ S_{(N+1, N+1)}(\{k_{i\perp}\};\{\ell'_{j\perp}\})\gamma_{(N+1, N+1)}(\{\ell'_{i\perp}\};\{\ell_{j\perp}\})\bigg).\nonumber.
 \end{align}
 Recalling the definition of  $S_{(N+1, N+1)}^{(0)}$
 \beq
S_{(i,j)}^{(0)}(l_{i \perp};l_{i\perp}^\prime)=2 \delta_{ij}i^j j!  \prod_{a=1}^j l_{i \perp}^{\prime 2}\prod_{n=1}^{j-1} \deltabar^{d^\prime}(l_{n \perp}-l_{n \perp}^\prime)
\eeq

  we find the anomalous dimension is given by
 \begin{align}
 \label{gamma}
     \gamma_{(N+1,N+1)} &= -i^{N+1}(N+1)!\bigg[\sum_{i<j}\frac{\kappa^2}{8\pi}K(k_{i\perp},k_{j\perp};\ell_{i\perp},\ell_{j\perp})\prod_{m\neq i,j} \ell_{m\perp}^2\deltabar^{d-2}(\ell_{m\perp}-k_{m\perp})\\
     &\qquad \qquad +\sum_j\omega_G(\ell_j)\prod_{m\neq j} \ell_{m\perp}^2\deltabar^{d-2}(\ell_{m\perp}-k_{m\perp}) \bigg].\nonumber
 \end{align}
 A few comments are in order.  Firstly, we note that although it appears that the anomalous dimension might be imaginary for even $N$, this is somewhat illusory, as the factor of $i^{N+1}$ drops out in the convolution.  This is also the case with the overall factor of $(N+1)!$.  Secondly, we note that this does return $\gamma_{(2,2)}$ in Eq. (\ref{gamma22}) when setting $N = 1$.  To see this, we apply $\perp$ momentum conservation to set $k_2 = q-k_1$ and $\ell_2 = q-\ell_1$.  This also reproduces $\gamma_{(1,1)}$ after setting $N = 0$.  We simply drop the terms involving $K$ since there is no convolution at the one Glauber level, and we have 
 \begin{equation}
     \gamma_{(1,1)} = i q_\perp^2 \omega_G(q_\perp).
 \end{equation}
 The reason for the discrepancy of a factor of $i q_\perp^2$ between this anomalous dimension and the Regge trajectory computed in Section \ref{S11} is that this factor comes from the convolution for $S_{(1,1)}$, and in Section \ref{S11} this factor has been absorbed into the anomalous dimension, as the convolution is trivial.  For $N\geq 2$, this cannot be consistently done, and so the factor from the convolution has been pulled out.  Lastly, we mention that the anomalous dimension is symmetric under $k_{i\perp}\leftrightarrow \ell_{i\perp}$.  This is not obvious given the definition of the kernel $K$ in Eq. (\ref{K}).  Under the support of the $\perp$ delta-functions in the convolutions, one can see that $\gamma_{(N,N)}$ is indeed symmetric.
\section{Extracting the Classical Logs}

\subsection{The 3PM Classical Log}\label{3PMLog}
As per our power counting  discussion the first classical logs that can appear are at 3PM (two loop) order since
we are looking for contributions that scale as $\alpha_C=  G^2st$ relative to the leading order Glauber exchange which starts at $O(G)$. The relevant logs can be extracted from the classical piece of the anomalous dimensions of the
soft function. 
At each PM order there will be one classical log.
We could equally as well calculate them from the collinear piece. By working in the EFT we can considerably reduce the amount of effort it takes to extract the log since we only need to calculate the $1/\eta$ pole, moreover to get the log (at any PM order) we
never need to calculate more than a one loop diagrams. 
The price to be paid is the need to iteratively solve the RRG equations to the necessary order.
At $(2n+1)$ order we need to iterate the $n-1$ times, so that there is no need to solve 
the RRG at all at 3PM.
  
  The eikonal form of the amplitude is given explicitly by
\begin{equation}
    \left(1 + i\Delta_Q\right)e^{i\delta_{\text{Cl}}}-1 = i\tilde{\mathcal{M}}(s,b),
\end{equation}
where $\tilde{\mathcal{M}}(s,b)$ is the Fourier transform of the amplitude,

As prevsiously mentioned there exist terms in the series expansion of   $\tilde{\mathcal{M}}(s,b)$ that scale classically which
arise from mixing between quantum and super-classical terms. However, these terms are easily discarded at the beginning
of the calculation as they are guaranteed not to contribute to the classical phase.  To see this explicitly we may consider a graph with  quantum loops with 
any number of  Glauber enhancements  that contributes to the amplitude at the classical level. Its Fourier transfrom will be equal to
the product of the Fourier transform of the purely quantum piece and of  (possibly iterated) Glauber box with
a symmetry factor.
This term will cancel with the aforemention mixed terms in eq(\ref{M}).
 For example, consider a purely quantum contribution   which
is down by a factor of $(\frac{t}{M_{pl}^2})^n$
$I(k)$. To bring it up to classical scaling we need  $n$ Glaubers.  
Peforming the light cone integrals  generates a factor of $1/(n+1)!$ and
the Fourier transform then just leads to
the products $\frac{1}{(n+1)!}I(b) \delta_0(b)^n$. We compare this to the cotribution which arises from  expanding out the exponential to order $\delta_0(b)^n$. The difference in the combinatorial
factors $1/(n+1)$ is compensated for by the fact that in the diagram we may insert $I(k)$ in any of $n+1$ places. 
The general rule that we need not worry about enhanced quantum corrections is violated by any quantum insertion
which gives non-trivial dependence on the Glauber light cone momentum, as this spoils the factorization in impact
parameter space. As an example of this are the power suppressed operators, mentioned in section \ref{power}.

Thus to get the 3PM log we need only calculate the $H$ graph rapidity divergent contribution which is given at one loop by
\begin{align}
    i\mathcal{M}_H^{(\log)} = -\log\left(\frac{\nu^2}{-t}\right)\,2G^3 s^3\left(\frac{\bar{\mu}^2}{-t}\right)^{2\epsilon}\left(-\frac{6-4\epsilon}{3}B(1,1)B(1,1 + \epsilon) + B(1,1)^2\right),
\end{align}

using  (\ref{M}) and (\ref{phase}) which leaves for the 3PM classical log in impact parameter space
\begin{align}
    \delta^{(2,\log)}_{\text{Cl}} 
    &= i \log\left(\nu^2 b^2 \right) \frac{G^3s^2}{b^2}\frac{\left(\bar{\mu}^2b^2\right)^{3\epsilon}}{\pi^{1-\epsilon}2^{4\epsilon}}\frac{\Gamma(1-3\epsilon)\Gamma(-\epsilon)^3}{3\Gamma(2\epsilon)}\bigg(\frac{3\Gamma(-\epsilon)\Gamma(1 + \epsilon)^2}{\Gamma(-2\epsilon)^2}-2\frac{(3-2\epsilon)\Gamma(1 + 2\epsilon)}{\Gamma(-3\epsilon)}\bigg),\nonumber\\
    & =i \log(s b^2)\frac{4 G^3 s^2}{b^2}\left(-\frac{1}{\epsilon} + 2\right)\frac{\left(\bar{\mu}^2b^2\right)^{3\epsilon}}{\pi^{1-\epsilon}2^{4\epsilon}} + O(\epsilon).
\end{align}
This reproduces the  result \cite{DiVecchia:2020ymx,DiVecchia:2021bdo}.  As a cross-check, this also reproduces the $O(\epsilon^3)$ of the eikonal phase given in \cite{Henn:2019rgj, Boucher-Veronneau:2011rlc,DiVecchia:2019kta} for $N= 8$ supergravity.  However, our result at order $\epsilon^4$ seems to disagree with the ``possible guess'' made for this term. 
Note the $Log(s)$ comes from the fact that to eliminate all of the large logs from the collinear sector, and place them into the soft sector, we choose $\nu^2=s$.

The phase is imaginary indicating that it is a consequence of real radiation. At next order (5PM) the leading
log will be real since it will arise from $S_{(3,3)}$ which has an additional Glauber, each of which generates a factor of $i$.
This pattern will continue as $N$ is increased.

\beq
i\pi^{\epsilon-2} 2^{-8 \epsilon} (G^5 s^3) b^{10 \epsilon-4} \exp (9 \gamma  \epsilon) \left(\frac{3 }{8 \pi ^4 \epsilon^2}-\frac{3 }{\pi ^4 \epsilon}-\frac{3 \zeta (3)}{4 \pi ^4}+\frac{5 }{32 \pi ^2}+\frac{15 }{2 \pi ^4}\right)  \log ^2(s);
\eeq


\subsection{Extracting Classical Logs to any PM Order}\label{Classical Logs Any Order}

This procedure may be generalized to extract classical logarithms at any PM order by solving the rapidity RGEs for higher Glauber soft functions.  
To see this
consider the $(2N+1)$PM term.  This contribution to the amplitude will scale as 
\begin{equation}
    \mathcal{M}^{(2N+1)\text{PM}} \sim \frac{Gs^2}{t}\alpha_{\text{C}}^{N}\sim\frac{Gs^2}{t}(Gs)^N\alpha_Q^N.
\end{equation}
Since each Glauber loop generates an enhancement of $\sim s G$, a classical term will generally involve $N$ Glauber loops and $N$ soft loops.  To 
obtain the $(2N+1)$PM term, we then need to calculate the $N$-loop correction to $S_{(N+1,N+1)}$, as this is the only operator in the EFT that has the appropriate number of $s/M_{pl}$ enhancements. That is, we only need to consider one of the soft operators at each order in the $PM$ expansion.  As a concrete example, we have already computed the one loop correction to $S_{(2,2)}$, which gave the 3PM correction to the amplitude.  To calculate the log  at 5PM, it seems that we need the two loop correction to $S_{(3,3)}$.
However we can get that log indirectly via the RRG.  By computing the one soft  loop correction to $S_{(N+1,N+1)}$, we can extract the lowest-order anomalous dimension and write down the leading RRGE.  The solution of this equation generates a series of logs in powers of $\alpha_Q \log(s)$, and so by picking out the $N$th order term in the series, we have selected the classical log generated by the RRG.  Moreover, this tells us that the $(2N+1)$PM contribution will generically contain $\log^N(s)$.  At 3PM, we see this with the single $\log(s)$, and at 5PM we can expect the logarithmic term to be a $\log^2(s)$.  These logs predicted by the one loop RRG's will also be the leading logs at each PM order.  Rapidity anomalous dimensions are independent of $\nu$ and therefore of $\log(s)$, so the RRG can only generate a single power of $\log(s)$ at each order.  An $m$-loop diagram can then at best generate an $\alpha_Q^m$ correction to the anomalous dimension; any $\alpha_Q^m\log^m(s)$ terms must then be predicted by the one loop RRG.  We may then predict the classical $\alpha_Q^N \log^N(s)$ contribution to $S_{(N+1, N+1)}$ just through solving the one loop RRGE. To get sub-leading logs at a given order we need to
calculate the two loop anomalous dimension but the order of necessary iterations is one less. To avoid having to
subtract out quantum interference terms we simply only include the classical contribution to the anomalous dimensions
as we did in the case of $3PM$. However, at
higher orders we would expect to have to
include sub-leading Glauber operators (as  discussed in section \ref{power}) to
reproduce the subleading logs.

We should mention that if we are interested in the classical problem of scattering objects with typical
size $r$, then this  scale introduces a  new set of
logarithms of the ratio $r/b$. In our theory, the scale $r$ fits into the hierarchy as follows
\beq
s\gg M^2_{pl}\gg 1/r^2 \gg t. 
\eeq
This scale shows up as a matching scales in the problem, the relevant log will be an RG and not an RRG log. The associated counter-term will correspond to
a higher dimensional operator of the form $\phi^\star \phi (E,B)^n$, where $(E,B)$ are the electric and magnetic pieces of the Weyl curvature \cite{Cheung:2020sdj,Bern:2020uwk,Haddad:2020que,AccettulliHuber:2020dal,Ivanov:2022qqt,Jakobsen:2023pvx}.

\section{Conclusions and Future Directions}

We have presented an effective field theory which is valid  for massless particles in the (super-Planckian) Regge regime.
To avoid sensitivity to the UV completion of GR we restrict ourselves to observables which get no contributions from, uncontrolled,
 local interactions.
 By utilizing a factorization theorem we have shown how to systematically resum large rapidity logs for the 
 scattering of massless particles. We have calculated the one loop graviton Regge trajectory, the BFKL equation 
 as well as the classical rapidity log at 3PM that is a consequence of radiation losses. The factorization theorem
 makes manifest the all orders form of the series.  At $2N+1$ order in the PM expansion one generates 
 a series of Logs starting at $\log^{N-1}$ down to $\log{N}$. The logs have complex/real coefficients  for $N$ even/odd. 
 This is a consequence of the fact that each Glauber loop gives an additional factor of $i$.
 The leading classical Log at each order can be calculated by utilizing the one loop anomalous dimensions shown in
 Eq.(\ref{gamma}) and by iterating the RRGE $N-1$ times. The next to leading logs will follow from the two loop
 anomalous dimensions and so on.

 While in this paper we have only considered massless particles, the leading logs we have calculated will
 also apply to the case of massive particles, as the log follows from the soft function which is insensitive
 to the partonic masses. 
 As discussed in the appendix, the couplings of soft gravitons to collinear particles is universal, and therefore the soft sector is independent of the particle species being scattered.  Furthermore, any logs computed via the RRGE will then be universal as well.  This can be seen explicitly via the equality of the 3PM eikonal phase in the high-energy limit computed in various gravitational theories with both massive and massless scalars and various degrees of supersymmetry \cite{Amati:1990xe, DiVecchia:2020ymx, DiVecchia:2021bdo} (see also \cite{Parra-Martinez:2020dzs}).
 In a future publication we will extend the formalism to the case of massive partons
 with $s\gg m \gg M_{pl}$.  We expect that other simplifications will arise once one accounts for
 unitarity constraints. In QCD is has been shown that unitarity imposes very strong constraints on the
 structure of the anomalous dimensions \cite{Rothstein:2023dgb}. In particular, by considering amplitudes of definite signature
 it was shown that anomalous dimensions (including Regge trajectories) are related to  cut amplitudes.
 Moreover, the full anomalous dimensions (including both the Regge pole and cut pieces) of the two Glauber operator anti-symmetric octet operator, can be determined from the anomalous dimension of the single Glauber exchange operator\cite{Rothstein:2024fpx}. We expect similar simplifications to arise in gravity.

 \vskip1in 
\noindent {\bf Acknowledgments:} 
The authors benefited from discussions with  Iain Stewart and Chris White, Ted Jacobson and Gabriele Veneziano.
This work is supported by the US Department of Energy (HEP) Award DE-SC0013528. IZR would like to thank the  Erwin-Schrödinger-Institute for Mathematics and Physics
for hospitality. Some of this work was done while at KITP UCSB during  the``What is Particle Theory" Workshop.

\appendix

\section{SCET Gravity Operator Building Blocks}
\label{AppendixA}

\subsection{Gauge Symmetry}

We will always be considering the active point of view of gauge transformations where the coordinate system is fixed, 
and diffeomorphisms are maps which move points, along with the geometric structures, to other points on the manifold, leading to an equivalence class 
of  states of the system.   While the  coordinates are left invariant under such transformations, a scalar   field ($\phi(x)$) transforms such that, under a  diffeomorphsim $f$  acts 
\beq
f:  \phi(x) \rightarrow \phi^\prime(f(x)),
\eeq
or
\beq
\phi^\prime(x)= \phi(f^{-1}(x)).
\eeq
We use the term ``diff. covariant" for such transformations. In this language passive transformations simply relabel
 points,  i.e. passive transformations are not gauge
transformations.
Under an active diff. a general tensor is pushed forward (vector indices) and pulled back (co-vector indices), resulting
in a Jacobian for each index
\beq
\label{J}
J_\alpha^\beta= \frac{\partial f^{ \beta}}{\partial x^\alpha}(f^{-1}(x)).
\eeq

In the active  picture the invariance of the Einstein-Hilbert action follows by first acting with the diffeomorphism 
such that the action transforms as
\beq
\label{A}
\int d^dx \sqrt{-g(x)}R(x) \rightarrow  \int d^dx \sqrt{-g^\prime(f(x))}R^\prime(f(x)).
\eeq
The invariance of the action then follows by changing coordinates from $x$ to $f(x)$ with the determinant compensating for 
the Jacobian in the usual way since the metric components transforms under the coordinate change.
We repeat this elementary definition
because we will consider  three types of gauge symmetries which, when we have operators connecting various sectors, 
necessitates careful treatment. 

As stated in the introduction, the amplitude is factorized into soft $k_s^\mu \sim \lambda$, collilnear $k_n^\mu \sim  (1,\lambda^2,\lambda)$ and anti-collinear $k_{\bar n}^\mu \sim  (\lambda^2,1,\lambda)$ sectors.
To aid in the factorization, the large momentum support of the fields is removed
by a phase redefinition.  For instance, a collinear field is decomposed as follows
\beq
\phi= \sum_{p_+,p_\perp} e^{i p_+ x_+ \vec p_\perp \cdot \vec x_\perp} \phi_{p_+,p_\perp}(x).
\eeq
The large pieces removed from the fields are called ``labels". Doing things in this way facilitate
power counting, as all derivative $\partial$ acting on the field scale like the ``residual momentum"  $\partial  \phi_{p_+,p_\perp}(x) \sim \lambda^2$.  One then introduces the label operator $\cal P$ which acts on the fields as follows
\bea
{\bar n \cdot \cal P } \phi_{p_+,p_\perp}(x)&=& {\bar n \cdot p } \phi_{p_+,p_\perp}(x) \nn \\
{\cal P_\perp } \phi_{p_+,p_\perp}(x)&=& p_\perp \phi_{p_+,p_\perp}(x).
\eea
For every composite operator label conservation is implied. Also, the conjugate is defined via,  $ \phi_{p_+} \bar n \cdot {\cal P}^\dagger= -\bar n \cdot p\phi_{p_+}$.


In the EFT each sector, soft, collinear and anti-collinear has its own diffeomorphism invariance, each of 
which is a subset of the full gauge group. Given a map in a fixed coordinate system $x^\mu \rightarrow x^\mu+\epsilon^\mu(x)$
the scalings of the derivatives of $\epsilon$ is restricted by the scalings of the associated components of the metric which
can be either read off from the form of the two point function in a covariant gauge \footnote{ Non-covariant gauge fixings complicates power counting. A classic example of this arise in HQET where if one naively chooses the gauge $v\cdot A=0$, then one might conclude that the heavy quark does not interact\cite{Manohar:2000dt}.}, or by imposing consistency with the Ward identity \cite{Okui:2017all} such that for the collinear sectors 
we have the scaling relation (\ref{pc})
\beq
h_n^{\mu \nu} \sim \frac{p_n^\mu p_n^\nu}{ \lambda}.
\eeq
Imposing that the gauge transformation does not destroy this scaling generates constraints on the momentum support of the gauge
parameters $\epsilon^\mu(x)$  such that
\beq
\label{colle}
\partial^\nu \epsilon_n^\mu +\partial^\mu \epsilon_n^\nu \sim \frac{p_n^\mu p_n^\nu}{\lambda}.
\eeq
so that we may further conclude that 
\beq
\epsilon_n^\mu \sim  \frac{p_n^\mu}{ \lambda},
\eeq
for a collinear gauge transformation.

It's important to keep in mind that these $\sim$ relations are only meant to equate scalings in $\lambda$.
All dimensions are in units of $\sqrt{s}$.
For soft gauge transformations all of the components of the gauge field scale as $\lambda$ so that  we have the constraint
\beq
\label{softe}
\partial_\mu \epsilon_s^\nu\sim  p^s_\mu \epsilon^\nu_s \sim \lambda,
\eeq
and $\epsilon_s \sim \lambda^0$.


We are interested in  building a diff. invariant Glauber operator of the form
\beq
S_{G}= \int d^4x O_n O_s O_{\bar n}.
\eeq
We can make it manifestly invariant if we make all of the operators invariant (i.e. not just covariant) with respect
to their respective symmetry group.  Clearly covariance would not be sufficient since the coordinate transformation
necessary to return the action to its original form will shift the arguments of the operators and violate momentum scaling.
Thus it would seem that  we should find operators within each sector which are diff. {\bf in}variant. We will see however, that the soft sector will
present challenges that imposing invariance under large gauge transformations will not allow for this.

\subsection{Gauge Invariant Building Blocks}

\subsection{The Collinear Sector}
In this section we discuss the gauge invariant operator building blocks for scalar and graviton collinear operators 
starting with a scalar field. 
This is accomplished (see \cite{Donnelly:2015hta}) by considering a (codimension one null hypersurface) ``platform" at minus null infinity (where diffs are no longer gauge redundancies but global symmetries) and
shooting out a geodesic in a direction $\mu=\kappa$ orthogonal to the   hypersurface which  is coordinatized by $\tilde x^\mu$. 
We assume that the points $x$ and $\tilde x$ are in the same convex neighborhood such that there is unique geodesic which
connects the two point. 
One then labels the points in the bulk as 
\beq
x^\mu= \tilde x^\mu+ \hat x ^\kappa s + X_n^\mu(s).
\eeq
where $X^\mu(-\infty)=0$.  $\hat x ^\kappa$ is the unit vector orthogonal to the platform. $X_n^\mu$ is a map from the platform to th space-timebulk.
The end-point of the geodesic in the bulk is taken at $s=0$. In flat space $X_n^\mu(0)=0$ and the subscript is there to remind us
that it is built from collinear fields. 
$X^\mu$ is a geometric object which is invariant under passive diffs. i.e. it pick out a point on the manifold, which have physical significance. e.g. it would be the position of a local observer who travelled on a rocket ship (free fall) from the platform.

 We may then  define a gauge invariant  quantity 
\beq
\chi(\tilde x^\mu =x^\mu- X_n^\mu(0)).
\eeq
$x^\mu- X_n^\mu(0)=\tilde x^\mu$   which is defined at infinity and thus a gauge invariant quantity.
 Under an active diff the shift in $x^\mu$  will be cancelled by the shift in $X_n$. Each point in the bulk has a unique gauge invariant image at infinity for small
 gauge transformations, i.e. those which dont act as global transformations at infinity.


Iterating the geodesic equation to  second order    we find
\begin{equation}
\label{X}
    X_n^\mu =  -\frac{1}{-\bar{n}\cdotp\mathcal{P}^2}\Gamma^{(1)\mu}_{++} -\frac{1}{-\bar{n}\cdotp\mathcal{P}^2}\Gamma^{(2)\mu}_{++} -\frac{1}{-\bar{n}\cdotp\mathcal{P}^2}\left(2\Gamma^{(1)\mu}_{\nu+}\frac{1}{-i\bar{n}\cdotp\mathcal{P}}\Gamma^{(1)\nu}_{++} + \left(\mathcal{D}^n_\nu\Gamma^{\mu(1)}_{++}\right)\frac{1}{-\bar{n}\cdotp\mathcal{P}^2}\Gamma^{(1)\nu}_{++}\right) + O(h_n^3).
\end{equation}
In the above, $\Gamma^{(1)}$ and $\Gamma^{(2)}$ are the one and two graviton terms in the Christoffel symbols. 
 $\mathcal{D}^n_\mu$ is the operator \footnote{The label operator which acts on the $O(\lambda^0)$ momentum obeys the following equivalence  ${\cal P}= i \partial $.}
\begin{equation}
    \mathcal{D}^n_\mu = -i\frac{n_\mu}{2}\bar{n}\cdotp\mathcal{P}  -i\mathcal{P}^\perp_\mu + \frac{\bar{n}_\mu}{2}n\cdotp  \partial,
\end{equation}
and  $\mathcal{D}^{\bar n}_\mu$ is defined analogously. The inverse  of the label operator in coordinate space operating on a general field is defined via 
\beq
\frac{1}{-i\bar n \cdot {\cal P}}  \phi = \int_{n \cdot x}^\infty \phi (x^\prime) dn \cdot x ^\prime,
\eeq
which is important to keep in mind when considering the action of large gauge transformations.
The exponentiation of these terms leads to the gravitational Wilson line for which there is no known closed form expression
to all order in the $\kappa$, unlike the QCD case.

One then  defines  the  translation operator $V_n^{-1}$  via
\begin{equation}
    V_n^{-1} = 1 + X_n^\mu \mathcal{D}^n_\mu + \frac12 X_n^\mu X_n^\nu  \mathcal{D}^n_\mu  \mathcal{D}^n_\nu+ ...,
\end{equation}
and the invariant building block is given by
\begin{align}
    \chi_n &= \left[V_n^{-1}\phi_n\right],\nonumber\\
\end{align}
and is  diff. invariant order by order in metric perturbations.  
 To leading order in the metric,
\begin{align}
\label{wilson}
    \chi_n &= \phi_n + \frac{\kappa}{2}\left[ \frac{1}{-i\bar{n}\cdotp\mathcal{P}}\left(h_{n+\mu} - \frac{\mathcal{D}_\mu^n h_{n++}}{-2i\bar{n}\cdotp\mathcal{P}}\right)\right] \mathcal{D}^{n \mu}\phi_n + O(\kappa^2).
\end{align}  
Notice that the dangerous polarization $h_{++}$ which scales as $1/\lambda$ is accompanied by a factor $k^2 \sim \lambda^2$. 
The operators $\cal P$ only act within the brackets.
 The second term will include pieces that do not involve an eikonal propagator.
The source of this behaviour can be traced   back to  the full theory  contributions (the second figure in the top line of figure 7)  arising from the $\sqrt{g}$ as well as from 
expanding the inverse metric in the full theory action,  when one does the explicit matching (as opposed to solving for the Wilson line via
the geodesic equation).
Physically, this is expected since the Wilson line is responsible for ensuring that   the diff. invariance of the full theory is inherited
by the effective theory.

 To build a invariant non-singlet tensor  \cite{Komar:1958ymq}  we need to additionally include a tensor to counter-act the transformation in eq.(\ref{J}).
One then defines 
  the gravitational Wilson line $\hat W_n$, whose action on  a  tensor $T$ is
\begin{equation}
\label{action}
    \left[\hat W_n^{-1}T\right]^{A_1...A_n}_{B_1...B_m} = {W_n}_{\mu_1'}^{\,\,\,A_1}...{W_n}_{\mu_n'}^{\,\,\,A_n}{(W_n^{-1})}^{\nu_1'}_{\,\,\,B_1}...{(W_n^{-1})}^{\nu_m'}_{\,\,\,B_m}\left(V_n^{-1}T^{\mu'_1...\mu'_n}_{\nu'_1...\nu'_m}\right).
\end{equation}
where we have defined

\begin{equation}
    {W_{n\nu}}^{~\!\!\!A}(\tilde x) = \mathcal{D}^n_\nu( X_n(\tilde x))^A.
\end{equation}
is the tangent map associated with $X^\mu$
and  $\hat W_n$  converts a singlet into a covariant tensor such that 

 \begin{equation}
\label{action}
    \left[\hat W_n T\right]_{\nu_1...\nu_m}^{\mu_1...\mu_n} = V_n ({W_n}^{-1})^{ \mu_1}_{\,\,\,A_1}...({W_n}^{-1})^{\mu_n}_{\,\,\,A_n}({W_n})_{\nu_1}\,^{B_1}...({W_n})_{\nu_m}^{\,\,B_m}\left(T^{A_1...A_n}_{B_1...B_m}\right).
\end{equation}
Here we have introduced some new notation, namely indices with capital Roman letters are active diffemorphism singlets, i.e. ``platform indices" so that
the platform coordinates are labelled as $\tilde x^A$.
Notice that we have written $X$ with a Roman index because we are mapping a point in the bulk back to the platform i.e. we are considering 
\beq
(X_n)^A(x) = \tilde x^A (x),
\eeq
which maps the bulk point back to the platform.
$X^A$ are four space-time scalar functions, i.e. the index $A$ does not transform under a diff.  Thus  only the $\nu$ index in  ${W_{n\nu}}^{~\!\!\!A}(\tilde x) = \mathcal{D}^n_\nu( X_n(\tilde x))^A$ will transform under a diff. \footnote{ At this point the fact that $X^A$ is determined by a geodesic fired from the platform is irrelevant.
We can simply declare that there exists a map from the bulk to the boundary and it will transform like a collection of scalars. However, this would introduce
new structure (like a Stuckelberg field if its dynamical) or would break general covariance. We avoid this complication by defining $X$ geometrically so that
it only depends upon the metric and not some new additional structures.}
so that
\beq
\label{Tr}
W_{n\nu}^{~\!\!\!A}(\tilde x) \rightarrow \frac{\partial x^{ \prime \mu }}{\partial x^\nu } W_{n\mu}^{~\!\!\!A}(\tilde x),
\eeq
\beq
(W_{n}^{-1})^\mu_{~\!\!\!A}(\tilde x) \rightarrow \frac{\partial x^\mu}{\partial x^{\prime\nu } } (W_{n}^{-1})^\nu_{~\!\!\!A}(\tilde x).
\eeq

The only $x$ dependence is through the metric which fixes the geodesic path and hence the image of the map. To determine how $X$ transforms we use
 $\Gamma_{++}^\mu=\frac{1}{2\kappa}(h^\mu_{n+,+}-\partial^\mu h_{n++})$ such  that under a diff. \footnote{This is an passive diff. but the net effect of 
an active diff. is equivalent.} $x^\mu\rightarrow x^\mu+\xi^\mu$,  at leading order we have,
\beq
\delta \Gamma_{++}^\mu= \partial_+^2 \xi^\mu,
\eeq
such that, using (\ref{X}), $\delta X^\mu= \xi^\mu$ and thus $\delta (W^{-1}_n)^{^\mu}_{n~B}= \partial_B \xi^\mu$.
Thus under a diff.  $(W_n^{-1})^{^\nu}_{~B}(\tilde x)$ transforms 
according to
\beq
(W_n^{-1})^{^\nu}_{~B} \rightarrow (W_n^{-1})^{^\nu}_{~B}+ \partial_B \xi^\nu+O(\kappa,\xi^2).
\eeq
i.e. its transforms by a jacobian under the greek (bulk) indices as dictated by (\ref{Tr}).
Notice that the light-cone reference vectors $n^\mu$ and $\bar n^\mu$ are flat space reference vectors that
are treated as singlets under diffs so from here on we will write them with  latin indices $n^A$, $\bar n^B$.

We define the  graviton building block via
\begin{align}
    \eta_{AB} + \frac{\kappa}{2}\mathfrak{h}_{nAB} &= \left[\hat W_n^{-1} g_n\right]_{AB},
    \label{Collinear Metric}
\end{align}
where 
$g_{n\mu \nu} = \eta_{\mu \nu} +\frac{\kappa}{2} h_{n\mu \nu}$ is the collinear metric. 
At linearized order we have 
 \begin{align}   
    \mathfrak{h}_{nAB}^{\,\perp\perp}&= \delta_A^{\mu_\perp} \delta_B^{\nu_\perp}(h_{n\mu\nu} -\frac{\mathcal{P}^\perp_\mu}{\bar{n}\cdotp \mathcal{P}}h_{n\nu+} -\frac{\mathcal{P}^\perp_\nu}{\bar{n}\cdotp \mathcal{P}}h_{n\mu+}  +\frac{\mathcal{P}^\perp_\mu\mathcal{P}^\perp_\nu}{\bar{n}\cdotp\mathcal{P}^2}h_{n++} )+ O(\kappa).
    \label{hpp}
\end{align}  
which all comes from the Jacobian factors, as the translational piece is quadratic in the field.
The transverse components of the metric have been picked out and scales as $O(\lambda)$.   In addition we have  \cite{Beneke:2021aip}  $ \mathfrak{h}_{n++}= \mathfrak{h}_{n+\mu}=0$
due to the tethering of the field to infinity \footnote{This is equivalent to working in light cone gauge in YM where the collinear Wilson lines vanish.}, as may be checked by using eqs.  (\ref{action})  and (\ref{X}).




In SCET, once the operator building blocks have been fixed, writing down the relevant Glauber operators is relatively simple in that 
each building block scales with a non-zero power of $\lambda$, but in gravity this is no longer the case.
  Every insertion of $\mathfrak{h}_n$   beyond the first is accompaniied by a factor of $\kappa$, 
  so the combination $\kappa \mathfrak{h}_n \sim \frac{\sqrt{t}}{{M_{pl}}}| \sim \alpha_Q^{1/2}$
  Thus  one can insert an arbitrary number of graviton blocks into an operator without changing the $\lambda$ scaling or mass-dimension.  
  In other words, we may consider any polynomial in $\mathfrak{h}_{nAB}$ as a building block.
At first this may not seem like a big concern since, as discussed in the introduction,  beyond two loops  in 
$\alpha_Q$  we run into model dependent counter-term contributions of the same size. Nonetheless we might be interested say in calculating in $N=8$, where it is perhaps true that there are no counter-terms \footnote{Presently $N=8$ is known to be finite up to five loops \cite{Bern:2018jmv}.}, in which case the calculation would remain predictive to all orders in $\alpha_Q$.

We will in fact show that, constraints from diffeomorphism invariance will fix the functional dependence of operators on   $\mathfrak{h}_n$ since in the full theory metric perturbations  arise either as a full metric tensor (i.e. $\eta+ h$) or inverse metric tensor; therefore, the graviton building blocks can only come in the combinations
\begin{align}
    \mathfrak{g}_{nAB} &\equiv \eta_{AB} + \frac{\kappa}{2}\mathfrak{h}_{nAB},\nonumber\\
    (\mathfrak{g}_{n}^{-1})^{AB} &\equiv \eta^{AB} -\frac{\kappa}{2}\mathfrak{h}^{AB} + O(\kappa^2).
\end{align}
Thus in the absence of counter-terms beyond the Einstein-Hilbert action, we will still maintain predictive power.

We also know that in the full theory we build invariant operators out of covariant derivatives. Thus we expect the
effective theory operators to be built from invariant forms of the connection.
  With this in mind we introduce the function of  the building block $\mathfrak{h}_{nAB}$
\begin{equation}
    {\mathfrak{B}_n}^{A}_{B} =( \frac{2}{\kappa}\frac{\bar{n}^D}{\bar{n}\cdotp \mathcal{P}}\left[\frac{1}{2}(\mathfrak{g}_n^{-1})^{AC}\left(\mathcal{D}_D\mathfrak{g}_{n C B} + \mathcal{D}_B\mathfrak{g}_{n CD} -\mathcal{D}_C\mathfrak{g}_{nD B} \right)\right]).
\end{equation}
This  is the simply the $+$ component of the Levi-Civita symbol built out of gauge-invariant metric building blocks.
Then using the fact that $\mathfrak{h}_{+\mu}=0$ we have.
\begin{equation}
    {\mathfrak{B}_n}^{A}_{B} = \frac{2}{\kappa}\frac{1}{\bar{n}\cdotp\mathcal{P}}(\mathfrak{g}_n^{-1})^{AC}\bar{n}\cdotp\mathcal{P}\mathfrak{g}_{nC B}.
\end{equation}

The $\mathfrak{B}_n$'s have the same $\lambda$ scaling and mass dimension as the $\mathfrak{h}_n$'s, and differ only at $O(\kappa)$.  They also inherit the useful properties of the $\mathfrak{h}_n$ building blocks.  In particular, ${\mathfrak{B}_n}^+_A ={\mathfrak{B}_n}^A_+ =0$, which follows from $\mathfrak{h}_{n+\mu} = 0$. 
   We will
see that this operator (thought of as a function of $\mathfrak{h}_{nAB}$) will naturally arise in the matching.

\subsection{The Soft Sector}
In the soft sector, we define the  soft Wilson lines $S_n$, which are defined similarly to Eq. (\ref{action}):
\begin{equation}
    \left[S_n^{-1}T\right]^{A_1...A_n}_{B_1...B_m} = {S_n}_{\mu_1'}^{\,\,\,A_1}...{S_n}_{\mu_n'}^{\,\,\,A_n}{S_n}^{\nu_1'}_{\,\,\,B_1}...{S_n}^{\nu_m'}_{\,\,\,B_m}\left(Z_n^{-1}T^{\mu'_1...\mu'_n}_{\nu'_1...\nu'_m}\right),
\end{equation}
with
\begin{eqnarray}
\label{Xs}
    Z_n^{-1} &=& 1 + X_{Sn}^\mu (-i\mathcal{P}_{S\mu}) + \frac12 X_{Sn}^\mu X_{Sn}^\nu  (-i\mathcal{P}_{S\mu}  )(-i\mathcal{P}_{S\nu})+ ...,\nonumber\\
    {S_n}^{~A}_{\nu} & = &(-i\mathcal{P}_{S\nu}) (x+X_{Sn})^{A},\label{Soft Wilson Lines}\\
    X^\mu_{Sn} &=&  -\frac{1}{-n\cdotp\partial_S^2}\Gamma^{(1)\mu}_{--} -\frac{1}{n\cdotp\partial_S^2}\Gamma^{(2)\mu}_{--} -\frac{1}{n\cdotp\partial_S^2}\left(2\Gamma^{(1)\mu}_{\nu-}\frac{1}{n\cdotp\partial_S}\Gamma^{(1)\nu}_{--} + \left(-i\mathcal{P}_{S\nu}\Gamma^{\mu(1)}_{--}\right)\frac{1}{n\cdotp\partial_S^2}\Gamma^{(1)\nu}_{--}\right) + O(h_s^3).\nonumber
\end{eqnarray}
The soft label operator $\mathcal{P}_S$ is written as
\begin{equation}
\mathcal{P}^\mu_S = \frac{\bar{n}^\mu}{2}n\cdotp i\partial_S + \frac{n^\mu}{2}\bar{n}\cdotp i\partial_S + \mathcal{P}^\mu_\perp,
\end{equation}
where all of the components  are $O(\lambda)$. Notice that when acting on collinear fields $\mathcal{P}^\mu_S$ will only
act upon the transverse momentum.

\subsubsection{Useful Wilson Line Identities}
Here we list some useful properties of the gravitational Wilson lines that will be utilized in the matching procedure.  These are presented in Appendix C of \cite{Beneke:2021aip}, but we reproduce them here for completeness.  Firstly, the Wilson lines satisfy a ``Product Rule":
\begin{equation}
    [\hat W_n\phi_1\phi_2] = [\hat W_n\phi_1][\hat W_n\phi_2],
    \label{prod}
\end{equation}
where the square brackets denote that the Wilson line operator only acts on the terms enclosed in the square brackets.
This result follows from the fact that the Wilson line is a translation operator and we may consuder the product of two tensors
an just a single tensor.
  Similarly, there is an integration by parts  (distributional) identity,
\begin{equation}
    [\hat W_n^{-1}T_1]\,T_2 = \det(W_n) \,T_1\,[\hat W_n(T_2)],
    \label{Wilson Line IBP}
\end{equation}
where $\det(W_n)$ is the determinant of the Jacobian matrix. 
This identity follows by shifting the coordinates such that $T_1$ on the l.h.s goes back to it initial position.
 These identities are obeyed in the soft sector as well.



\section{Matching the Glauber Lagrangian }\label{Glauber Lagrangian}

We begin by
presenting the matching of the Glauber operator at tree level.  The Glauber operators are constructed by considering the scattering of projectiles in two distinct rapidity sectors of $\{n,\, \bar{n},\, s\}$, and the projectiles may be taken to be either scalars or gravitons.  

\subsection{Collinear Glauber Operators}

We start by considering $n$-$\bar{n}$ scattering, collinear soft scattering will follow in a similar fashion.  We perform the matching using the same conventions as \cite{Rothstein:2016bsq}, where the external lines are chosen to be $\phi(p_2^n)  + \phi(p_1^{\bar{n}})\rightarrow \phi(p_3^n) + \phi(p_4^{\bar{n}})$, where the superscript denotes the collinear sector of each momentum.  All calculations are in the de Donder gauge, and we write the polarization tensors for $h=\pm 2$ as products of spin-1 polarization vectors on shell
\begin{equation}
    \epsilon_{\pm}^{\mu\nu}(p_i) = \epsilon_\pm^\mu(p_i)\epsilon_\pm^\nu(p_i),
\end{equation}
and  we will suppress the $\pm$ label.  To simplify notation further, we write $\epsilon^\mu(p_i) \equiv \epsilon_i^\mu$.  For the on-shell states we are considering, we also have $\epsilon_i^2 = 0$ and $p_i\cdotp \epsilon_i = 0$.  

For the chosen kinematics, momentum conservation gives $p_1  + p_2 = p_3 + p_4$.  The momentum transfer is given by $q = p_3 - p_2 = p_1 - p_4$.  For $n$-$\bar{n}$ scattering the  Glauber momentum scales as  $q^\mu\sim ( \lambda^2,\lambda^2, \lambda)$.  This then implies that the large $\sim \lambda^0$ components of the collinear momenta are conserved, giving $p_2^+ = p_3^+$ and $p_1^- = p_4^-$.  We choose to work in a frame where $q^+ = q^- = 0$, which allows us to write the momenta as
\begin{align}
    p_1^\mu &= \frac{\bar{n}^\mu}{2}p_1^- + \frac{n^\mu}{2} p_1^+  + \frac{1}{2}q_\perp^\mu,\qquad  p_2^\mu = \frac{\bar{n}^\mu}{2}p_2^- + \frac{n^\mu}{2} p_2^+  - \frac{1}{2}q_\perp^\mu,\label{FS Kinematics}\\
    p_3^\mu &= \frac{\bar{n}^\mu}{2}p_3^- + \frac{n^\mu}{2} p_2^+  + \frac{1}{2}q_\perp^\mu,\qquad  p_4^\mu = \frac{\bar{n}^\mu}{2}p_1^- + \frac{n^\mu}{2} p_4^+  - \frac{1}{2}q_\perp^\mu.\nonumber
\end{align}
The on-shell condition $p_i^2=0$ also lets us fix the small $\sim \lambda^2$ component of the momenta, $p_1^+ = p_4^+ = -q_\perp^2/p_1^-$ and $p_2^- = p_3^- = -q_\perp^2/p_2^+$.  The Mandelstam variables $s$ and $t$ may then be written in terms of these variables as
\begin{equation}
    s = p_1^- p_2^+ + O(\lambda^2), \qquad t = q_\perp^2.
\end{equation}
Note that the expression for $s$ has corrections which are subleading in the power-expansion, whereas $t$ is exact in the chosen frame of $q = q_\perp$.  We have $s\sim \lambda^0$ and $t\sim \lambda^2$, and for physical kinematics $s>0$ and $t<0$.  

\begin{figure}
\begin{subfigure}[b]{0.23\textwidth}
\subcaption{\qquad \qquad \qquad \qquad \qquad }
\centering
\scalebox{.9}{
\begin{tikzpicture}
\begin{feynman}
	\node[dot] (f1) at (0,1);
	\node[dot] (f2) at (0,-1);
    \vertex(p3) at (1.5, 1);
	\vertex (p2) at (-1.5, 1);
	\vertex (p1) at (-1.5, -1);
	\vertex (p4) at (1.5, -1);
	\vertex [label= \(n\)] at (-1.7, .75);
	\vertex [label= \(\bar{n}\)] at (-1.7, -1.25);
    \vertex [label= \(n\)] at (1.7, .75);
	\vertex [label= \(\bar{n}\)] at (1.7, -1.25);
\diagram*{
	(f2)--[gluon, line width = 0.3mm, momentum = \(q\)	](f1),
	(p2)--[ line width = 0.3mm](f1)--[  line width = 0.3mm](p3),
	(p4)--[  line width = 0.3mm](f2)--[ line width = 0.3mm](p1),
};
\end{feynman}
\end{tikzpicture}
}
\end{subfigure}
\begin{subfigure}[b]{0.23\textwidth}
\centering
\scalebox{.9}{
\begin{tikzpicture}
\begin{feynman}
	\node[dot] (f1) at (0,1);
	\node[dot] (f2) at (0,-1);
    \vertex(p3) at (1.5, 1);
	\vertex (p2) at (-1.5, 1);
	\vertex (p1) at (-1.5, -1);
	\vertex (p4) at (1.5, -1);
	\vertex [label= \(n\)] at (-1.7, .75);
	\vertex [label= \(\bar{n}\)] at (-1.7, -1.25);
    \vertex [label= \(n\)] at (1.7, .75);
	\vertex [label= \(\bar{n}\)] at (1.7, -1.25);
\diagram*{
	(f2)--[gluon, line width = 0.3mm, momentum = \(q\)	](f1),
	(p3)--[gluon, line width = 0.3mm](p2),
	(p4)--[  line width = 0.3mm](f2)--[ line width = 0.3mm](p1),
};
\end{feynman}
\end{tikzpicture}
}
\end{subfigure}
\begin{subfigure}[b]{0.23\textwidth}
\centering
\scalebox{.9}{
\begin{tikzpicture}
\begin{feynman}
	\node[dot] (f1) at (0,1);
	\node[dot] (f2) at (0,-1);
    \vertex(p3) at (1.5, 1);
	\vertex (p2) at (-1.5, 1);
	\vertex (p1) at (-1.5, -1);
	\vertex (p4) at (1.5, -1);
	\vertex [label= \(n\)] at (-1.7, .75);
	\vertex [label= \(\bar{n}\)] at (-1.7, -1.25);
    \vertex [label= \(n\)] at (1.7, .75);
	\vertex [label= \(\bar{n}\)] at (1.7, -1.25);
\diagram*{
	(f2)--[gluon, line width = 0.3mm, momentum = \(q\)	](f1),
	(p2)--[ line width = 0.3mm](f1)--[  line width = 0.3mm](p3),
	(p1)--[gluon,  line width = 0.3mm](p4),
};
\end{feynman}
\end{tikzpicture}
}
\end{subfigure}
\begin{subfigure}[b]{0.23\textwidth}
\centering
\scalebox{.9}{
\begin{tikzpicture}
\begin{feynman}
	\node[dot] (f1) at (0,1);
	\node[dot] (f2) at (0,-1);
    \vertex(p3) at (1.5, 1);
	\vertex (p2) at (-1.5, 1);
	\vertex (p1) at (-1.5, -1);
	\vertex (p4) at (1.5, -1);
	\vertex [label= \(n\)] at (-1.7, .75);
	\vertex [label= \(\bar{n}\)] at (-1.7, -1.25);
    \vertex [label= \(n\)] at (1.7, .75);
	\vertex [label= \(\bar{n}\)] at (1.7, -1.25);
\diagram*{
	(f2)--[gluon, line width = 0.3mm, momentum = \(q\)	](f1),
	(p3)--[gluon, line width = 0.3mm](p2),
	(p1)--[gluon,  line width = 0.3mm](p4),
};
\end{feynman}
\end{tikzpicture}
}
\end{subfigure}
\caption{Tree level matching for $n$-$\bar{n}$ Glauber operators.   These are the full theory diagrams with a $t$-channel pole.  For scalar-scalar scattering this is sufficient to extract the Glauber operator, but for scalar-graviton scattering, one must also include $s$- and $u$-channel graphs, as well as the 4-point contact term.  These additional contributions will be automatically accounted for
order by order in expansion parameters in the EFT given that we build operators out of gauge invariant building blocks. 
}
\label{nnb Glaubers}
\end{figure}

We now match the Glauber operators onto the tree-level graphs shown in Fig. (\ref{nnb Glaubers}).  We expand each diagram to leading power in $\lambda$, and we have
\begin{align}
 iM=   &-i\left[\frac{\kappa}{2}\bar{n}\cdotp p_2^2\right]\frac{2}{q_\perp^2}\left[\frac{\kappa}{2}n\cdotp p_1^2\right],\nonumber\\
    &-i\left[\frac{\kappa}{2}\frac{\left(\bar{n}\cdotp p_2^2  \,\epsilon_2\cdotp\epsilon_3 - \bar{n}\cdotp p_2\, p_3\cdotp\epsilon_2\,\bar{n}\cdotp\epsilon_3- \bar{n}\cdotp p_2\, p_2\cdotp\epsilon_3\,\bar{n}\cdotp\epsilon_2 + p_2\cdotp p_3\,\bar{n}\cdotp \epsilon_2\,\bar{n}\cdotp \epsilon_3\right)^2}{\bar{n}\cdotp p_2^2 }\right]\frac{2}{q_\perp^2}\left[\frac{\kappa}{2}n\cdotp p_1^2\right],\nonumber\\
    &-i\left[\frac{\kappa}{2}\bar{n}\cdotp p_2^2\right]\frac{2}{q_\perp^2}\left[\frac{\kappa}{2}\frac{\left(n\cdotp p_1^2  \,\epsilon_1\cdotp\epsilon_4 - n\cdotp p_1\, p_4\cdotp\epsilon_1\,n\cdotp\epsilon_4- n\cdotp p_1\, p_1\cdotp\epsilon_4\,n\cdotp\epsilon_1 + p_1\cdotp p_4\,n\cdotp \epsilon_1\,n\cdotp \epsilon_4\right)^2}{n\cdotp p_1^2 }\right],\nonumber\\
    &-i\left[\frac{\kappa}{2}\frac{\left(\bar{n}\cdotp p_2^2  \,\epsilon_2\cdotp\epsilon_3 - \bar{n}\cdotp p_2\, p_3\cdotp\epsilon_2\,\bar{n}\cdotp\epsilon_3- \bar{n}\cdotp p_2\, p_2\cdotp\epsilon_3\,\bar{n}\cdotp\epsilon_2 + p_2\cdotp p_3\,\bar{n}\cdotp \epsilon_2\,\bar{n}\cdotp \epsilon_3\right)^2}{\bar{n}\cdotp p_2^2 }\right]\frac{2}{q_\perp^2}\nonumber\\
    &\qquad\qquad \times \left[\frac{\kappa}{2}\frac{\left(n\cdotp p_1^2  \,\epsilon_1\cdotp\epsilon_4 - n\cdotp p_1\, p_4\cdotp\epsilon_1\,n\cdotp\epsilon_4- n\cdotp p_1\, p_1\cdotp\epsilon_4\,n\cdotp\epsilon_1 + p_1\cdotp p_4\,n\cdotp \epsilon_1\,n\cdotp \epsilon_4\right)^2}{n\cdotp p_1^2 }\right].
    \label{FT Tree}
\end{align}
The double copy relation \cite{Bern:2010ue}  are manifested in the last four lines.

One may then write down the Lagrangian for Glauber operators to match the amplitudes
\begin{align}
    \mathcal{L}_G^{ns\bar{n}} = \sum_{n,\bar{n}}\sum_{i,j} \mathcal{O}_n^i\frac{1}{\mathcal{P}_\perp^2}\mathcal{O}_S\frac{1}{\mathcal{P}_\perp^2}\mathcal{O}_{\bar{n}}^j .
    \label{Collinear operators}
\end{align}
Here $i$ and $j$ run over the particle species of the projectiles, which in this case is just scalars and gravitons.  To match onto the full-theory diagrams in Eq. (\ref{FT Tree}), where there are no soft-graviton emissions.  We find the collinear operators to be
\begin{align}
    \mathcal{O}_n^\phi &= \frac{\kappa}{2}\chi^\dagger_n\left[\frac{\bar{n}}{2}\cdotp\left(\mathcal{P}  +\mathcal{P}^\dag\right)\right]^2\chi_n(1+F[\mathfrak{h},\mathfrak{B}]),\nonumber\\
    \mathcal{O}_n^h &= \frac{\kappa}{2} \left[\left((\bar{n}\cdotp\mathcal{P}{(\mathfrak{B}_n,\mathfrak{h}_n})^A_B\right)^2 +A\left((\bar{n}\cdotp\mathcal{P}{(\mathfrak{B}_n,\mathfrak{h}_n)}^A_A\right)^2\right]+O[\mathfrak{h}_n^3]...\nonumber \\
    \mathcal{O}_S &= 2\mathcal{P}_\perp^2
    \label{Collinear Ops}
\end{align}
with $\mathcal{O}_{\bar{n}}^i$ given by swapping $n\leftrightarrow \bar{n}$.
Note that $\phi_n^\dagger \phi_n$ is NOT a collinear diff. invariant object, so that neither is $\phi_n^\dagger  W_n^{-1}W_n\phi_n$.
Thus when we write the collinear bilinear, we define $\chi^\dagger= \phi^\dagger W_n^{T}$, as $W_n/W_n^{-1}$ shift the argument by $(+/-) X^\mu$.

$A$  is an unknown constant that can be fixed by going to higher orders in the metric. This trace term vanishes on shell at leading order in $\mathfrak{h}$ \cite{Beneke:2021aip} so was not detected in the on shell matching result (\ref{FT Tree}).
Notice that for the operators we have written $(\mathfrak{B}_n,\mathfrak{h}_n)$, this is because we have only matched
at quadratic order in the field and at this order this is no distinction between $\mathfrak{h}_n$ and $\mathfrak{B}_n$.
In the next section we will show that constraints from  the full theory will fix the field to be $\mathfrak{B}_n$, that $A=-1,F=0$
and that there are no powers of the $\mathfrak{h}$ beyond quadratic.

  The Glauber Lagrangian exactly reproduces the full-theory diagrams to leading power in the $\lambda$-expansion.  Since $\chi_n\sim(\mathfrak{B}_n,\mathfrak{h}_n)\sim\lambda$, $\bar{n}\cdotp\mathcal{P}\sim\lambda^0$, the collinear operators scale as $\mathcal{O}_n^i\sim \kappa \lambda^2$ while for the soft operator  $\mathcal{O}_S = 2\mathcal{P}_\perp^2\sim \lambda^2$ for zero soft graviton emissions, and thus $\mathcal{L}_G^{ns\bar{n}}\sim \kappa^2  \lambda^2$.  
Given that the measure scales as $1/\lambda^2$ we see that $S_G \sim s/M_{pl}^2 \sim \frac{\alpha_c }{\alpha_Q}$.  
Thus for each Glauber we will need an additional quantum piece to keep the scaling classical.
  The matching and construction of the full soft operator is more involved, and  will be discussed in the next section.  

It is interesting to see how the Ward indentities are satsified in the EFT given that  when we matched we did not bother
with the contact interactions that in the full theory are required to ensure they are satisified. The terms which
would arise from contact terms in the full theory, arise in the EFT Ward identity from the last term in \ref{hpp} which
end up killing the factor of $1/q_\perp^2$ in the ampltiude.

\subsection{Soft-Collinear Glauber Operators}

We can perform analogous matching calculations for $n$-$s$ and $\bar{n}-s$ scattering. 
 We focus here on $n$-$s$ scattering, as the results for $\bar{n}s$ scattering are given simply by replacing $n\leftrightarrow \bar{n}$.  We take the $n$-$s$ scattering to be given by $\phi(p_2^n)  + \phi(p_1^s)\rightarrow \phi(p_3^n) + \phi(p_4^s)$.  The momentum transfer $q$ is defined identically as $q = p_3-p_2 = p_1 - p_4$, but $q$ now has scaling $q\sim(\lambda, \lambda^2,\lambda)$.  
   Expanding the full-theory diagrams in these kinematics, the result is identical to Eq.(\ref{nnb Glaubers}).
For the Glauber operators, we may write the Glauber Lagrangian as
\begin{equation}
    \mathcal{L}^{ns} = \sum_n\sum_{i,j}\mathcal{O}_n^i\frac{1}{\mathcal{P}_\perp^2}\mathcal{O}_{Sn}^j.
\end{equation}
The collinear operators in $\mathcal{L}^{ns}$ are identical to those in $\mathcal{L}^{ns\bar{n}}$.  The soft operators meanwhile are identical to the collinear case with the replacement of collinear fields with soft fields $\chi_{Sn},\mathfrak{B}_{Sn}$ and    $\mathfrak{h}_{Sn}$.

\begin{figure}
\begin{subfigure}[b]{0.23\textwidth}
\subcaption{\qquad \qquad \qquad \qquad \qquad }
\centering
\scalebox{.9}{
\begin{tikzpicture}
\begin{feynman}
	\node[dot] (f1) at (0,1);
	\node[dot] (f2) at (0,-1);
    \vertex(p3) at (1.5, 1);
	\vertex (p2) at (-1.5, 1);
	\vertex (p1) at (-1.5, -1);
	\vertex (p4) at (1.5, -1);
	\vertex [label= \(n\)] at (-1.7, .75);
	\vertex [label= \(s\)] at (-1.7, -1.25);
    \vertex [label= \(n\)] at (1.7, .75);
	\vertex [label= \(s\)] at (1.7, -1.25);
\diagram*{
	(f2)--[gluon, line width = 0.3mm, momentum = \(q\)	](f1),
	(p2)--[ line width = 0.3mm](f1)--[  line width = 0.3mm](p3),
	(p4)--[  line width = 0.3mm](f2)--[ line width = 0.3mm](p1),
};
\end{feynman}
\end{tikzpicture}
}
\end{subfigure}
\begin{subfigure}[b]{0.23\textwidth}
\centering
\scalebox{.9}{
\begin{tikzpicture}
\begin{feynman}
	\node[dot] (f1) at (0,1);
	\node[dot] (f2) at (0,-1);
    \vertex(p3) at (1.5, 1);
	\vertex (p2) at (-1.5, 1);
	\vertex (p1) at (-1.5, -1);
	\vertex (p4) at (1.5, -1);
	\vertex [label= \(n\)] at (-1.7, .75);
	\vertex [label= \(s\)] at (-1.7, -1.25);
    \vertex [label= \(n\)] at (1.7, .75);
	\vertex [label= \(s\)] at (1.7, -1.25);
\diagram*{
	(f2)--[gluon, line width = 0.3mm, momentum = \(q\)	](f1),
	(p3)--[gluon, line width = 0.3mm](p2),
	(p4)--[  line width = 0.3mm](f2)--[ line width = 0.3mm](p1),
};
\end{feynman}
\end{tikzpicture}
}
\end{subfigure}
\begin{subfigure}[b]{0.23\textwidth}
\centering
\scalebox{.9}{
\begin{tikzpicture}
\begin{feynman}
	\node[dot] (f1) at (0,1);
	\node[dot] (f2) at (0,-1);
    \vertex(p3) at (1.5, 1);
	\vertex (p2) at (-1.5, 1);
	\vertex (p1) at (-1.5, -1);
	\vertex (p4) at (1.5, -1);
	\vertex [label= \(n\)] at (-1.7, .75);
	\vertex [label= \(s\)] at (-1.7, -1.25);
    \vertex [label= \(n\)] at (1.7, .75);
	\vertex [label= \(s\)] at (1.7, -1.25);
\diagram*{
	(f2)--[gluon, line width = 0.3mm, momentum = \(q\)	](f1),
	(p2)--[ line width = 0.3mm](f1)--[  line width = 0.3mm](p3),
	(p1)--[gluon,  line width = 0.3mm](p4),
};
\end{feynman}
\end{tikzpicture}
}
\end{subfigure}
\begin{subfigure}[b]{0.23\textwidth}
\centering
\scalebox{.9}{
\begin{tikzpicture}
\begin{feynman}
	\node[dot] (f1) at (0,1);
	\node[dot] (f2) at (0,-1);
    \vertex(p3) at (1.5, 1);
	\vertex (p2) at (-1.5, 1);
	\vertex (p1) at (-1.5, -1);
	\vertex (p4) at (1.5, -1);
	\vertex [label= \(n\)] at (-1.7, .75);
	\vertex [label= \(s\)] at (-1.7, -1.25);
    \vertex [label= \(n\)] at (1.7, .75);
	\vertex [label= \(s\)] at (1.7, -1.25);
\diagram*{
	(f2)--[gluon, line width = 0.3mm, momentum = \(q\)	](f1),
	(p3)--[gluon, line width = 0.3mm](p2),
	(p1)--[gluon,  line width = 0.3mm](p4),
};
\end{feynman}
\end{tikzpicture}
}
\end{subfigure}
\label{ns Glaubers}
\caption{Tree level matching for $n$-$s$ Glauber operators.  These are the full theory diagrams with a $t$-channel pole.   We can obtain the matching for $\bar{n}$-$s$ scattering by taking $n\leftrightarrow \bar{n}$.}
\end{figure}

 The soft fields $\chi_{Sn},\mathfrak{B}_{Sn}$ and    $\mathfrak{h}_{Sn}$  scale as $\sim\lambda$, while the soft momenta scale as $i\partial_S\sim \lambda$.  The soft operators $\mathcal{O}_{Sn}$ then scale as $\sim\kappa \lambda^4$.  Since $O_n^i\sim \kappa \lambda^2$, we find the $n$-$s$ Glauber Lagrangian scales as $\sim \kappa^2 \lambda^4$.  
 The scaling of the measure $d^4x \sim \lambda^{-3}$ for the soft Glauber operator 
 since the soft momenta are all order $\lambda$.  Therefore, the Glauber actions will scale as
\begin{align}
    \mathcal{S}_G^{ns\bar{n}} &= \int d^4 x\mathcal{L}_G^{ns\bar{n}}\sim \alpha_Q \lambda^{-2},\\
    \mathcal{S}_G^{ns} &= \int d^4 x\mathcal{L}_G^{ns}\sim \alpha_Q \lambda^{-1}.  
\end{align}
Given that the actions for the kinetic terms in the soft and collinear Lagrangians are normalized to scale as $\sim\lambda^0$, we can clearly see that the Glauber operators are enhanced, as discussed in the main body of the text.  We also see that the action for $n$-$s$ Glaubers is down by $\lambda$ compared to the $n$-$\bar{n}$ action.  However, time-ordered products of $n$-$s$ and $\bar{n}$-$s$ Glaubers have the same enhancement as $n$-$\bar{n}$ Glaubers.

\subsection{Collinear Operators to All Orders}
\label{coll}

So far we have matched the operators $\mathcal{O}_{n}^i$ and $\mathcal{O}_{Sn}^i$ at lowest order in $\alpha_Q$ and $\lambda$, we must still show that there are no higher order corrections in $\alpha_Q$.  However, it is important to recall that we  can not claim 
to be able to calculate to all orders in $\alpha_Q$, since as stressed in the introduction, higher dimensional operators which
correct Einstein gravity will pollute any calcuation at $O(\alpha_Q^3$) and beyond by unknown counter term coefficients.
If $N=8$ SUGRA were truly finite, then perhaps one can make this claim. Thus  it behooves us to see if we can indeed formulate
the theory which is valid to all orders in $\alpha_Q$ for Einstein gravity. This same comment will apply when we  write down
the all orders action for the soft part of the theory. 

Matching to all orders in the non-forward version of SCET is relatively simple since gauge symmetry fixes all of the positions of the Wilson lines.
Moreover, there are no building blocks that are not power suppressed. In the case of forward scattering we can only
fix the collinear Wilson lines by symmetry arguments, but at this time we do not have an argument that fixes the positions of the soft Wilson lines, so that
 an explicit matching must be done. In the gravitational case we have the additional problem of 
having building blocks which are not power suppressed \footnote{Compared to QCD where all building blocks scale as powers of $\lambda$, the gravitational case is complicated by the existence of powers of $M_{pl}$ in the denominator, which leads to {\it enhancements} in the Super-Planckian regime.}. Let's address these two issues in turn.

\begin{figure}
\begin{subfigure}[b]{0.23\textwidth}
\subcaption{\qquad \qquad \qquad \qquad \qquad }
\centering
\scalebox{0.8}{
\begin{tikzpicture}
\begin{feynman}
	\node[dot] (f1) at (0,1);
	\node[dot] (f2) at (0,-1);
    \node[dot] (g1) at (-0.75, 1);
    \vertex[label = right: \(n\)] (h1) at(.25,1.75);
    \vertex(p3) at (1.5, 1);
	\vertex (p2) at (-1.5, 1);
	\vertex (p1) at (-1.5, -1);
	\vertex (p4) at (1.5, -1);
	\vertex [label= \(n\)] at (-1.7, .75);
	\vertex [label= \(\bar{n}\)] at (-1.7, -1.25);
    \vertex [label= \(n\)] at (1.7, .75);
	\vertex [label= \(\bar{n}\)] at (1.7, -1.25);
\diagram*{
	(f2)--[gluon, line width = 0.3mm	](f1),
	(p2)--[ line width = 0.3mm](f1)--[  line width = 0.3mm](p3),
	(p4)--[  line width = 0.3mm](f2)--[ line width = 0.3mm](p1),
    (h1)--[gluon, line width = 0.3 mm](g1),
};
\end{feynman}
\end{tikzpicture}
}
\end{subfigure}
\begin{subfigure}[b]{0.23\textwidth}
\centering
\scalebox{0.8}{
\begin{tikzpicture}
\begin{feynman}
	\node[dot] (f1) at (0,1);
	\node[dot] (f2) at (0,-1);
    \node[dot] (g1) at (.5, 1);
    \vertex[label = right: \(n\)] (h1) at(1.5,1.75);
	\vertex(p3) at (1.5, 1);
	\vertex (p2) at (-1.5, 1);
	\vertex (p1) at (-1.5, -1);
	\vertex (p4) at (1.5, -1);
	\vertex [label= \(n\)] at (-1.7, .75);
	\vertex [label= \(\bar{n}\)] at (-1.7, -1.25);
    \vertex [label= \(n\)] at (1.7, .75);
	\vertex [label= \(\bar{n}\)] at (1.7, -1.25);
\diagram*{
	(f2)--[gluon, line width = 0.3mm	](f1),
	(p2)--[ line width = 0.3mm](f1)--[  line width = 0.3mm](p3),
	(p4)--[  line width = 0.3mm](f2)--[ line width = 0.3mm](p1),
    (h1)--[gluon, line width = 0.3 mm](g1),
};
\end{feynman}
\end{tikzpicture}
}
\end{subfigure}
\begin{subfigure}[b]{0.23\textwidth}
\centering
\scalebox{0.8}{
\begin{tikzpicture}
\begin{feynman}
	\node[dot] (f1) at (0,1);
	\node[dot] (f2) at (0,-1);
    \node[dot] (g1) at (0, 1);
    \vertex[label = right: \(n\)] (h1) at(1,1.75);
	\vertex(p3) at (1.5, 1);
	\vertex (p2) at (-1.5, 1);
	\vertex (p1) at (-1.5, -1);
	\vertex (p4) at (1.5, -1);
	\vertex [label= \(n\)] at (-1.7, .75);
	\vertex [label= \(\bar{n}\)] at (-1.7, -1.25);
    \vertex [label= \(n\)] at (1.7, .75);
	\vertex [label= \(\bar{n}\)] at (1.7, -1.25);
\diagram*{
	(f2)--[gluon, line width = 0.3mm	](f1),
	(p2)--[ line width = 0.3mm](f1)--[  line width = 0.3mm](p3),
	(p4)--[  line width = 0.3mm](f2)--[ line width = 0.3mm](p1),
    (h1)--[gluon, line width = 0.3 mm](g1),
};
\end{feynman}
\end{tikzpicture}
}
\end{subfigure}

\begin{subfigure}[b]{0.23\textwidth}
\centering
\scalebox{0.8}{
\begin{tikzpicture}
\begin{feynman}
	\node[dot] (f1) at (0,1);
	\node[dot] (f2) at (0,-1);
    \node[dot] (g1) at (0, 0);
    \vertex[label = right: \(n\)] (h1) at(1.25,0);
	\vertex(p3) at (1.5, 1);
	\vertex (p2) at (-1.5, 1);
	\vertex (p1) at (-1.5, -1);
	\vertex (p4) at (1.5, -1);
	\vertex [label= \(n\)] at (-1.7, .75);
	\vertex [label= \(\bar{n}\)] at (-1.7, -1.25);
    \vertex [label= \(n\)] at (1.7, .75);
	\vertex [label= \(\bar{n}\)] at (1.7, -1.25);
\diagram*{
	(f2)--[gluon, line width = 0.3mm	](f1),
	(p2)--[ line width = 0.3mm](f1)--[  line width = 0.3mm](p3),
	(p4)--[  line width = 0.3mm](f2)--[ line width = 0.3mm](p1),
    (h1)--[gluon, line width = 0.3 mm](g1),
};
\end{feynman}
\end{tikzpicture}
}
\end{subfigure}
\begin{subfigure}[b]{0.23\textwidth}

\centering
\scalebox{0.8}{
\begin{tikzpicture}
\begin{feynman}
	\node[dot] (f1) at (0,0.25);
	\node[dot] (f2) at (0,-1.75);
    \node[dot] (g1) at (.5, -1.75);
    \vertex[label = right: \(n\)] (h1) at(1.75,-1.25);
	\vertex(p3) at (1.5, .25);
	\vertex (p2) at (-1.5, .25);
	\vertex (p1) at (-1.5, -1.75);
	\vertex (p4) at (1.7, -1.75);
	\vertex [label= \(n\)] at (-1.75, 0);
	\vertex [label= \(\bar{n}\)] at (-1.7, -2);
    \vertex [label= \(n\)] at (1.7, 0);
	\vertex [label= \(\bar{n}\)] at (1.7, -2);
\diagram*{
	(f2)--[gluon, line width = 0.3mm	](f1),
	(p2)--[ line width = 0.3mm](f1)--[  line width = 0.3mm](p3),
	(p4)--[  line width = 0.3mm](f2)--[ line width = 0.3mm](p1),
    (h1)--[gluon, line width = 0.3 mm](g1),
};
\end{feynman}
\end{tikzpicture}
}
\end{subfigure}
\begin{subfigure}[b]{0.23\textwidth}

\centering
\scalebox{0.8}{
\begin{tikzpicture}
\begin{feynman}
	\node[dot] (f1) at (0,0.25);
	\node[dot] (f2) at (0,-1.75);
    \node[dot] (g1) at (-0.5, -1.75);
    \vertex[label = left: \(n\)] (h1) at(-1.75,-1.25);
	\vertex(p3) at (1.5, .25);
	\vertex (p2) at (-1.5, .25);
	\vertex (p1) at (-1.5, -1.75);
	\vertex (p4) at (1.7, -1.75);
	\vertex [label= \(n\)] at (-1.75, 0);
	\vertex [label= \(\bar{n}\)] at (-1.7, -2);
    \vertex [label= \(n\)] at (1.7, 0);
	\vertex [label= \(\bar{n}\)] at (1.7, -2);
\diagram*{
	(f2)--[gluon, line width = 0.3mm	](f1),
	(p2)--[ line width = 0.3mm](f1)--[  line width = 0.3mm](p3),
	(p4)--[  line width = 0.3mm](f2)--[ line width = 0.3mm](g1)--[ line width = 0.3mm](p1),
    (h1)--[gluon, line width = 0.3 mm](g1),
};
\end{feynman}
\end{tikzpicture}
}
\end{subfigure}

\caption{The matching for one collinear graviton emission.   The first two diagrams are reproduced exactly by time ordered products in the EFT and do not contribute to the matching. The last diagram involves and highly off shell line $p^2 \sim s$ which
builds up part of the Wilson line.}
\end{figure}


\subsubsection{Matching Collinear Wilson Lines}

As previously  stated, the position of the collinear Wilson lines is fixed by collinear diff. invariance. Nonetheless it is worth while to understand how
these Wilson lineas are built up.  In fig(7) we show all the relevant collinear emissions in the $n$ direction in the full theory.
The first two diagrams are trivially reproduces in the EFT via a Lagrangian insertions.
 The last two diagrams involve
off-shell (eikonal) lines that build up the Wilson line in the EFT.   The full bottom row involves the t-channel graviton
exchange has collinear scaling.   This non-anayticity is reproduced by a time ordered product in the EFT. 
After subtracting the time ordered products in the EFT 
 what is left over
in the matching is generated by the  collinear Wilson line diagram in the EFT (see section 6.1 of \cite{Rothstein:2016bsq}).

\subsubsection{Fixing the Dependence on Order One Building Blocks in Glauber Operators}

Recall that to this point we dont know whether to choose  $\mathfrak{B}$ or $\mathfrak{h}$ in eq.(\ref{Collinear Ops}).
 As discussed in Appendix \ref{AppendixA}, we can in principle freely add arbitrarily many collinear graviton building blocks $(\mathfrak{h},\mathfrak{B})$ to any operator, as these are not constrained by $\lambda$  power-counting.  
However, as we will show,  all terms at higher order in $(\mathfrak{h},\mathfrak{B})$ are fixed by symmetry.

We recall that, in isolation,  the $n$-collinear sector is equivalent to full GR. 
Thus we may match by building operators in the full theory that are diff. covariant,  and
then lifting them to their diff. invariant form in the EFT.
This means that we should match before linearizing, since the metric transforms covariantly, but the metric pertubration
does not. In this way we immediately eliminate the possibility of having polynomials in $h$ since the full metric is an irreducilbe polynomtial
on its own.
We also restrict the form of the operators by using two symmetries of the EFT. The first is the shift symmetry on the massless scalar field inhereted 
from the full theory and the other is 
RPI-III \cite{Manohar:2002fd} which  is the symmetry of the EFT under the rescaling $\bar{n}\rightarrow \alpha \bar{n}$ and $n\rightarrow 1/\alpha n$.  From the tree-matching, we can see that each collinear operator transforms as $\mathcal{O}_n\rightarrow \alpha^2 \mathcal{O}^i_n$, and so each collinear operator must have two lightcone vectors $\bar{n}$ contracted with it.  
Finally the operators must start off bilinear in the fields.
With all this, we may write down the most general form of the collinear operator \footnote{$\bar n^\nu$ is a diff. scalar so we write it as $\bar n^A$.}, at tree level matching:
\begin{equation}
    \mathcal{O}_n^i = \left[\left(\hat W_n^{-1}\right)^{\rho\sigma}_{AB}f^i_{\rho\sigma}(g_n,\grad_n, {\cal D}_\mu\phi_n)\right]\bar{n}^A\bar{n}^B,\label{Collinear Full}
\end{equation}
where $f^i_{\rho\sigma}$ has mass-dimension 3 and transforms covariantly under diffeomorphisms.    There is only one possibility for $f$ which we can write down:
\begin{equation}
    f^\phi_{\rho\sigma} = \frac{\kappa}{2}{\cal D}_\rho\phi^\dagger_n {\cal D}_\sigma \phi_n.
\end{equation}
The corresonding invariant operator is
\begin{equation}
    \mathcal{O}_n^\phi = \frac{\kappa}{2}\mid\bar{n}\cdotp \mathcal{P}\chi_n\mid^2.
\end{equation}
Now we can address the question of the function $F[\mathfrak{h},\mathfrak{B}]$ in  Eq. (\ref{Collinear Ops}).
As we have already discussed pure polynomials in $\mathfrak{h}$ are not allowed since only the combination that forms $\mathfrak{g}$ can appear.
Whereas polynomial in $\mathfrak{B}$ can not appear on dimensional grounds since 
they  whould need to be accompanied
by a factor of $\kappa^n$. However, we know from the full theory that factors of $\kappa$ only appear in combination with $h$ (i.e. within the
connection), and thus any additional factors of $\kappa$ in the matching would have to come from higher dimensional operators which are 
outside of Einstein Hilbert gravity.
Thus we may conclude that  $F=0$.  For the gravitons, there are two operators which we can write down:
\begin{equation}
    f^h_{1,\rho\sigma} = \frac{2}{\kappa}g^n_{\rho\sigma}(\grad^n)^2,\qquad  f^h_{2,\rho\sigma} =  \frac{2}{\kappa}\grad^n_\rho \grad^n_\sigma.
\end{equation}
We note that since $f_{\rho\sigma}$ is symmetric, only the symmetric piece of $\grad^n_\rho \grad^n_\sigma$ will contribute.  $f^h_1$ vanishes when plugged into Eq. (\ref{Collinear Full}), as we have
\begin{align}
   \bar{n}^A\bar{n}^B\left[\left(W_n^{-1}\right) g^n_{\rho\sigma}\right]_{AB} = \bar{n}^A\bar{n}^B(\eta_{AB} + \mathfrak{h}^n_{AB}) = 0,
\end{align}
where we have used  $\mathfrak{h}^n$, $\mathfrak{h}^n_{++} = 0$.  Thus the only non-vanishing operator we can construct is $f_2^g$ which is equivalent to the graviton operator in Eq. (\ref{Collinear operators}).  Writing out the covariant derivatives in terms of the connection, we may write the graviton operator as
\begin{align}
   \bar{n}^A\bar{n}^B  \frac{2}{\kappa}\left[\left(W_n^{-1}\right)\grad^n_\rho \grad^n_\sigma \right]_{AB}&= \frac{2}{\kappa}\left(\bar{n}\cdotp \mathcal{P} + \frac{\kappa}{2}\bar{n}\cdotp\mathcal{P}{\mathfrak{B}_n}^A_A\right)\left(\bar{n}\cdotp \mathcal{P} +\frac{\kappa}{2}\bar{n}\cdotp\mathcal{P} {\mathfrak{B}_n}^A_A\right),\nonumber\\
    &=\bar{n}\cdotp\mathcal{P}^2{\mathfrak{B}_n}^A_A + \frac{\kappa}{2}\left((\bar{n}\cdotp\mathcal{P}{\mathfrak{B}_n}^A_A\right)^2,
    \label{Collinear Graviton Operator Pre EOM}
\end{align}
where in the first line we have used ${\mathfrak{B}_n}^A_+ = 0$ and in the second line we have used the fact that $\bar{n}\cdotp\mathcal{P} = 0$ when not acting on collinear operators.  We now  invoke the equations of motion for collinear gravitons.  By acting with a Wilson line and restricting to the $++$ component, the equations of motion become
\begin{equation}
    \bar{n}\cdotp\mathcal{P}^2{\mathfrak{B}_n}^A_A  + \frac{\kappa}{2} \left((\bar{n}\cdotp\mathcal{P}{\mathfrak{B}_n}^A_B\right)^2 = -\kappa(\bar{n}\cdotp \mathcal{P}\chi_n)^2.
\end{equation}
Using this, we may remove the term in Eq. (\ref{Collinear Graviton Operator Pre EOM}) linear in $\mathfrak{B}_n$, and we obtain

\begin{eqnarray}
    \frac{2}{\kappa}\left[\left(W_n^{-1}\right)^{\rho\sigma}_{AB}\grad^n_\rho \grad^n_\sigma\right]\bar{n}^A\bar{n}^B &=& - \frac{\kappa}{2} \left[\left((\bar{n}\cdotp\mathcal{P}{\mathfrak{B}_n}^A_B\right)^2 -\left((\bar{n}\cdotp\mathcal{P}{\mathfrak{B}_n}^A_A\right)^2\right] - 2\mathcal{O}_n^\phi.
\end{eqnarray}
Despite the fact that the equations of motion introduce scalar fields, we see that this is the same scalar operator that we have already introduced.  Therefore, the graviton and collinear operators are uniquely determined to be, up to overall numerical factors,
\begin{equation}
    \mathcal{O}_n^\phi = \frac{\kappa}{2}(\bar{n}\cdot \mathcal{P}\chi_n)^2, \qquad
    \mathcal{O}_n^g = \frac{\kappa}{2} \left[\left((\bar{n}\cdot \mathcal{P} \mathfrak{B}^A_{n B})\right)^2 - \left((\bar{n}\cdot \mathcal{P} \mathfrak{B}_{nA}^A)\right)^2\right],
\end{equation}
supports the conclusion reached after  Eq. (\ref{Collinear operators}).  A similar procedure gives the $\bar{n}$-collinear as well as the $S_n$ and $S_{\bar{n}}$ operators as well.



\section{The Graviton Soft Operator to All Orders }

In this section, we shall describe the construction and matching of the gravitational mid-rapidity Glauber soft operator.  Using the observations made in the previous section, we shall show that the operator basis has a finite number of terms, and that the matching can be performed at the level of a single soft graviton emission.

\subsection{Soft Gauge Symmetry in Soft-Collinear Gravity}
\label{soft}

Let us now consider the constructing the  full soft-collinear Glauber  operator.
In so doing we are confonted with the fact that 
  the soft operator is not necessary from the point of view of gauge symmetry. Thus in \cite{Rothstein:2016bsq} the operator was build up making and ansatz for its form and then fixing
its coefficient for the matching. It has since been realized \cite{tobepublished}, that by imposing the invariance of the  action under large gauge transformations, 
 the soft operator is indeed necessary.  Each collinear operator needs to be capped by a soft Wilson line in order for the action to be invariant under the BMS group.
In YM theory this understandable in a straight forward way, but in gravity the soft wilson shifts the argument of the collinear fields upon which it operates and thus  destroys the invariance of those collinear fields under the collinear gauge symmetry. To cure this problem we will choose
the soft operator to transform covariantly, instead of invariantly. Then the coordinate transformartion needed to restore the form of the soft operator will shift
the collinear operator back to the its invariant form. The final invariance of full operator will then necessitate the inclusion of a factor of $\sqrt{-g_s}$.

More explicitly,  the action of a soft Wilson line on a  collinear invariant
\begin{equation}
   \mathcal{O}_{\bar{n}} \rightarrow\left[  S_{\bar{n}}\mathcal{O}_{\bar{n}}(x)\right].
\end{equation}
 is to translate the collinear operator from a point $x$ to the point $Y_{S\bar{n}}(x)$, where $Y_{S\bar{n}}(x)$ is related to $X_{S\bar{n}}$ (defined in \ref{Xs}) by
\begin{equation}
    X_{S\bar{n}}\left(Y_{S\bar{n}}(x)\right) = Y_{S\bar{n}}\left(X_{S\bar{n}}(x)\right) = x.
\end{equation}
The operator evaluated at $Y_{S\bar{n}}$ then transforms covariantly under soft diffeomorphisms.

Schematically, we may then decompose the soft operator as
\begin{equation}
    \mathcal{O}_S = \sum_{i}f_i(\mathcal{P}_S)S_{n}^T C_i\,\sqrt{-g_S}O_i S_{\bar{n}} g_i(\mathcal{P}_S).
\end{equation}
In the above, the operators $O_i$ are soft diffeomorphism scalars built out of covariant derivatives and soft fields, the $C_i$ are some numerical coefficients, and the functions $f_i$ and $g_i$ are scalar functions of the soft label operators.  We have also included an explicit factor of the determinant of the metric, which is required by gauge-invariance.  The Wilson line $S_n^T$ denotes the transpose, in the sense that $S_n$ acts on fields to the left, while $S_{\bar{n}}$ acts on fields to the right.  In the next section, we will discuss constraints on the functions $f_i$, $g_i$, and the operators $O_i$.


\subsection{The Basis of Soft Graviton Operators}

We now describe the construction of the soft operator basis.  These operators must have mass dimension 2 and scale as $\sim\lambda^2$, and must be consistent with soft diffeomorphism symmetry.  To make operators which are consistent with large gauge transformations, as discussed in the previous section, every term must contain one $n$ inverse Wilson line $S_n$ and one $\bar{n}$ inverse Wilson line.
Also, as discussed in the last section, each operator
 must include  a factor of the soft metric determinant, $\sqrt{-g_s}$.  We then build our operators out of soft label operators $\mathcal{P}^S_\mu$ and covariant derivatives $\grad_{S\mu}$.  The soft label operators do not transform under soft diffeomorphisms, and so they can only appear outside the Wilson line pair; similarly, the soft covariant derivatives can only appear between the two Wilson lines.

Constraints from reparameterization invariance are crucial here.
   As can be seen from their definitions in Eq. (\ref{Collinear operators}), the collinear operators each have RPI weight 2 in their respective direction.
Thus it would seem  that the soft operator should be RPI invariant on its own, which leads to an infinite number of possible tensor structures.
However, we can consider forward scattering where the incoming beams are no longer parallel and thus we have two independent
RPI transformations which must leave the operator invariant. This uniquely fixes the operator to have the form
%
\begin{equation}
    \mathcal{O}_{ns\bar{n}} = \mathcal{O}_n\frac{1}{\mathcal{P}_\perp^2}n_A n_B\mathcal{O}_S^{AB,CD}\bar{n}_C \bar{n}_D\frac{1}{\mathcal{P}_\perp^2}\mathcal{{O}}_{\bar{n}}.
    \label{Glauber RPI}
\end{equation}

Next, we have constraints from the hermiticity of the full Glauber operator.  As described in Section 6.3 of \cite{Rothstein:2016bsq}, equality of $\mathcal{L}_G$ and $(\mathcal{L}_G)^\dag$ requires the soft operator to satisfy
\begin{equation}
    (\mathcal{O}_S)^\dag = \left.\mathcal{O}_S\right|_{n\leftrightarrow \bar{n}}.
\end{equation}
This is a slight variation on the statement that there is a symmetry between the $n$ and $\bar{n}$ sectors, given that swapping $n$ and $\bar{n}$ is equivalent to taking an adjoint.  In the context of the full Glauber operator, this reduces to the usual symmetry under exchanging $n$ and $\bar{n}$. An additional constraint is that the total label momentum flowing through each term in the Glauber Lagrangian is conserved.  Therefore we have equality between $\mathcal{P}_S$ and $\mathcal{P}_S^\dag$, and we may interchange them freely.

 We find it useful then to introduce the notation
\begin{align}
    (S_n)_{\nu_1...\nu_n}^{\mu_1...\mu_n} &\equiv \det[\left(S_n^{-1}\right)^\mu_\nu]Z_n {S_n}_{\,\,\,\nu_1}^{\mu_1}... {S_n}_{\,\,\,\nu_2}^{\mu_2},\\
    (S_n^T)_{\nu_1...\nu_n}^{\mu_1...\mu_n} &\equiv {S_n}_{\,\,\,\nu_1}^{\mu_1}... {S_n}_{\,\,\,\nu_2}^{\mu_2}Z^T_n \det[\left(S_n^{-1}\right)^\mu_\nu].
\end{align}

In the above, $Z_n$ acts on all fields to the right, including the Jacobian factors, and similarly $Z^T_n$ acts on all fields to the left.  $(S_n)_{\nu_1...\nu_n}^{\mu_1...\mu_n}$ is then an inverse Wilson line in the sense of Eq. (\ref{Wilson Line IBP}), as it satisfies
\begin{equation}
    \left[S_n^{-1}T\right]^{\mu_1...\mu_n}\phi_1 = T^{\nu_1...\nu_n}\left[(S_n)_{\nu_1...\nu_n}^{\mu_1...\mu_n}\phi\right].
\end{equation}

Lastly, there are two useful identities which will be used to simplify the operator basis.  The first follows from the properties of the gauge invariant metric building blocks $\mathfrak{h}^{S_n}$, which is defined analogously to the collinear metric building blocks in Eq. (\ref{Collinear Metric}).  Using $\mathfrak{h}^{S_n}_{-\mu} = 0$, we have
\begin{equation}
    n^{\nu_1}n^{\nu_2} (S_{n}^T)^{\mu_1\mu_2,\alpha_1...\alpha_n}_{\nu_1\nu_2,\beta_1...\beta_n }g_{\mu_1\mu2} =  n^{\mu}n^{\nu}(\eta_{\mu\nu}  + \mathfrak{h}^{S_n}_{\mu\nu})\,(S_{n}^T)^{\alpha_1...\alpha_n}_{\beta_1...\beta_n} = 0.
    \label{Soft Metric Identity 1}
\end{equation}
Similarly, replacing a light cone vector $n$ with a derivative also leads to a vanishing operator,
\begin{equation}
    \mathcal{P}_S^{\nu_1}n^{\nu_2} (S_{n}^T)^{\mu_1\mu_2,\alpha_1...\alpha_n}_{\nu_1\nu_2,\beta_1...\beta_n }g_{\mu_1\mu2} =  \mathcal{P}_S^{\mu}n^{\nu}(\eta_{\mu\nu}  + \mathfrak{h}^{S_n}_{\mu\nu})\,(S_{n}^T)^{\alpha_1...\alpha_n}_{\beta_1...\beta_n} = 0,
\end{equation}
where in the final equality we have used $n\cdotp \mathcal{P}_S = 0$ when acting to the left of the Wilson lines, as soft $n\cdot k$ momenta cannot flow into $\mathcal{O}_n^i$.  

With these constraints, we can now write down a list of all possible operators that satisfy them.  There are eight such operators:
\begin{align}
    O_{1} &= \mathcal{P}_S^2(S_{n}^T)_{--}^{\mu\nu}g_{\mu\rho}g_{\nu\sigma}(S_{\bar{n}})_{++}^{\rho\sigma} + (S_{n}^T)_{--}^{\mu\nu}g_{\mu\rho}g_{\nu\sigma}(S_{\bar{n}})_{++}^{\rho\sigma} \mathcal{P}_S^2,\nonumber\\
    O_{2} &= (S_{n}^T)_{--}^{\mu\nu}g_{\mu\rho}g_{\nu\sigma}g^{\alpha\beta}\grad_{S\alpha}\grad_{S\beta}(S_{\bar{n}})_{++}^{\rho\sigma},\nonumber\\
    O_{3} & = (S_{n}^T)_{--}^{\mu\nu}R^S_{\mu\rho\nu\sigma}(S_{\bar{n}})_{++}^{\rho\sigma},\nonumber\\
    O_{4} & = \mathcal{P}_S^A(S_{n}^T)_{--A}^{\mu\,\nu\,\beta}g_{\beta\rho}g_{\nu\sigma}g_{\mu\lambda}(S_{\bar{n}})_{++B}^{\rho\,\sigma\,\lambda}\mathcal{P}_S^B,\label{Soft Operator List}\\
    O_{5} & = \mathcal{P}_S^A(S_{n}^T)_{--A}^{\mu\,\nu\,\beta}g_{\mu\rho}g_{\beta\lambda}g_{\nu\sigma}(S_{\bar{n}})_{++B}^{\rho\,\sigma\,\lambda}\mathcal{P}_S^B,\nonumber\\
    O_{6} & = \mathcal{P}_S^A\mathcal{P}_S^B(S_{n}^T)_{--AB}^{\mu\,\nu\,\gamma\lambda}g_{\gamma\lambda}g_{\mu\rho}g_{\nu\sigma}(S_{\bar{n}})_{++}^{\rho\sigma} + (S_{n}^T)_{--}^{\mu\nu}g_{\mu\rho}g_{\nu\sigma}g_{\gamma\lambda}(S_{\bar{n}})_{++AB}^{\rho\,\sigma\,\gamma\lambda} \mathcal{P}_S^A\mathcal{P}_S^B,\nonumber\\
    O_{7} & = \mathcal{P}_S^A(S_{n}^T)_{--A}^{\mu\,\nu\,\beta}g_{\mu\rho}g_{\nu\sigma}i\grad_{S\beta}(S_{\bar{n}})_{++}^{\rho\sigma} + (S_{n}^T)_{--}^{\mu\nu}g_{\mu\rho}g_{\nu\sigma}i\grad_{S\beta}(S_{\bar{n}})_{++A}^{\rho\,\sigma\,\beta}\mathcal{P}_S^A,\nonumber\\
    O_{8} & =  \mathcal{P}_S^A(S_{n}^T)_{--A}^{\mu\,\nu\,\beta}g_{\beta\rho}g_{\nu\sigma}i\grad_{S\mu}(S_{\bar{n}})_{++}^{\rho\sigma} + (S_{n}^T)_{--}^{\mu\nu}g_{\mu\beta}g_{\nu\sigma}i\grad_{S\rho}(S_{\bar{n}})_{++A}^{\rho\,\sigma\,\beta}\mathcal{P}_S^A.\nonumber
\end{align}
In the above, we are implicitly assuming the Lorentz indices are contracted with lightcone vectors as in Eq. (\ref{Glauber RPI}).  Not making this assumption can lead to several more allowed operators, as identities such as Eqs. (\ref{Soft Metric Identity 1}) would no longer apply.  This could be an important consideration when trying to construct the operator basis to subleading level, but for the current purposes it is enough to consider those in Eq. (\ref{Soft Operator List}).

\subsection{Matching}

We now perform the matching of the Wilson coefficients for the soft operator.  It will be sufficient to match at 0, 1, or 2 soft graviton emissions.  Moreover, we may perform this matching taking the external collinear projectiles to be scalars.  We can in principle replace one or both of the scalars by collinear gravitons, but we will obtain the same result for the soft operator.  This is due to the universal nature of the coupling of Glauber gravitons to either soft or collinear particles, as well as the universal eikonal coupling of soft particles.

At zero soft graviton emissions, the Glauber operator must reproduce the tree scalar-scalar amplitude given in Eq. (\ref{FT Tree}).  The soft operator in this case must reduce to $\mathcal{O}_S = 2\mathcal{P}_\perp^2$.  From their definitions in Eq. (\ref{Soft Wilson Lines}), the soft Wilson lines simply become the identity, the covariant derivative becomes $\mathcal{P}_S$, and $\mathcal{P}_S^2 = \mathcal{P}_\perp^2$ since no soft $k^\pm$ flows through the operator.  This then places the constraint of
\begin{equation}
    2 = 2 C_1 - C_2 +C_5 + 2 C_6 + 2 C_7 .
\end{equation}

\begin{figure}
\begin{subfigure}[b]{0.23\textwidth}
\subcaption{\qquad \qquad \qquad \qquad \qquad }
\centering
\scalebox{0.8}{
\begin{tikzpicture}
\begin{feynman}
	\node[dot] (f1) at (0,1);
	\node[dot] (f2) at (0,-1);
    \node[dot] (g1) at (-0.75, 1);
    \vertex[label = right: \(s\)] (h1) at(.25,1.75);
    \vertex(p3) at (1.5, 1);
	\vertex (p2) at (-1.5, 1);
	\vertex (p1) at (-1.5, -1);
	\vertex (p4) at (1.5, -1);
	\vertex [label= \(n\)] at (-1.7, .75);
	\vertex [label= \(\bar{n}\)] at (-1.7, -1.25);
    \vertex [label= \(n\)] at (1.7, .75);
	\vertex [label= \(\bar{n}\)] at (1.7, -1.25);
\diagram*{
	(f2)--[gluon, line width = 0.3mm	](f1),
	(p2)--[ line width = 0.3mm](f1)--[  line width = 0.3mm](p3),
	(p4)--[  line width = 0.3mm](f2)--[ line width = 0.3mm](p1),
    (h1)--[gluon, line width = 0.3 mm](g1),
};
\end{feynman}
\end{tikzpicture}
}
\end{subfigure}
\begin{subfigure}[b]{0.23\textwidth}
\centering
\scalebox{0.8}{
\begin{tikzpicture}
\begin{feynman}
	\node[dot] (f1) at (0,1);
	\node[dot] (f2) at (0,-1);
    \node[dot] (g1) at (0, 1);
    \vertex[label = right: \(s\)] (h1) at(1,1.75);
	\vertex(p3) at (1.5, 1);
	\vertex (p2) at (-1.5, 1);
	\vertex (p1) at (-1.5, -1);
	\vertex (p4) at (1.5, -1);
	\vertex [label= \(n\)] at (-1.7, .75);
	\vertex [label= \(\bar{n}\)] at (-1.7, -1.25);
    \vertex [label= \(n\)] at (1.7, .75);
	\vertex [label= \(\bar{n}\)] at (1.7, -1.25);
\diagram*{
	(f2)--[gluon, line width = 0.3mm	](f1),
	(p2)--[ line width = 0.3mm](f1)--[  line width = 0.3mm](p3),
	(p4)--[  line width = 0.3mm](f2)--[ line width = 0.3mm](p1),
    (h1)--[gluon, line width = 0.3 mm](g1),
};
\end{feynman}
\end{tikzpicture}
}
\end{subfigure}
\begin{subfigure}[b]{0.23\textwidth}
\centering
\scalebox{0.8}{
\begin{tikzpicture}
\begin{feynman}
	\node[dot] (f1) at (0,1);
	\node[dot] (f2) at (0,-1);
    \node[dot] (g1) at (.5, 1);
    \vertex[label = right: \(s\)] (h1) at(1.5,1.75);
	\vertex(p3) at (1.5, 1);
	\vertex (p2) at (-1.5, 1);
	\vertex (p1) at (-1.5, -1);
	\vertex (p4) at (1.5, -1);
	\vertex [label= \(n\)] at (-1.7, .75);
	\vertex [label= \(\bar{n}\)] at (-1.7, -1.25);
    \vertex [label= \(n\)] at (1.7, .75);
	\vertex [label= \(\bar{n}\)] at (1.7, -1.25);
\diagram*{
	(f2)--[gluon, line width = 0.3mm	](f1),
	(p2)--[ line width = 0.3mm](f1)--[  line width = 0.3mm](p3),
	(p4)--[  line width = 0.3mm](f2)--[ line width = 0.3mm](p1),
    (h1)--[gluon, line width = 0.3 mm](g1),
};
\end{feynman}
\end{tikzpicture}
}
\end{subfigure}
\begin{subfigure}[b]{0.23\textwidth}
\centering
\scalebox{0.8}{
\begin{tikzpicture}
\begin{feynman}
	\node[dot] (f1) at (0,1);
	\node[dot] (f2) at (0,-1);
    \node[dot] (g1) at (0, 0);
    \vertex[label = right: \(s\)] (h1) at(1.25,0);
	\vertex(p3) at (1.5, 1);
	\vertex (p2) at (-1.5, 1);
	\vertex (p1) at (-1.5, -1);
	\vertex (p4) at (1.5, -1);
	\vertex [label= \(n\)] at (-1.7, .75);
	\vertex [label= \(\bar{n}\)] at (-1.7, -1.25);
    \vertex [label= \(n\)] at (1.7, .75);
	\vertex [label= \(\bar{n}\)] at (1.7, -1.25);
\diagram*{
	(f2)--[gluon, line width = 0.3mm	](f1),
	(p2)--[ line width = 0.3mm](f1)--[  line width = 0.3mm](p3),
	(p4)--[  line width = 0.3mm](f2)--[ line width = 0.3mm](p1),
    (h1)--[gluon, line width = 0.3 mm](g1),
};
\end{feynman}
\end{tikzpicture}
}
\end{subfigure}
\begin{subfigure}[b]{0.23\textwidth}
\centering
\scalebox{0.8}{
\begin{tikzpicture}
\begin{feynman}
	\node[dot] (f1) at (0,1);
	\node[dot] (f2) at (0,-1);
    \node[dot] (g1) at (-0.75, -1);
    \vertex[label = right: \(s\)] (h1) at(.25,-1.75);
    \vertex(p3) at (1.5, 1);
	\vertex (p2) at (-1.5, 1);
	\vertex (p1) at (-1.5, -1);
	\vertex (p4) at (1.5, -1);
	\vertex [label= \(n\)] at (-1.7, .75);
	\vertex [label= \(\bar{n}\)] at (-1.7, -1.25);
    \vertex [label= \(n\)] at (1.7, .75);
	\vertex [label= \(\bar{n}\)] at (1.7, -1.25);
\diagram*{
	(f2)--[gluon, line width = 0.3mm	](f1),
	(p2)--[ line width = 0.3mm](f1)--[  line width = 0.3mm](p3),
	(p4)--[  line width = 0.3mm](f2)--[ line width = 0.3mm](p1),
    (h1)--[gluon, line width = 0.3 mm](g1),
};
\end{feynman}
\end{tikzpicture}
}
\end{subfigure}
\begin{subfigure}[b]{0.23\textwidth}
\centering
\scalebox{0.8}{
\begin{tikzpicture}
\begin{feynman}
	\node[dot] (f1) at (0,1);
	\node[dot] (f2) at (0,-1);
    \node[dot] (g1) at (0,- 1);
    \vertex[label = right: \(s\)] (h1) at(1,-1.75);
	\vertex(p3) at (1.5, 1);
	\vertex (p2) at (-1.5, 1);
	\vertex (p1) at (-1.5, -1);
	\vertex (p4) at (1.5, -1);
	\vertex [label= \(n\)] at (-1.7, .75);
	\vertex [label= \(\bar{n}\)] at (-1.7, -1.25);
    \vertex [label= \(n\)] at (1.7, .75);
	\vertex [label= \(\bar{n}\)] at (1.7, -1.25);
\diagram*{
	(f2)--[gluon, line width = 0.3mm	](f1),
	(p2)--[ line width = 0.3mm](f1)--[  line width = 0.3mm](p3),
	(p4)--[  line width = 0.3mm](f2)--[ line width = 0.3mm](p1),
    (h1)--[gluon, line width = 0.3 mm](g1),
};
\end{feynman}
\end{tikzpicture}
}
\end{subfigure}
\begin{subfigure}[b]{0.23\textwidth}
\centering
\scalebox{0.8}{
\begin{tikzpicture}
\begin{feynman}
	\node[dot] (f1) at (0,1);
	\node[dot] (f2) at (0,-1);
    \node[dot] (g1) at (.5, -1);
    \vertex[label = right: \(s\)] (h1) at(1.5,-1.75);
	\vertex(p3) at (1.5, 1);
	\vertex (p2) at (-1.5, 1);
	\vertex (p1) at (-1.5, -1);
	\vertex (p4) at (1.5, -1);
	\vertex [label= \(n\)] at (-1.7, .75);
	\vertex [label= \(\bar{n}\)] at (-1.7, -1.25);
    \vertex [label= \(n\)] at (1.7, .75);
	\vertex [label= \(\bar{n}\)] at (1.7, -1.25);
\diagram*{
	(f2)--[gluon, line width = 0.3mm	](f1),
	(p2)--[ line width = 0.3mm](f1)--[  line width = 0.3mm](p3),
	(p4)--[  line width = 0.3mm](f2)--[ line width = 0.3mm](p1),
    (h1)--[gluon, line width = 0.3 mm](g1),
};
\end{feynman}
\end{tikzpicture}
}
\end{subfigure}
\newline
\begin{subfigure}[b]{1\textwidth}
\subcaption{\hspace{1\textwidth} }\label{Lipatov Vertex}
\centering
\begin{align}
\begin{gathered}
\begin{tikzpicture}
\begin{feynman}
	\vertex (f1) at (0,1);
	\vertex (f2) at (0,-1);
	\vertex(p3) at (1.5, 1);
	\vertex (p2) at (-1.5, 1);
	\vertex (p1) at (-1.5, -1);
	\vertex (p4) at (1.5, -1);
	\vertex [label= \(n\)] at (-1.7, .75);
	\vertex [label= \(\bar{n}\)] at (-1.7, -1.25);
    \vertex [label= \(n\)] at (1.7, .75);
	\vertex [label= \(\bar{n}\)] at (1.7, -1.25);
	\vertex (s1) at (0,0);
    \vertex[label = above right: \(\textcolor{SGreen}{s}\)] (s2) at ( 1.5,0);
        \vertex[label= \(\textcolor{SGreen}{\mu,\nu}\)] at ( 1.8,-.45);
    \vertex[label = \(q\)] at (-.6,.2);
    \vertex[label = \(q^\prime\)] at (-.6,-.8);
\diagram*{
	(s1)--[scalar, red, line width = 0.3mm, momentum = \(\)](f1),
	(f2)--[scalar, red, line width = 0.3mm, momentum = \(\)](s1),
	(p2)--[scalar,  line width = 0.3mm](f1)--[ scalar,  line width = 0.3mm](p3),
	(p4)--[ scalar,  line width = 0.3mm](f2)--[ scalar,  line width = 0.3mm](p1),
    (s2)--[SGreen,gluon, line width = 0.3mm](s1),
};
\end{feynman}
	\filldraw[red] (0,1) ellipse (0.6mm and 1.2mm);
	\filldraw[red] (0, -1) ellipse (0.6mm and 1.2mm);
	\filldraw[red] (0,0) ellipse (0.6mm and 1.2mm);
\end{tikzpicture}
\end{gathered}
& =i \left[\frac{\kappa}{2}\bar{n}\cdotp p_2^2\right]\left[\frac{\kappa}{2}n\cdotp p_1^2\right]\left(\frac{\kappa}{\sqrt{2}q_\perp^2 q^{\prime2}_\perp}\right)\bigg(
2\frac{\bar{n}^\mu\bar{n}^\nu}{\bar{n}\cdotp q^2}q_\perp^{\prime2}q\cdotp(q^\prime-q) + 2\frac{n^\mu n^\nu}{n\cdotp q^2}q_\perp^{2}q^\prime\cdotp(q-q^\prime) \nonumber\\
&- 2q_\perp^{\prime2}\frac{n\cdotp q^\prime\bar{n}^\mu\bar{n}^\nu - \bar{n}^\mu q^\nu- \bar{n}^\nu q^\mu}{\bar{n}\cdotp q}- 2q_\perp^{2}\frac{\bar{n}\cdotp q\, n^\mu n^\nu - n^\mu q^{\prime\nu}- n^\nu q^{\prime\mu}}{n\cdotp q^\prime}\nonumber\\
&+ 2 (q^\mu q^{\prime\nu} + q^\nu q^{\prime\mu})  - (q^\mu + q^{\prime\mu})(n\cdotp q^{\prime}\bar{n}^\nu + \bar{n}\cdotp q\,    n^\nu) - (q^\nu + q^{\prime\nu})(n\cdotp q^{\prime}\bar{n}^\mu + \bar{n}\nonumber\cdotp q\, n^\mu) \nonumber\\
&+ (n\cdotp q^{\prime}\bar{n}^\mu + \bar{n}\nonumber\cdotp q\, n^\mu)(n\cdotp q^{\prime}\bar{n}^\nu + \bar{n}\nonumber\cdotp q \,n^\nu) - (q_\perp^2 + q_\perp^{\prime 2})(n^\mu\bar{n}^\nu + n^\nu\bar{n}^\mu)-2q\cdotp q^\prime \eta^{\mu\nu}
\bigg).\nonumber
\end{align}
\end{subfigure}
\caption{The matching for one soft graviton emission.  In (a), we show the 7 full-theory diagrams which can contribute.  In (b), we have the lone SCET diagram, which reproduces the gravitational Lipatov vertex, along with its Feynman rule.}
\end{figure}

At one soft graviton emission, we have 7 full theory diagrams which contribute.  We calculate on-shell, with arbitrary graviton polarization tensors, and soft graviton momentum $k$.  Using momentum conservation to write $k = q' - q$, the amplitude contains several momentum structures which generate matching conditions.  Several of the momentum structures generate degenerate matching conditions, and in the end the one graviton matching yields 5 constraints:
\begin{align}
    2 &= C_1 -C_5 + 2 C_6 - C_7,\nonumber\\
    0 &= C_1 -C_2 + C_7,\nonumber\\
    0 & = 4 C_1 - 4 C_2 +  C_4 +2 C_5 - 4 C_6 - 4 C_7 + C_8,\\
    8 & = 4 C_1 + C_4 -2 C_5 + 4 C_6 + C_8,\nonumber\\
    4 &= C_4- C_3.\nonumber
\end{align}
Combining this with the constraint from zero graviton emissions, we are able to fix 6 of the eight coefficients:
\begin{equation}
    C_1 = 2,\quad C_4 = 4 + C_3,\quad C_5 = 0,\quad C_6 = 1-C_2/2,\quad C_7 = 0,\quad C_8 =-2 C_2-C_3.
\end{equation}

\begin{figure}
\begin{subfigure}[b]{0.19\textwidth}
\subcaption{\qquad \qquad \qquad \qquad \qquad }
\centering
\scalebox{0.6}{
\begin{tikzpicture}
\begin{feynman}
	\node[dot] (f1) at (0,.75);
	\node[dot] (f2) at (0,-.75);
    \node[dot] (g1) at (-0.75, .75);
    \node[dot] (g2) at (.25,1.5);
    \vertex[label = right: \(s\)] (h1) at(1,1.75);
    \vertex[label = right: \(s\)] (h2) at(1,1.25);
    \vertex(p3) at (1.5, .75);
	\vertex (p2) at (-1.5, .75);
	\vertex (p1) at (-1.5, -.75);
	\vertex (p4) at (1.5, -.75);
	\vertex [label= \(n\)] at (-1.7, .5);
	\vertex [label= \(\bar{n}\)] at (-1.7, -1);
    \vertex [label= \(n\)] at (1.7, .5);
	\vertex [label= \(\bar{n}\)] at (1.7, -1);
\diagram*{
	(f2)--[gluon, line width = 0.3mm	](f1),
	(p2)--[ line width = 0.3mm](f1)--[  line width = 0.3mm](p3),
	(p4)--[  line width = 0.3mm](f2)--[ line width = 0.3mm](p1),
    (h1)--[gluon, line width = 0.3 mm](g2),
    (h2)--[gluon, line width = 0.3 mm](g2),
    (g2)--[gluon, line width = 0.3 mm](g1),
};
\end{feynman}
\end{tikzpicture}
}
\end{subfigure}
\begin{subfigure}[b]{0.19\textwidth}
\centering
\scalebox{0.6}{
\begin{tikzpicture}
\begin{feynman}
	\node[dot] (f1) at (0,.75);
	\node[dot] (f2) at (0,-.75);
    \node[dot] (g1) at (0, .75);
    \node[dot] (g2) at (1,1.5);
    \vertex[label = right: \(s\)] (h2) at(1.75,1.25);
    \vertex[label = right: \(s\)] (h1) at(1.75,1.75);
	\vertex(p3) at (1.5, .75);
	\vertex (p2) at (-1.5, .75);
	\vertex (p1) at (-1.5, -.75);
	\vertex (p4) at (1.5, -.75);
	\vertex [label= \(n\)] at (-1.7, .5);
	\vertex [label= \(\bar{n}\)] at (-1.7, -1);
    \vertex [label= \(n\)] at (1.7, .5);
	\vertex [label= \(\bar{n}\)] at (1.7, -1);
\diagram*{
	(f2)--[gluon, line width = 0.3mm	](f1),
	(p2)--[ line width = 0.3mm](f1)--[  line width = 0.3mm](p3),
	(p4)--[  line width = 0.3mm](f2)--[ line width = 0.3mm](p1),
    (h1)--[gluon, line width = 0.3 mm](g2),
    (h2)--[gluon, line width = 0.3 mm](g2),
    (g2)--[gluon, line width = 0.3 mm](g1),
};
\end{feynman}
\end{tikzpicture}
}
\end{subfigure}
\begin{subfigure}[b]{0.19\textwidth}
\centering
\scalebox{0.6}{
\begin{tikzpicture}
\begin{feynman}
	\node[dot] (f1) at (0,.75);
	\node[dot] (f2) at (0,-.75);
    \node[dot] (g1) at (.5, .75);
    \node[dot] (g2) at (1.5,1.5);
    \vertex[label = right: \(s\)] (h2) at(2.25,1.25);
    \vertex[label = right: \(s\)] (h1) at(2.25,1.75);
	\vertex(p3) at (1.5, .75);
	\vertex (p2) at (-1.5, .75);
	\vertex (p1) at (-1.5, -.75);
	\vertex (p4) at (1.5, -.75);
	\vertex [label= \(n\)] at (-1.7, .5);
	\vertex [label= \(\bar{n}\)] at (-1.7, -1);
    \vertex [label= \(n\)] at (1.7, .5);
	\vertex [label= \(\bar{n}\)] at (1.7, -1);
\diagram*{
	(f2)--[gluon, line width = 0.3mm	](f1),
	(p2)--[ line width = 0.3mm](f1)--[  line width = 0.3mm](p3),
	(p4)--[  line width = 0.3mm](f2)--[ line width = 0.3mm](p1),
    (h1)--[gluon, line width = 0.3 mm](g2),
    (h2)--[gluon, line width = 0.3 mm](g2),
    (g2)--[gluon, line width = 0.3 mm](g1),
};
\end{feynman}
\end{tikzpicture}
}
\end{subfigure}
\begin{subfigure}[b]{0.19\textwidth}
\centering
\scalebox{0.6}{
\begin{tikzpicture}
\begin{feynman}
	\node[dot] (f1) at (0,.75);
	\node[dot] (f2) at (0,-.75);
    \node[dot] (g1) at (0, 0);
    \node[dot] (g2) at (1.25,0);
    \vertex[label = right: \(s\)] (h2) at(2,-.25);
    \vertex[label = right: \(s\)] (h1) at(2,.25);
	\vertex(p3) at (1.5, .75);
	\vertex (p2) at (-1.5, .75);
	\vertex (p1) at (-1.5, -.75);
	\vertex (p4) at (1.5, -.75);
	\vertex [label= \(n\)] at (-1.7, .5);
	\vertex [label= \(\bar{n}\)] at (-1.7, -1);
    \vertex [label= \(n\)] at (1.7, .5);
	\vertex [label= \(\bar{n}\)] at (1.7, -1);
\diagram*{
	(f2)--[gluon, line width = 0.3mm	](f1),
	(p2)--[ line width = 0.3mm](f1)--[  line width = 0.3mm](p3),
	(p4)--[  line width = 0.3mm](f2)--[ line width = 0.3mm](p1),
    (h1)--[gluon, line width = 0.3 mm](g2),
    (h2)--[gluon, line width = 0.3 mm](g2),
    (g2)--[gluon, line width = 0.3 mm](g1),
};
\end{feynman}
\end{tikzpicture}
}
\end{subfigure}
\begin{subfigure}[b]{0.19\textwidth}
\centering
\scalebox{0.6}{
\begin{tikzpicture}
\begin{feynman}
	\node[dot] (f1) at (0,.75);
	\node[dot] (f2) at (0,-.75);
    \node[dot] (g1) at (-0.75, -.75);
    \node[dot] (g2) at (.25,-1.5);
    \vertex[label = right: \(s\)] (h1) at(1,-1.75);
    \vertex[label = right: \(s\)] (h2) at(1,-1.25);
    \vertex(p3) at (1.5, .75);
	\vertex (p2) at (-1.5, .75);
	\vertex (p1) at (-1.5, -.75);
	\vertex (p4) at (1.5, -.75);
	\vertex [label= \(n\)] at (-1.7, .5);
	\vertex [label= \(\bar{n}\)] at (-1.7, -1);
    \vertex [label= \(n\)] at (1.7, .5);
	\vertex [label= \(\bar{n}\)] at (1.7, -1);
\diagram*{
	(f2)--[gluon, line width = 0.3mm	](f1),
	(p2)--[ line width = 0.3mm](f1)--[  line width = 0.3mm](p3),
	(p4)--[  line width = 0.3mm](f2)--[ line width = 0.3mm](p1),
    (h1)--[gluon, line width = 0.3 mm](g2),
    (h2)--[gluon, line width = 0.3 mm](g2),
    (g2)--[gluon, line width = 0.3 mm](g1),
};
\end{feynman}
\end{tikzpicture}
}
\end{subfigure}
\begin{subfigure}[b]{0.19\textwidth}
\centering
\scalebox{0.6}{
\begin{tikzpicture}
\begin{feynman}
	\node[dot] (f1) at (0,.75);
	\node[dot] (f2) at (0,-.75);
    \node[dot] (g1) at (0, -.75);
    \node[dot] (g2) at (1,-1.5);
    \vertex[label = right: \(s\)] (h2) at(1.75,-1.25);
    \vertex[label = right: \(s\)] (h1) at(1.75,-1.75);
	\vertex(p3) at (1.5, .75);
	\vertex (p2) at (-1.5, .75);
	\vertex (p1) at (-1.5, -.75);
	\vertex (p4) at (1.5, -.75);
	\vertex [label= \(n\)] at (-1.7, .5);
	\vertex [label= \(\bar{n}\)] at (-1.7, -1);
    \vertex [label= \(n\)] at (1.7, .5);
	\vertex [label= \(\bar{n}\)] at (1.7, -1);
\diagram*{
	(f2)--[gluon, line width = 0.3mm	](f1),
	(p2)--[ line width = 0.3mm](f1)--[  line width = 0.3mm](p3),
	(p4)--[  line width = 0.3mm](f2)--[ line width = 0.3mm](p1),
    (h1)--[gluon, line width = 0.3 mm](g2),
    (h2)--[gluon, line width = 0.3 mm](g2),
    (g2)--[gluon, line width = 0.3 mm](g1),
};
\end{feynman}
\end{tikzpicture}
}
\end{subfigure}
\begin{subfigure}[b]{0.19\textwidth}
\centering
\scalebox{0.6}{
\begin{tikzpicture}
\begin{feynman}
	\node[dot] (f1) at (0,.75);
	\node[dot] (f2) at (0,-.75);
    \node[dot] (g1) at (.5, -.75);
    \node[dot] (g2) at (1.5,-1.5);
    \vertex[label = right: \(s\)] (h2) at(2.25,-1.25);
    \vertex[label = right: \(s\)] (h1) at(2.25,-1.75);
	\vertex(p3) at (1.5, .75);
	\vertex (p2) at (-1.5, .75);
	\vertex (p1) at (-1.5, -.75);
	\vertex (p4) at (1.5, -.75);
	\vertex [label= \(n\)] at (-1.7, .5);
	\vertex [label= \(\bar{n}\)] at (-1.7, -1);
    \vertex [label= \(n\)] at (1.7, .5);
	\vertex [label= \(\bar{n}\)] at (1.7, -1);
\diagram*{
	(f2)--[gluon, line width = 0.3mm	](f1),
	(p2)--[ line width = 0.3mm](f1)--[  line width = 0.3mm](p3),
	(p4)--[  line width = 0.3mm](f2)--[ line width = 0.3mm](p1),
    (h1)--[gluon, line width = 0.3 mm](g2),
    (h2)--[gluon, line width = 0.3 mm](g2),
    (g2)--[gluon, line width = 0.3 mm](g1),
};
\end{feynman}
\end{tikzpicture}
}
\end{subfigure}
\begin{subfigure}[b]{0.19\textwidth}
\centering
\scalebox{0.6}{
\begin{tikzpicture}
\begin{feynman}
	\node[dot] (f1) at (0,.75);
	\node[dot] (f2) at (0,-.75);
    \node[dot] (g1) at (0, 0);
    \vertex[label = right: \(s\)] (h1) at(1.25,0);
    \node[dot] (g2) at (-.75,.75);
    \vertex[label = right: \(s\)] (h2) at(.25,1.5);
	\vertex(p3) at (1.5, .75);
	\vertex (p2) at (-1.5, .75);
	\vertex (p1) at (-1.5, -.75);
	\vertex (p4) at (1.5, -.75);
	\vertex [label= \(n\)] at (-1.7, .5);
	\vertex [label= \(\bar{n}\)] at (-1.7, -1);
    \vertex [label= \(n\)] at (1.7, .5);
	\vertex [label= \(\bar{n}\)] at (1.7, -1);
\diagram*{
	(f2)--[gluon, line width = 0.3mm	](f1),
	(p2)--[ line width = 0.3mm](f1)--[  line width = 0.3mm](p3),
	(p4)--[  line width = 0.3mm](f2)--[ line width = 0.3mm](p1),
    (h1)--[gluon, line width = 0.3 mm](g1),
    (h2)--[gluon, line width = 0.3 mm](g2),
};
\end{feynman}
\end{tikzpicture}
}
\end{subfigure}
\begin{subfigure}[b]{0.19\textwidth}
\centering
\scalebox{0.6}{
\begin{tikzpicture}
\begin{feynman}
	\node[dot] (f1) at (0,.75);
	\node[dot] (f2) at (0,-.75);
    \node[dot] (g1) at (0, 0);
    \vertex[label = right: \(s\)] (h1) at(1.25,0);
    \node[dot] (g2) at (0,.75);
    \vertex[label = right: \(s\)] (h2) at(1,1.5);
	\vertex(p3) at (1.5, .75);
	\vertex (p2) at (-1.5, .75);
	\vertex (p1) at (-1.5, -.75);
	\vertex (p4) at (1.5, -.75);
	\vertex [label= \(n\)] at (-1.7, .5);
	\vertex [label= \(\bar{n}\)] at (-1.7, -1);
    \vertex [label= \(n\)] at (1.7, .5);
	\vertex [label= \(\bar{n}\)] at (1.7, -1);
\diagram*{
	(f2)--[gluon, line width = 0.3mm	](f1),
	(p2)--[ line width = 0.3mm](f1)--[  line width = 0.3mm](p3),
	(p4)--[  line width = 0.3mm](f2)--[ line width = 0.3mm](p1),
    (h1)--[gluon, line width = 0.3 mm](g1),
    (h2)--[gluon, line width = 0.3 mm](g2),
};
\end{feynman}
\end{tikzpicture}
}
\end{subfigure}
\begin{subfigure}[b]{0.19\textwidth}
\centering
\scalebox{0.6}{
\begin{tikzpicture}
\begin{feynman}
	\node[dot] (f1) at (0,.75);
	\node[dot] (f2) at (0,-.75);
    \node[dot] (g1) at (0, 0);
    \vertex[label = right: \(s\)] (h1) at(1.25,0);
    \node[dot] (g2) at (.5,.75);
    \vertex[label = right: \(s\)] (h2) at(1.5,1.5);
	\vertex(p3) at (1.5, .75);
	\vertex (p2) at (-1.5, .75);
	\vertex (p1) at (-1.5, -.75);
	\vertex (p4) at (1.5, -.75);
	\vertex [label= \(n\)] at (-1.7, .5);
	\vertex [label= \(\bar{n}\)] at (-1.7, -1);
    \vertex [label= \(n\)] at (1.7, .5);
	\vertex [label= \(\bar{n}\)] at (1.7, -1);
\diagram*{
	(f2)--[gluon, line width = 0.3mm	](f1),
	(p2)--[ line width = 0.3mm](f1)--[  line width = 0.3mm](p3),
	(p4)--[  line width = 0.3mm](f2)--[ line width = 0.3mm](p1),
    (h1)--[gluon, line width = 0.3 mm](g1),
    (h2)--[gluon, line width = 0.3 mm](g2),
};
\end{feynman}
\end{tikzpicture}
}
\end{subfigure}
\begin{subfigure}[b]{0.19\textwidth}
\centering
\scalebox{0.6}{
\begin{tikzpicture}
\begin{feynman}
	\node[dot] (f1) at (0,.75);
	\node[dot] (f2) at (0,-.75);
    \node[dot] (g1) at (0, .25);
    \vertex[label = right: \(s\)] (h1) at(1.5,.25);
    \node[dot] (g2) at (0,-.25);
    \vertex[label = right: \(s\)] (h2) at(1.5,-.25);
	\vertex(p3) at (1.5, .75);
	\vertex (p2) at (-1.5, .75);
	\vertex (p1) at (-1.5, -.75);
	\vertex (p4) at (1.5, -.75);
	\vertex [label= \(n\)] at (-1.7, .5);
	\vertex [label= \(\bar{n}\)] at (-1.7, -1);
    \vertex [label= \(n\)] at (1.7, .5);
	\vertex [label= \(\bar{n}\)] at (1.7, -1);
\diagram*{
	(f2)--[gluon, line width = 0.3mm	](f1),
	(p2)--[ line width = 0.3mm](f1)--[  line width = 0.3mm](p3),
	(p4)--[  line width = 0.3mm](f2)--[ line width = 0.3mm](p1),
    (h1)--[gluon, line width = 0.3 mm](g1),
    (h2)--[gluon, line width = 0.3 mm](g2),
};
\end{feynman}
\end{tikzpicture}
}
\end{subfigure}
\begin{subfigure}[b]{0.19\textwidth}
\centering
\scalebox{0.6}{
\begin{tikzpicture}
\begin{feynman}
	\node[dot] (f1) at (0,.75);
	\node[dot] (f2) at (0,-.75);
    \node[dot] (g1) at (0, 0);
    \vertex[label = right: \(s\)] (h1) at(1.5,-.25);
    \node[dot] (g2) at (0,0);
    \vertex[label = right: \(s\)] (h2) at(1.5,.25);
	\vertex(p3) at (1.5, .75);
	\vertex (p2) at (-1.5, .75);
	\vertex (p1) at (-1.5, -.75);
	\vertex (p4) at (1.5, -.75);
	\vertex [label= \(n\)] at (-1.7, .5);
	\vertex [label= \(\bar{n}\)] at (-1.7, -1);
    \vertex [label= \(n\)] at (1.7, .5);
	\vertex [label= \(\bar{n}\)] at (1.7, -1);
\diagram*{
	(f2)--[gluon, line width = 0.3mm	](f1),
	(p2)--[ line width = 0.3mm](f1)--[  line width = 0.3mm](p3),
	(p4)--[  line width = 0.3mm](f2)--[ line width = 0.3mm](p1),
    (h1)--[gluon, line width = 0.3 mm](g1),
    (h2)--[gluon, line width = 0.3 mm](g2),
};
\end{feynman}
\end{tikzpicture}
}
\end{subfigure}
\begin{subfigure}[b]{0.19\textwidth}
\centering
\scalebox{0.6}{
\begin{tikzpicture}
\begin{feynman}
	\node[dot] (f1) at (0,.75);
	\node[dot] (f2) at (0,-.75);
    \node[dot] (g1) at (0, 0);
    \vertex[label = right: \(s\)] (h1) at(1.25,0);
    \node[dot] (g2) at (-.75,-.75);
    \vertex[label = right: \(s\)] (h2) at(.25,-1.5);
	\vertex(p3) at (1.5, .75);
	\vertex (p2) at (-1.5, .75);
	\vertex (p1) at (-1.5, -.75);
	\vertex (p4) at (1.5, -.75);
	\vertex [label= \(n\)] at (-1.7, .5);
	\vertex [label= \(\bar{n}\)] at (-1.7, -1);
    \vertex [label= \(n\)] at (1.7, .5);
	\vertex [label= \(\bar{n}\)] at (1.7, -1);
\diagram*{
	(f2)--[gluon, line width = 0.3mm	](f1),
	(p2)--[ line width = 0.3mm](f1)--[  line width = 0.3mm](p3),
	(p4)--[  line width = 0.3mm](f2)--[ line width = 0.3mm](p1),
    (h1)--[gluon, line width = 0.3 mm](g1),
    (h2)--[gluon, line width = 0.3 mm](g2),
};
\end{feynman}
\end{tikzpicture}
}
\end{subfigure}
\begin{subfigure}[b]{0.19\textwidth}
\centering
\scalebox{0.6}{
\begin{tikzpicture}
\begin{feynman}
	\node[dot] (f1) at (0,.75);
	\node[dot] (f2) at (0,-.75);
    \node[dot] (g1) at (0, 0);
    \vertex[label = right: \(s\)] (h1) at(1.25,0);
    \node[dot] (g2) at (0,-.75);
    \vertex[label = right: \(s\)] (h2) at(1,-1.5);
	\vertex(p3) at (1.5, .75);
	\vertex (p2) at (-1.5, .75);
	\vertex (p1) at (-1.5, -.75);
	\vertex (p4) at (1.5, -.75);
	\vertex [label= \(n\)] at (-1.7, .5);
	\vertex [label= \(\bar{n}\)] at (-1.7, -1);
    \vertex [label= \(n\)] at (1.7, .5);
	\vertex [label= \(\bar{n}\)] at (1.7, -1);
\diagram*{
	(f2)--[gluon, line width = 0.3mm	](f1),
	(p2)--[ line width = 0.3mm](f1)--[  line width = 0.3mm](p3),
	(p4)--[  line width = 0.3mm](f2)--[ line width = 0.3mm](p1),
    (h1)--[gluon, line width = 0.3 mm](g1),
    (h2)--[gluon, line width = 0.3 mm](g2),
};
\end{feynman}
\end{tikzpicture}
}
\end{subfigure}
\begin{subfigure}[b]{0.19\textwidth}
\centering
\scalebox{0.6}{
\begin{tikzpicture}
\begin{feynman}
	\node[dot] (f1) at (0,.75);
	\node[dot] (f2) at (0,-.75);
    \node[dot] (g1) at (0, 0);
    \vertex[label = right: \(s\)] (h1) at(1.25,0);
    \node[dot] (g2) at (.5,-.75);
    \vertex[label = right: \(s\)] (h2) at(1.5,-1.5);
	\vertex(p3) at (1.5, .75);
	\vertex (p2) at (-1.5, .75);
	\vertex (p1) at (-1.5, -.75);
	\vertex (p4) at (1.5, -.75);
	\vertex [label= \(n\)] at (-1.7, .5);
	\vertex [label= \(\bar{n}\)] at (-1.7, -1);
    \vertex [label= \(n\)] at (1.7, .5);
	\vertex [label= \(\bar{n}\)] at (1.7, -1);
\diagram*{
	(f2)--[gluon, line width = 0.3mm	](f1),
	(p2)--[ line width = 0.3mm](f1)--[  line width = 0.3mm](p3),
	(p4)--[  line width = 0.3mm](f2)--[ line width = 0.3mm](p1),
    (h1)--[gluon, line width = 0.3 mm](g1),
    (h2)--[gluon, line width = 0.3 mm](g2),
};
\end{feynman}
\end{tikzpicture}
}
\end{subfigure}
\begin{subfigure}[b]{0.19\textwidth}
\centering
\scalebox{0.6}{
\begin{tikzpicture}
\begin{feynman}
	\node[dot] (f1) at (0,.75);
	\node[dot] (f2) at (0,-.75);
    \node[dot] (g1) at (-.75,.75);
    \vertex[label = right: \(s\)] (h1) at(.25,1.5);
    \node[dot] (g2) at (-.75,-.75);
    \vertex[label = right: \(s\)] (h2) at(.25,-1.5);
	\vertex(p3) at (1.5, .75);
	\vertex (p2) at (-1.5, .75);
	\vertex (p1) at (-1.5, -.75);
	\vertex (p4) at (1.5, -.75);
	\vertex [label= \(n\)] at (-1.7, .5);
	\vertex [label= \(\bar{n}\)] at (-1.7, -1);
    \vertex [label= \(n\)] at (1.7, .5);
	\vertex [label= \(\bar{n}\)] at (1.7, -1);
\diagram*{
	(f2)--[gluon, line width = 0.3mm	](f1),
	(p2)--[ line width = 0.3mm](f1)--[  line width = 0.3mm](p3),
	(p4)--[  line width = 0.3mm](f2)--[ line width = 0.3mm](p1),
    (h1)--[gluon, line width = 0.3 mm](g1),
    (h2)--[gluon, line width = 0.3 mm](g2),
};
\end{feynman}
\end{tikzpicture}
}
\end{subfigure}
\begin{subfigure}[b]{0.19\textwidth}
\centering
\scalebox{0.6}{
\begin{tikzpicture}
\begin{feynman}
	\node[dot] (f1) at (0,.75);
	\node[dot] (f2) at (0,-.75);
    \node[dot] (g1) at (0,.75);
    \vertex[label = right: \(s\)] (h1) at(1,1.5);
    \node[dot] (g2) at (-.75,-.75);
    \vertex[label = right: \(s\)] (h2) at(.25,-1.5);
	\vertex(p3) at (1.5, .75);
	\vertex (p2) at (-1.5, .75);
	\vertex (p1) at (-1.5, -.75);
	\vertex (p4) at (1.5, -.75);
	\vertex [label= \(n\)] at (-1.7, .5);
	\vertex [label= \(\bar{n}\)] at (-1.7, -1);
    \vertex [label= \(n\)] at (1.7, .5);
	\vertex [label= \(\bar{n}\)] at (1.7, -1);
\diagram*{
	(f2)--[gluon, line width = 0.3mm	](f1),
	(p2)--[ line width = 0.3mm](f1)--[  line width = 0.3mm](p3),
	(p4)--[  line width = 0.3mm](f2)--[ line width = 0.3mm](p1),
    (h1)--[gluon, line width = 0.3 mm](g1),
    (h2)--[gluon, line width = 0.3 mm](g2),
};
\end{feynman}
\end{tikzpicture}
}
\end{subfigure}
\begin{subfigure}[b]{0.19\textwidth}
\centering
\scalebox{0.6}{
\begin{tikzpicture}
\begin{feynman}
	\node[dot] (f1) at (0,.75);
	\node[dot] (f2) at (0,-.75);
    \node[dot] (g1) at (.5,.75);
    \vertex[label = right: \(s\)] (h1) at(1.5,1.5);
    \node[dot] (g2) at (-.75,-.75);
    \vertex[label = right: \(s\)] (h2) at(.25,-1.5);
	\vertex(p3) at (1.5, .75);
	\vertex (p2) at (-1.5, .75);
	\vertex (p1) at (-1.5, -.75);
	\vertex (p4) at (1.5, -.75);
	\vertex [label= \(n\)] at (-1.7, .5);
	\vertex [label= \(\bar{n}\)] at (-1.7, -1);
    \vertex [label= \(n\)] at (1.7, .5);
	\vertex [label= \(\bar{n}\)] at (1.7, -1);
\diagram*{
	(f2)--[gluon, line width = 0.3mm	](f1),
	(p2)--[ line width = 0.3mm](f1)--[  line width = 0.3mm](p3),
	(p4)--[  line width = 0.3mm](f2)--[ line width = 0.3mm](p1),
    (h1)--[gluon, line width = 0.3 mm](g1),
    (h2)--[gluon, line width = 0.3 mm](g2),
};
\end{feynman}
\end{tikzpicture}
}
\end{subfigure}
\begin{subfigure}[b]{0.19\textwidth}
\centering
\scalebox{0.6}{
\begin{tikzpicture}
\begin{feynman}
	\node[dot] (f1) at (0,.75);
	\node[dot] (f2) at (0,-.75);
    \node[dot] (g1) at (-.75,.75);
    \vertex[label = right: \(s\)] (h1) at(.25,1.5);
    \node[dot] (g2) at (0,-.75);
    \vertex[label = right: \(s\)] (h2) at(1,-1.5);
	\vertex(p3) at (1.5, .75);
	\vertex (p2) at (-1.5, .75);
	\vertex (p1) at (-1.5, -.75);
	\vertex (p4) at (1.5, -.75);
	\vertex [label= \(n\)] at (-1.7, .5);
	\vertex [label= \(\bar{n}\)] at (-1.7, -1);
    \vertex [label= \(n\)] at (1.7, .5);
	\vertex [label= \(\bar{n}\)] at (1.7, -1);
\diagram*{
	(f2)--[gluon, line width = 0.3mm	](f1),
	(p2)--[ line width = 0.3mm](f1)--[  line width = 0.3mm](p3),
	(p4)--[  line width = 0.3mm](f2)--[ line width = 0.3mm](p1),
    (h1)--[gluon, line width = 0.3 mm](g1),
    (h2)--[gluon, line width = 0.3 mm](g2),
};
\end{feynman}
\end{tikzpicture}
}
\end{subfigure}
\begin{subfigure}[b]{0.19\textwidth}
\centering
\scalebox{0.6}{
\begin{tikzpicture}
\begin{feynman}
	\node[dot] (f1) at (0,.75);
	\node[dot] (f2) at (0,-.75);
    \node[dot] (g1) at (0,.75);
    \vertex[label = right: \(s\)] (h1) at(1,1.5);
    \node[dot] (g2) at (0,-.75);
    \vertex[label = right: \(s\)] (h2) at(1,-1.5);
	\vertex(p3) at (1.5, .75);
	\vertex (p2) at (-1.5, .75);
	\vertex (p1) at (-1.5, -.75);
	\vertex (p4) at (1.5, -.75);
	\vertex [label= \(n\)] at (-1.7, .5);
	\vertex [label= \(\bar{n}\)] at (-1.7, -1);
    \vertex [label= \(n\)] at (1.7, .5);
	\vertex [label= \(\bar{n}\)] at (1.7, -1);
\diagram*{
	(f2)--[gluon, line width = 0.3mm	](f1),
	(p2)--[ line width = 0.3mm](f1)--[  line width = 0.3mm](p3),
	(p4)--[  line width = 0.3mm](f2)--[ line width = 0.3mm](p1),
    (h1)--[gluon, line width = 0.3 mm](g1),
    (h2)--[gluon, line width = 0.3 mm](g2),
};
\end{feynman}
\end{tikzpicture}
}
\end{subfigure}
\begin{subfigure}[b]{0.23\textwidth}
\centering
\scalebox{0.6}{
\begin{tikzpicture}
\begin{feynman}
	\node[dot] (f1) at (0,.75);
	\node[dot] (f2) at (0,-.75);
    \node[dot] (g1) at (.5,.75);
    \vertex[label = right: \(s\)] (h1) at(1.5,1.5);
    \node[dot] (g2) at (0,-.75);
    \vertex[label = right: \(s\)] (h2) at(1,-1.5);
	\vertex(p3) at (1.5, .75);
	\vertex (p2) at (-1.5, .75);
	\vertex (p1) at (-1.5, -.75);
	\vertex (p4) at (1.5, -.75);
	\vertex [label= \(n\)] at (-1.7, .5);
	\vertex [label= \(\bar{n}\)] at (-1.7, -1);
    \vertex [label= \(n\)] at (1.7, .5);
	\vertex [label= \(\bar{n}\)] at (1.7, -1);
\diagram*{
	(f2)--[gluon, line width = 0.3mm	](f1),
	(p2)--[ line width = 0.3mm](f1)--[  line width = 0.3mm](p3),
	(p4)--[  line width = 0.3mm](f2)--[ line width = 0.3mm](p1),
    (h1)--[gluon, line width = 0.3 mm](g1),
    (h2)--[gluon, line width = 0.3 mm](g2),
};
\end{feynman}
\end{tikzpicture}
}
\end{subfigure}
\begin{subfigure}[b]{0.23\textwidth}
\centering
\scalebox{0.6}{
\begin{tikzpicture}
\begin{feynman}
	\node[dot] (f1) at (0,.75);
	\node[dot] (f2) at (0,-.75);
    \node[dot] (g1) at (-.75,.75);
    \vertex[label = right: \(s\)] (h1) at(.25,1.5);
    \node[dot] (g2) at (.5,-.75);
    \vertex[label = right: \(s\)] (h2) at(1.5,-1.5);
	\vertex(p3) at (1.5, .75);
	\vertex (p2) at (-1.5, .75);
	\vertex (p1) at (-1.5, -.75);
	\vertex (p4) at (1.5, -.75);
	\vertex [label= \(n\)] at (-1.7, .5);
	\vertex [label= \(\bar{n}\)] at (-1.7, -1);
    \vertex [label= \(n\)] at (1.7, .5);
	\vertex [label= \(\bar{n}\)] at (1.7, -1);
\diagram*{
	(f2)--[gluon, line width = 0.3mm	](f1),
	(p2)--[ line width = 0.3mm](f1)--[  line width = 0.3mm](p3),
	(p4)--[  line width = 0.3mm](f2)--[ line width = 0.3mm](p1),
    (h1)--[gluon, line width = 0.3 mm](g1),
    (h2)--[gluon, line width = 0.3 mm](g2),
};
\end{feynman}
\end{tikzpicture}
}
\end{subfigure}
\begin{subfigure}[b]{0.23\textwidth}
\centering
\scalebox{0.6}{
\begin{tikzpicture}
\begin{feynman}
	\node[dot] (f1) at (0,.75);
	\node[dot] (f2) at (0,-.75);
    \node[dot] (g1) at (0,.75);
    \vertex[label = right: \(s\)] (h1) at(1,1.5);
    \node[dot] (g2) at (.5,-.75);
    \vertex[label = right: \(s\)] (h2) at(1.5,-1.5);
	\vertex(p3) at (1.5, .75);
	\vertex (p2) at (-1.5, .75);
	\vertex (p1) at (-1.5, -.75);
	\vertex (p4) at (1.5, -.75);
	\vertex [label= \(n\)] at (-1.7, .5);
	\vertex [label= \(\bar{n}\)] at (-1.7, -1);
    \vertex [label= \(n\)] at (1.7, .5);
	\vertex [label= \(\bar{n}\)] at (1.7, -1);
\diagram*{
	(f2)--[gluon, line width = 0.3mm	](f1),
	(p2)--[ line width = 0.3mm](f1)--[  line width = 0.3mm](p3),
	(p4)--[  line width = 0.3mm](f2)--[ line width = 0.3mm](p1),
    (h1)--[gluon, line width = 0.3 mm](g1),
    (h2)--[gluon, line width = 0.3 mm](g2),
};
\end{feynman}
\end{tikzpicture}
}
\end{subfigure}
\begin{subfigure}[b]{0.23\textwidth}
\centering
\scalebox{0.6}{
\begin{tikzpicture}
\begin{feynman}
	\node[dot] (f1) at (0,.75);
	\node[dot] (f2) at (0,-.75);
    \node[dot] (g1) at (.5,.75);
    \vertex[label = right: \(s\)] (h1) at(1.5,1.5);
     \node[dot] (g2) at (.5,-.75);
    \vertex[label = right: \(s\)] (h2) at(1.5,-1.5);
	\vertex(p3) at (1.5, .75);
	\vertex (p2) at (-1.5, .75);
	\vertex (p1) at (-1.5, -.75);
	\vertex (p4) at (1.5, -.75);
	\vertex [label= \(n\)] at (-1.7, .5);
	\vertex [label= \(\bar{n}\)] at (-1.7, -1);
    \vertex [label= \(n\)] at (1.7, .5);
	\vertex [label= \(\bar{n}\)] at (1.7, -1);
\diagram*{
	(f2)--[gluon, line width = 0.3mm	](f1),
	(p2)--[ line width = 0.3mm](f1)--[  line width = 0.3mm](p3),
	(p4)--[  line width = 0.3mm](f2)--[ line width = 0.3mm](p1),
    (h1)--[gluon, line width = 0.3 mm](g1),
    (h2)--[gluon, line width = 0.3 mm](g2),
};
\end{feynman}
\end{tikzpicture}
}
\end{subfigure}
\begin{subfigure}[b]{0.19\textwidth}
\centering
\scalebox{0.6}{
\begin{tikzpicture}
\begin{feynman}
	\node[dot] (f1) at (0,.75);
	\node[dot] (f2) at (0,-.75);
    \node[dot] (g1) at (-1,.75);
    \vertex[label = above right: \(s\)] (h1) at(-.25,1.5);
     \node[dot] (g2) at (-.5,.75);
    \vertex[label = above right: \(s\)] (h2) at(.25,1.5);
	\vertex(p3) at (1.5, .75);
	\vertex (p2) at (-1.5, .75);
	\vertex (p1) at (-1.5, -.75);
	\vertex (p4) at (1.5, -.75);
	\vertex [label= \(n\)] at (-1.7, .5);
	\vertex [label= \(\bar{n}\)] at (-1.7, -1);
    \vertex [label= \(n\)] at (1.7, .5);
	\vertex [label= \(\bar{n}\)] at (1.7, -1);
\diagram*{
	(f2)--[gluon, line width = 0.3mm	](f1),
	(p2)--[ line width = 0.3mm](f1)--[  line width = 0.3mm](p3),
	(p4)--[  line width = 0.3mm](f2)--[ line width = 0.3mm](p1),
    (h1)--[gluon, line width = 0.3 mm](g1),
    (h2)--[gluon, line width = 0.3 mm](g2),
};
\end{feynman}
\end{tikzpicture}
}
\end{subfigure}
\begin{subfigure}[b]{0.19\textwidth}
\centering
\scalebox{0.6}{
\begin{tikzpicture}
\begin{feynman}
	\node[dot] (f1) at (0,.75);
	\node[dot] (f2) at (0,-.75);
    \node[dot] (g1) at (-1,.75);
    \vertex[label = above right: \(s\)] (h1) at(-.25,1.5);
     \node[dot] (g2) at (0,.75);
    \vertex[label = above right: \(s\)] (h2) at(.75,1.5);
	\vertex(p3) at (1.5, .75);
	\vertex (p2) at (-1.5, .75);
	\vertex (p1) at (-1.5, -.75);
	\vertex (p4) at (1.5, -.75);
	\vertex [label= \(n\)] at (-1.7, .5);
	\vertex [label= \(\bar{n}\)] at (-1.7, -1);
    \vertex [label= \(n\)] at (1.7, .5);
	\vertex [label= \(\bar{n}\)] at (1.7, -1);
\diagram*{
	(f2)--[gluon, line width = 0.3mm	](f1),
	(p2)--[ line width = 0.3mm](f1)--[  line width = 0.3mm](p3),
	(p4)--[  line width = 0.3mm](f2)--[ line width = 0.3mm](p1),
    (h1)--[gluon, line width = 0.3 mm](g1),
    (h2)--[gluon, line width = 0.3 mm](g2),
};
\end{feynman}
\end{tikzpicture}
}
\end{subfigure}
\begin{subfigure}[b]{0.19\textwidth}
\centering
\scalebox{0.6}{
\begin{tikzpicture}
\begin{feynman}
	\node[dot] (f1) at (0,.75);
	\node[dot] (f2) at (0,-.75);
    \node[dot] (g1) at (-1,.75);
    \vertex[label = above right: \(s\)] (h1) at(-.25,1.5);
     \node[dot] (g2) at (1,.75);
    \vertex[label = above right: \(s\)] (h2) at(1.75,1.5);
	\vertex(p3) at (1.5, .75);
	\vertex (p2) at (-1.5, .75);
	\vertex (p1) at (-1.5, -.75);
	\vertex (p4) at (1.5, -.75);
	\vertex [label= \(n\)] at (-1.7, .5);
	\vertex [label= \(\bar{n}\)] at (-1.7, -1);
    \vertex [label= \(n\)] at (1.7, .5);
	\vertex [label= \(\bar{n}\)] at (1.7, -1);
\diagram*{
	(f2)--[gluon, line width = 0.3mm	](f1),
	(p2)--[ line width = 0.3mm](f1)--[  line width = 0.3mm](p3),
	(p4)--[  line width = 0.3mm](f2)--[ line width = 0.3mm](p1),
    (h1)--[gluon, line width = 0.3 mm](g1),
    (h2)--[gluon, line width = 0.3 mm](g2),
};
\end{feynman}
\end{tikzpicture}
}
\end{subfigure}
\begin{subfigure}[b]{0.19\textwidth}
\centering
\scalebox{0.6}{
\begin{tikzpicture}
\begin{feynman}
	\node[dot] (f1) at (0,.75);
	\node[dot] (f2) at (0,-.75);
    \node[dot] (g1) at (0,.75);
    \vertex[label = above right: \(s\)] (h1) at(.75,1.5);
     \node[dot] (g2) at (1,.75);
    \vertex[label = above right: \(s\)] (h2) at(1.75,1.5);
	\vertex(p3) at (1.5, .75);
	\vertex (p2) at (-1.5, .75);
	\vertex (p1) at (-1.5, -.75);
	\vertex (p4) at (1.5, -.75);
	\vertex [label= \(n\)] at (-1.7, .5);
	\vertex [label= \(\bar{n}\)] at (-1.7, -1);
    \vertex [label= \(n\)] at (1.7, .5);
	\vertex [label= \(\bar{n}\)] at (1.7, -1);
\diagram*{
	(f2)--[gluon, line width = 0.3mm	](f1),
	(p2)--[ line width = 0.3mm](f1)--[  line width = 0.3mm](p3),
	(p4)--[  line width = 0.3mm](f2)--[ line width = 0.3mm](p1),
    (h1)--[gluon, line width = 0.3 mm](g1),
    (h2)--[gluon, line width = 0.3 mm](g2),
};
\end{feynman}
\end{tikzpicture}
}
\end{subfigure}
\begin{subfigure}[b]{0.19\textwidth}
\centering
\scalebox{0.6}{
\begin{tikzpicture}
\begin{feynman}
	\node[dot] (f1) at (0,.75);
	\node[dot] (f2) at (0,-.75);
    \node[dot] (g1) at (.5,.75);
    \vertex[label = above right: \(s\)] (h1) at(1.25,1.5);
     \node[dot] (g2) at (1,.75);
    \vertex[label = above right: \(s\)] (h2) at(1.75,1.5);
	\vertex(p3) at (1.5, .75);
	\vertex (p2) at (-1.5, .75);
	\vertex (p1) at (-1.5, -.75);
	\vertex (p4) at (1.5, -.75);
	\vertex [label= \(n\)] at (-1.7, .5);
	\vertex [label= \(\bar{n}\)] at (-1.7, -1);
    \vertex [label= \(n\)] at (1.7, .5);
	\vertex [label= \(\bar{n}\)] at (1.7, -1);
\diagram*{
	(f2)--[gluon, line width = 0.3mm	](f1),
	(p2)--[ line width = 0.3mm](f1)--[  line width = 0.3mm](p3),
	(p4)--[  line width = 0.3mm](f2)--[ line width = 0.3mm](p1),
    (h1)--[gluon, line width = 0.3 mm](g1),
    (h2)--[gluon, line width = 0.3 mm](g2),
};
\end{feynman}
\end{tikzpicture}
}
\end{subfigure}
\begin{subfigure}[b]{0.19\textwidth}
\centering
\scalebox{0.6}{
\begin{tikzpicture}
\begin{feynman}
	\node[dot] (f1) at (0,.75);
	\node[dot] (f2) at (0,-.75);
    \node[dot] (g1) at (-1,-.75);
    \vertex[label = above right: \(s\)] (h1) at(-.25,-1.5);
     \node[dot] (g2) at (-.5,-.75);
    \vertex[label = above right: \(s\)] (h2) at(.25,-1.5);
	\vertex(p3) at (1.5, .75);
	\vertex (p2) at (-1.5, .75);
	\vertex (p1) at (-1.5, -.75);
	\vertex (p4) at (1.5, -.75);
	\vertex [label= \(n\)] at (-1.7, .5);
	\vertex [label= \(\bar{n}\)] at (-1.7, -1);
    \vertex [label= \(n\)] at (1.7, .5);
	\vertex [label= \(\bar{n}\)] at (1.7, -1);
\diagram*{
	(f2)--[gluon, line width = 0.3mm	](f1),
	(p2)--[ line width = 0.3mm](f1)--[  line width = 0.3mm](p3),
	(p4)--[  line width = 0.3mm](f2)--[ line width = 0.3mm](p1),
    (h1)--[gluon, line width = 0.3 mm](g1),
    (h2)--[gluon, line width = 0.3 mm](g2),
};
\end{feynman}
\end{tikzpicture}
}
\end{subfigure}
\begin{subfigure}[b]{0.19\textwidth}
\centering
\scalebox{0.6}{
\begin{tikzpicture}
\begin{feynman}
	\node[dot] (f1) at (0,.75);
	\node[dot] (f2) at (0,-.75);
    \node[dot] (g1) at (-1,-.75);
    \vertex[label = above right: \(s\)] (h1) at(-.25,-1.5);
     \node[dot] (g2) at (0,-.75);
    \vertex[label = above right: \(s\)] (h2) at(.75,-1.5);
	\vertex(p3) at (1.5, .75);
	\vertex (p2) at (-1.5, .75);
	\vertex (p1) at (-1.5, -.75);
	\vertex (p4) at (1.5, -.75);
	\vertex [label= \(n\)] at (-1.7, .5);
	\vertex [label= \(\bar{n}\)] at (-1.7, -1);
    \vertex [label= \(n\)] at (1.7, .5);
	\vertex [label= \(\bar{n}\)] at (1.7, -1);
\diagram*{
	(f2)--[gluon, line width = 0.3mm	](f1),
	(p2)--[ line width = 0.3mm](f1)--[  line width = 0.3mm](p3),
	(p4)--[  line width = 0.3mm](f2)--[ line width = 0.3mm](p1),
    (h1)--[gluon, line width = 0.3 mm](g1),
    (h2)--[gluon, line width = 0.3 mm](g2),
};
\end{feynman}
\end{tikzpicture}
}
\end{subfigure}
\begin{subfigure}[b]{0.19\textwidth}
\centering
\scalebox{0.6}{
\begin{tikzpicture}
\begin{feynman}
	\node[dot] (f1) at (0,.75);
	\node[dot] (f2) at (0,-.75);
    \node[dot] (g1) at (-1,-.75);
    \vertex[label = above right: \(s\)] (h1) at(-.25,-1.5);
     \node[dot] (g2) at (1,-.75);
    \vertex[label = above right: \(s\)] (h2) at(1.75,-1.5);
	\vertex(p3) at (1.5, .75);
	\vertex (p2) at (-1.5, .75);
	\vertex (p1) at (-1.5, -.75);
	\vertex (p4) at (1.5, -.75);
	\vertex [label= \(n\)] at (-1.7, .5);
	\vertex [label= \(\bar{n}\)] at (-1.7, -1);
    \vertex [label= \(n\)] at (1.7, .5);
	\vertex [label= \(\bar{n}\)] at (1.7, -1);
\diagram*{
	(f2)--[gluon, line width = 0.3mm	](f1),
	(p2)--[ line width = 0.3mm](f1)--[  line width = 0.3mm](p3),
	(p4)--[  line width = 0.3mm](f2)--[ line width = 0.3mm](p1),
    (h1)--[gluon, line width = 0.3 mm](g1),
    (h2)--[gluon, line width = 0.3 mm](g2),
};
\end{feynman}
\end{tikzpicture}
}
\end{subfigure}
\begin{subfigure}[b]{0.19\textwidth}
\centering
\scalebox{0.6}{
\begin{tikzpicture}
\begin{feynman}
	\node[dot] (f1) at (0,.75);
	\node[dot] (f2) at (0,-.75);
    \node[dot] (g1) at (0,-.75);
    \vertex[label = above right: \(s\)] (h1) at(.75,-1.5);
     \node[dot] (g2) at (1,-.75);
    \vertex[label = above right: \(s\)] (h2) at(1.75,-1.5);
	\vertex(p3) at (1.5, .75);
	\vertex (p2) at (-1.5, .75);
	\vertex (p1) at (-1.5, -.75);
	\vertex (p4) at (1.5, -.75);
	\vertex [label= \(n\)] at (-1.7, .5);
	\vertex [label= \(\bar{n}\)] at (-1.7, -1);
    \vertex [label= \(n\)] at (1.7, .5);
	\vertex [label= \(\bar{n}\)] at (1.7, -1);
\diagram*{
	(f2)--[gluon, line width = 0.3mm	](f1),
	(p2)--[ line width = 0.3mm](f1)--[  line width = 0.3mm](p3),
	(p4)--[  line width = 0.3mm](f2)--[ line width = 0.3mm](p1),
    (h1)--[gluon, line width = 0.3 mm](g1),
    (h2)--[gluon, line width = 0.3 mm](g2),
};
\end{feynman}
\end{tikzpicture}
}
\end{subfigure}
\begin{subfigure}[b]{0.19\textwidth}
\centering
\scalebox{0.6}{
\begin{tikzpicture}
\begin{feynman}
	\node[dot] (f1) at (0,.75);
	\node[dot] (f2) at (0,-.75);
    \node[dot] (g1) at (.5,-.75);
    \vertex[label = above right: \(s\)] (h1) at(1.25,-1.5);
     \node[dot] (g2) at (1,-.75);
    \vertex[label = above right: \(s\)] (h2) at(1.75,-1.5);
	\vertex(p3) at (1.5, .75);
	\vertex (p2) at (-1.5, .75);
	\vertex (p1) at (-1.5, -.75);
	\vertex (p4) at (1.5, -.75);
	\vertex [label= \(n\)] at (-1.7, .5);
	\vertex [label= \(\bar{n}\)] at (-1.7, -1);
    \vertex [label= \(n\)] at (1.7, .5);
	\vertex [label= \(\bar{n}\)] at (1.7, -1);
\diagram*{
	(f2)--[gluon, line width = 0.3mm	](f1),
	(p2)--[ line width = 0.3mm](f1)--[  line width = 0.3mm](p3),
	(p4)--[  line width = 0.3mm](f2)--[ line width = 0.3mm](p1),
    (h1)--[gluon, line width = 0.3 mm](g1),
    (h2)--[gluon, line width = 0.3 mm](g2),
};
\end{feynman}
\end{tikzpicture}
}
\end{subfigure}
\begin{subfigure}[b]{0.16\textwidth}
\centering
\scalebox{0.6}{
\begin{tikzpicture}
\begin{feynman}
	\node[dot] (f1) at (0,.75);
	\node[dot] (f2) at (0,-.75);
    \node[dot] (g1) at (-.75,.75);
    \vertex[label = left: \(s\)] (h1) at(-1.25,1.5);
     \node[dot] (g2) at (-.75,.75);
    \vertex[label = right: \(s\)] (h2) at(-.25,1.5);
	\vertex(p3) at (1.5, .75);
	\vertex (p2) at (-1.5, .75);
	\vertex (p1) at (-1.5, -.75);
	\vertex (p4) at (1.5, -.75);
	\vertex [label= \(n\)] at (-1.7, .5);
	\vertex [label= \(\bar{n}\)] at (-1.7, -1);
    \vertex [label= \(n\)] at (1.7, .5);
	\vertex [label= \(\bar{n}\)] at (1.7, -1);
\diagram*{
	(f2)--[gluon, line width = 0.3mm	](f1),
	(p2)--[ line width = 0.3mm](f1)--[  line width = 0.3mm](p3),
	(p4)--[  line width = 0.3mm](f2)--[ line width = 0.3mm](p1),
    (h1)--[gluon, line width = 0.3 mm](g1),
    (h2)--[gluon, line width = 0.3 mm](g2),
};
\end{feynman}
\end{tikzpicture}
}
\end{subfigure}
\begin{subfigure}[b]{0.16\textwidth}
\centering
\scalebox{0.6}{
\begin{tikzpicture}
\begin{feynman}
	\node[dot] (f1) at (0,.75);
	\node[dot] (f2) at (0,-.75);
    \node[dot] (g1) at (0,.75);
    \vertex[label = left: \(s\)] (h1) at(-.5,1.5);
     \node[dot] (g2) at (0,.75);
    \vertex[label = right: \(s\)] (h2) at(.5,1.5);
	\vertex(p3) at (1.5, .75);
	\vertex (p2) at (-1.5, .75);
	\vertex (p1) at (-1.5, -.75);
	\vertex (p4) at (1.5, -.75);
	\vertex [label= \(n\)] at (-1.7, .5);
	\vertex [label= \(\bar{n}\)] at (-1.7, -1);
    \vertex [label= \(n\)] at (1.7, .5);
	\vertex [label= \(\bar{n}\)] at (1.7, -1);
\diagram*{
	(f2)--[gluon, line width = 0.3mm	](f1),
	(p2)--[ line width = 0.3mm](f1)--[  line width = 0.3mm](p3),
	(p4)--[  line width = 0.3mm](f2)--[ line width = 0.3mm](p1),
    (h1)--[gluon, line width = 0.3 mm](g1),
    (h2)--[gluon, line width = 0.3 mm](g2),
};
\end{feynman}
\end{tikzpicture}
}
\end{subfigure}
\begin{subfigure}[b]{0.16\textwidth}
\centering
\scalebox{0.6}{
\begin{tikzpicture}
\begin{feynman}
	\node[dot] (f1) at (0,.75);
	\node[dot] (f2) at (0,-.75);
    \node[dot] (g1) at (.75,.75);
    \vertex[label = right: \(s\)] (h1) at(1.25,1.5);
     \node[dot] (g2) at (.75,.75);
    \vertex[label = left: \(s\)] (h2) at(.25,1.5);
	\vertex(p3) at (1.5, .75);
	\vertex (p2) at (-1.5, .75);
	\vertex (p1) at (-1.5, -.75);
	\vertex (p4) at (1.5, -.75);
	\vertex [label= \(n\)] at (-1.7, .5);
	\vertex [label= \(\bar{n}\)] at (-1.7, -1);
    \vertex [label= \(n\)] at (1.7, .5);
	\vertex [label= \(\bar{n}\)] at (1.7, -1);
\diagram*{
	(f2)--[gluon, line width = 0.3mm	](f1),
	(p2)--[ line width = 0.3mm](f1)--[  line width = 0.3mm](p3),
	(p4)--[  line width = 0.3mm](f2)--[ line width = 0.3mm](p1),
    (h1)--[gluon, line width = 0.3 mm](g1),
    (h2)--[gluon, line width = 0.3 mm](g2),
};
\end{feynman}
\end{tikzpicture}
}
\end{subfigure}
\begin{subfigure}[b]{0.16\textwidth}
\centering
\scalebox{0.6}{
\begin{tikzpicture}
\begin{feynman}
	\node[dot] (f1) at (0,.75);
	\node[dot] (f2) at (0,-.75);
    \node[dot] (g1) at (-.75,-.75);
    \vertex[label = left: \(s\)] (h1) at(-1.25,-1.5);
     \node[dot] (g2) at (-.75,-.75);
    \vertex[label = right: \(s\)] (h2) at(-.25,-1.5);
	\vertex(p3) at (1.5, .75);
	\vertex (p2) at (-1.5, .75);
	\vertex (p1) at (-1.5, -.75);
	\vertex (p4) at (1.5, -.75);
	\vertex [label= \(n\)] at (-1.7, .5);
	\vertex [label= \(\bar{n}\)] at (-1.7, -1);
    \vertex [label= \(n\)] at (1.7, .5);
	\vertex [label= \(\bar{n}\)] at (1.7, -1);
\diagram*{
	(f2)--[gluon, line width = 0.3mm	](f1),
	(p2)--[ line width = 0.3mm](f1)--[  line width = 0.3mm](p3),
	(p4)--[  line width = 0.3mm](f2)--[ line width = 0.3mm](p1),
    (h1)--[gluon, line width = 0.3 mm](g1),
    (h2)--[gluon, line width = 0.3 mm](g2),
};
\end{feynman}
\end{tikzpicture}
}
\end{subfigure}
\begin{subfigure}[b]{0.16\textwidth}
\centering
\scalebox{0.6}{
\begin{tikzpicture}
\begin{feynman}
	\node[dot] (f1) at (0,.75);
	\node[dot] (f2) at (0,-.75);
    \node[dot] (g1) at (0,-.75);
    \vertex[label = left: \(s\)] (h1) at(-.5,-1.5);
     \node[dot] (g2) at (0,-.75);
    \vertex[label = right: \(s\)] (h2) at(.5,-1.5);
	\vertex(p3) at (1.5, .75);
	\vertex (p2) at (-1.5, .75);
	\vertex (p1) at (-1.5, -.75);
	\vertex (p4) at (1.5, -.75);
	\vertex [label= \(n\)] at (-1.7, .5);
	\vertex [label= \(\bar{n}\)] at (-1.7, -1);
    \vertex [label= \(n\)] at (1.7, .5);
	\vertex [label= \(\bar{n}\)] at (1.7, -1);
\diagram*{
	(f2)--[gluon, line width = 0.3mm	](f1),
	(p2)--[ line width = 0.3mm](f1)--[  line width = 0.3mm](p3),
	(p4)--[  line width = 0.3mm](f2)--[ line width = 0.3mm](p1),
    (h1)--[gluon, line width = 0.3 mm](g1),
    (h2)--[gluon, line width = 0.3 mm](g2),
};
\end{feynman}
\end{tikzpicture}
}
\end{subfigure}
\begin{subfigure}[b]{0.16\textwidth}
\centering
\scalebox{0.6}{
\begin{tikzpicture}
\begin{feynman}
	\node[dot] (f1) at (0,.75);
	\node[dot] (f2) at (0,-.75);
    \node[dot] (g1) at (.75,-.75);
    \vertex[label = right: \(s\)] (h1) at(1.25,-1.5);
     \node[dot] (g2) at (.75,-.75);
    \vertex[label = left: \(s\)] (h2) at(.25,-1.5);
	\vertex(p3) at (1.5, .75);
	\vertex (p2) at (-1.5, .75);
	\vertex (p1) at (-1.5, -.75);
	\vertex (p4) at (1.5, -.75);
	\vertex [label= \(n\)] at (-1.7, .5);
	\vertex [label= \(\bar{n}\)] at (-1.7, -1);
    \vertex [label= \(n\)] at (1.7, .5);
	\vertex [label= \(\bar{n}\)] at (1.7, -1);
\diagram*{
	(f2)--[gluon, line width = 0.3mm	](f1),
	(p2)--[ line width = 0.3mm](f1)--[  line width = 0.3mm](p3),
	(p4)--[  line width = 0.3mm](f2)--[ line width = 0.3mm](p1),
    (h1)--[gluon, line width = 0.3 mm](g1),
    (h2)--[gluon, line width = 0.3 mm](g2),
};
\end{feynman}
\end{tikzpicture}
}
\end{subfigure}
\newline
\newline
\newline
\centering
\begin{subfigure}[b]{0.28\textwidth}
\subcaption{\qquad \qquad \qquad \qquad \qquad \qquad }
\centering
\begin{tikzpicture}
\begin{feynman}
	\vertex (f1) at (0,1);
	\vertex (f2) at (0,-1);
	\vertex(p3) at (1.5, 1);
	\vertex (p2) at (-1.5, 1);
	\vertex (p1) at (-1.5, -1);
	\vertex (p4) at (1.5, -1);
	\vertex [label= \(n\)] at (-1.7, .75);
	\vertex [label= \(\bar{n}\)] at (-1.7, -1.25);
    \vertex [label= \(n\)] at (1.7, .75);
	\vertex [label= \(\bar{n}\)] at (1.7, -1.25);
	\vertex (s1) at (0,0);
    \node[dot, color = SGreen](s2) at ( 1,0);
    \vertex[label = right: \(\textcolor{SGreen}{s}\)] (h2) at ( 1.5,.5);
    \vertex[label = right: \(\textcolor{SGreen}{s}\)] (h1) at ( 1.5,-.5);
\diagram*{
	(f1)--[scalar, red, line width = 0.3mm](s1),
	(f2)--[scalar, red, line width = 0.3mm](s1),
	(p2)--[scalar,  line width = 0.3mm](f1)--[ scalar,  line width = 0.3mm](p3),
	(p4)--[ scalar,  line width = 0.3mm](f2)--[ scalar,  line width = 0.3mm](p1),
    (s2)--[SGreen,gluon, line width = 0.3mm](s1),
    (h2)--[SGreen,gluon, line width = 0.3mm](s2),
    (h1)--[SGreen,gluon, line width = 0.3mm](s2),
};
\end{feynman}
	\filldraw[red] (0,1) ellipse (0.6mm and 1.2mm);
	\filldraw[red] (0, -1) ellipse (0.6mm and 1.2mm);
	\filldraw[red] (0,0) ellipse (0.6mm and 1.2mm);
\end{tikzpicture}
\end{subfigure}
\begin{subfigure}[b]{0.28\textwidth}
\centering
\begin{tikzpicture}
\begin{feynman}
	\vertex (f1) at (0,1);
	\vertex (f2) at (0,-1);
	\vertex(p3) at (1.5, 1);
	\vertex (p2) at (-1.5, 1);
	\vertex (p1) at (-1.5, -1);
	\vertex (p4) at (1.5, -1);
	\vertex [label= \(n\)] at (-1.7, .75);
	\vertex [label= \(\bar{n}\)] at (-1.7, -1.25);
    \vertex [label= \(n\)] at (1.7, .75);
	\vertex [label= \(\bar{n}\)] at (1.7, -1.25);
	\vertex (s1) at (0, .5);
    \vertex(s2) at (0, -.5);
    \vertex[label = right: \(\textcolor{SGreen}{s}\)] (h1) at ( 1.5,.333);
    \vertex[label = right: \(\textcolor{SGreen}{s}\)] (h2) at ( 1.5,-.333);
\diagram*{
	(f1)--[scalar, red, line width = 0.3mm](s1),
	(f2)--[scalar, red, line width = 0.3mm](s2),
	(p2)--[scalar,  line width = 0.3mm](f1)--[ scalar,  line width = 0.3mm](p3),
	(p4)--[ scalar,  line width = 0.3mm](f2)--[ scalar,  line width = 0.3mm](p1),
    (s2)--[SGreen,gluon, line width = 0.3mm](s1),
    (h2)--[SGreen,gluon, line width = 0.3mm](s2),
    (h1)--[SGreen,gluon, line width = 0.3mm](s1),
};
\end{feynman}
	\filldraw[red] (0,1) ellipse (0.6mm and 1.2mm);
	\filldraw[red] (0, -1) ellipse (0.6mm and 1.2mm);
	\filldraw[red] (0,.5) ellipse (0.6mm and 1.2mm);
    \filldraw[red] (0,-.5) ellipse (0.6mm and 1.2mm);
\end{tikzpicture}
\end{subfigure}
\begin{subfigure}[b]{0.28\textwidth}
\centering
\begin{tikzpicture}
\begin{feynman}
	\vertex (f1) at (0,1);
	\vertex (f2) at (0,-1);
	\vertex(p3) at (1.5, 1);
	\vertex (p2) at (-1.5, 1);
	\vertex (p1) at (-1.5, -1);
	\vertex (p4) at (1.5, -1);
	\vertex [label= \(n\)] at (-1.7, .75);
	\vertex [label= \(\bar{n}\)] at (-1.7, -1.25);
    \vertex [label= \(n\)] at (1.7, .75);
	\vertex [label= \(\bar{n}\)] at (1.7, -1.25);
	\vertex (s1) at (0,0);
    \vertex(s2) at ( 0,0);
    \vertex[label = right: \(\textcolor{SGreen}{s}\)] (h2) at ( 1.5,.5);
    \vertex[label = right: \(\textcolor{SGreen}{s}\)] (h1) at ( 1.5,-.5);
\diagram*{
	(f1)--[scalar, red, line width = 0.3mm](s1),
	(f2)--[scalar, red, line width = 0.3mm](s1),
	(p2)--[scalar,  line width = 0.3mm](f1)--[ scalar,  line width = 0.3mm](p3),
	(p4)--[ scalar,  line width = 0.3mm](f2)--[ scalar,  line width = 0.3mm](p1),
    (s2)--[SGreen,gluon, line width = 0.3mm](s1),
    (h2)--[SGreen,gluon, line width = 0.3mm](s2),
    (h1)--[SGreen,gluon, line width = 0.3mm](s2),
};
\end{feynman}
	\filldraw[red] (0,1) ellipse (0.6mm and 1.2mm);
	\filldraw[red] (0, -1) ellipse (0.6mm and 1.2mm);
	\filldraw[red] (0,0) ellipse (0.6mm and 1.2mm);
\end{tikzpicture}
\end{subfigure}
\caption{Diagrams for matching two soft graviton emissions.  In (a) we show the 40 full-theory diagrams.  In (b) we have the SCET diagrams.  The first two are time-ordered products of known EFT operators, while the third is an insertion of the two-graviton term in the soft operator.}
\label{2 Graviton Matching}
\end{figure}

At two soft graviton emissions, there are 40 full-theory Feynman diagrams which contribute.  We calculate all such diagrams directly, using Feyncalc \cite{Mertig:1990an, Shtabovenko:2016sxi, Shtabovenko:2020gxv, Shtabovenko:2023idz} to streamline the computation.  We performed the calculation using harmonic gauge, and we used the Feynman rules for the three and four graviton vertices \cite{DeWitt:1967uc}.  The calculation may be streamlined using other choices of gauge-fixing or choice of interpolating fields \cite{Cheung:2017kzx}, but given that the soft operator is gauge-invariant by construction, we would expect the result to be identical (up to field redefinitions).  As a non-trivial cross-check of the calculation, we verified that the result for the full amplitude satisfies the graviton Ward identity in both external graviton polarization tensors.

In the EFT, we have 3 contributions to the amplitude; one from the two-graviton contribution in the soft operator, and two involving T-products of EFT operators, including the one-graviton emission in a T-product with a Lagrangian insertion.  Because we used the graviton equations of motion to simplify the basis of soft operators, the first two rows of full-theory diagrams do not exactly match the contribution from the single soft graviton emission from the EFT.  However, we do cancel the non-local graviton propagator generated by the T-product.  Similarly, the full soft propagator in the remaining diagrams on the second row and those on the third match the soft propagator in the EFT T-product of the $n$-$s$ and $\bar{n}S$ Glauber operators.  The difference between the full amplitude and the EFT T-product contributions is then local, containing only Glauber $\perp$ propagators and eikonal $1/n\cdotp k$ and $1/\bar{n}\cdotp k$ terms.

It is then enough to match to the eikonal propagators.  From the $1/n\cdotp( k_1 + k_2)^2$ terms we are able to fix $C_2 =2$, and the remaining eikonal contributions of the form $1/n\cdotp k_1^2$ sets the remaining coefficient $C_3 =-4$.  Thus we have the full set of coefficients for the operator basis in Eq. (\ref{Soft Operator List}):
\begin{align}
    \boxed{
    C_1 = 2,\quad
    C_2 = 2,\quad 
    C_3 = -4,\quad 
    C_4 = 0,\quad
    C_5 = 0,\quad
    C_6 = 0,\quad
    C_7 = 0,\quad 
    C_8 = 0.
    }
\end{align}
This gives the full soft operator of
\begin{align}
    \mathcal{O}_S =& 2\mathcal{P}_S^2(S_{n}^T)_{--}^{\mu\nu}g_{\mu\rho}g_{\nu\sigma}(S_{\bar{n}})_{++}^{\rho\sigma} + (S_{n}^T)_{--}^{\mu\nu}g_{\mu\rho}g_{\nu\sigma}(S_{\bar{n}})_{++}^{\rho\sigma} \mathcal{P}_S^2 + 2(S_{n}^T)_{--}^{\mu\nu}g_{\mu\rho}g_{\nu\sigma}\square_S(S_{\bar{n}})_{++}^{\rho\sigma}\nonumber\\
    &- 4(S_{n}^T)_{--}^{\mu\nu}R^S_{\mu\rho\nu\sigma}(S_{\bar{n}})_{++}^{\rho\sigma}.
    \label{Soft Graviton Operator}
\end{align}
There are a few interesting points worth mentioning about this soft operator.  Firstly, the only operators with non-zero Wilson coefficients all have Wilson lines with only two Lorentz indices; all operators $O_{4-8}$ have at least one Wilson line with three or more indices in each term.  One way to potentially understand this is that only Wilson lines which have the same transformation under diffeomorphisms as the metric are allowed in the soft operator (i.e. traceless symmetric rank-2 tensors).  This is motivated by the QCD soft operator, only soft Wilson lines in the adjoint representation appear.  The soft graviton operator also shows striking parallels to the QCD soft operator, which can be written as
\begin{align}
    \mathcal{O}_{S,\,\text{QCD}}^{BC} = 4\pi \alpha_sn^\mu\bar{n}^\nu\left\{\mathcal{P}_S^2 \eta_{\mu\nu}\mathcal{S}_n^T \mathcal{S}_{\bar{n}} + \mathcal{S}_n^T \mathcal{S}_{\bar{n}}\eta_{\mu\nu}\mathcal{P}_S^2  + \mathcal{S}_n^T g_{\mu\nu}(i D_S)^2 \mathcal{S}_{\bar{n}} -2 \mathcal{S}_n^T ig\tilde{G}_{S\,\mu\nu} \mathcal{S}_{\bar{n}}\right\}^{BC},
\end{align}
where in the above $\mathcal{S}_n$ and $\mathcal{S}_{\bar{n}}$ are soft gluon Wilson lines, $D_S$ is the soft gluon covariant derivative, and $\tilde{G}$ is the gluon field strength tensor in the adjoint representation.  Comparing the soft graviton operator with the soft gluon operator, we can see that term-by-term we can obtain the soft graviton operator by replacing gluon Wilson lines with graviton Wilson lines, gluon field strength with the Riemann tensor, and adjoint color indices with Lorentz indices, contracted with an external $n$ and $\bar{n}$ vectors.  Some similarity might have been expected simply from double copy considerations, but it is somewhat surprising that this manifests at the level of the operators.  It could be interesting to explore this correspondence further in the future.

\subsection{Matching the Scalar Soft Function}

\begin{figure}
\renewcommand\thesubfigure{\arabic{subfigure}}
\begin{subfigure}[b]{0.19\textwidth}
\subcaption{\qquad \qquad \qquad \qquad \qquad }
\centering
\scalebox{0.6}{
\begin{tikzpicture}
\begin{feynman}
	\node[dot] (f1) at (0,.75);
	\node[dot] (f2) at (0,-.75);
    \node[dot] (g1) at (-0.75, .75);
    \node[dot] (g2) at (.25,1.5);
    \vertex[label = right: \(s\)] (h1) at(1,1.75);
    \vertex[label = right: \(s\)] (h2) at(1,1.25);
    \vertex(p3) at (1.5, .75);
	\vertex (p2) at (-1.5, .75);
	\vertex (p1) at (-1.5, -.75);
	\vertex (p4) at (1.5, -.75);
	\vertex [label= \(n\)] at (-1.7, .5);
	\vertex [label= \(\bar{n}\)] at (-1.7, -1);
    \vertex [label= \(n\)] at (1.7, .5);
	\vertex [label= \(\bar{n}\)] at (1.7, -1);
\diagram*{
	(f2)--[gluon, line width = 0.3mm	](f1),
	(p2)--[ line width = 0.3mm](f1)--[  line width = 0.3mm](p3),
	(p4)--[  line width = 0.3mm](f2)--[ line width = 0.3mm](p1),
    (h1)--[ line width = 0.3 mm](g2),
    (h2)--[ line width = 0.3 mm](g2),
    (g2)--[gluon, line width = 0.3 mm](g1),
};
\end{feynman}
\end{tikzpicture}
}
\end{subfigure}
\begin{subfigure}[b]{0.19\textwidth}
\centering
\scalebox{0.6}{
\begin{tikzpicture}
\begin{feynman}
	\node[dot] (f1) at (0,.75);
	\node[dot] (f2) at (0,-.75);
    \node[dot] (g1) at (0, .75);
    \node[dot] (g2) at (1,1.5);
    \vertex[label = right: \(s\)] (h2) at(1.75,1.25);
    \vertex[label = right: \(s\)] (h1) at(1.75,1.75);
	\vertex(p3) at (1.5, .75);
	\vertex (p2) at (-1.5, .75);
	\vertex (p1) at (-1.5, -.75);
	\vertex (p4) at (1.5, -.75);
	\vertex [label= \(n\)] at (-1.7, .5);
	\vertex [label= \(\bar{n}\)] at (-1.7, -1);
    \vertex [label= \(n\)] at (1.7, .5);
	\vertex [label= \(\bar{n}\)] at (1.7, -1);
\diagram*{
	(f2)--[gluon, line width = 0.3mm	](f1),
	(p2)--[ line width = 0.3mm](f1)--[  line width = 0.3mm](p3),
	(p4)--[  line width = 0.3mm](f2)--[ line width = 0.3mm](p1),
    (h1)--[ line width = 0.3 mm](g2),
    (h2)--[line width = 0.3 mm](g2),
    (g2)--[gluon, line width = 0.3 mm](g1),
};
\end{feynman}
\end{tikzpicture}
}
\end{subfigure}
\begin{subfigure}[b]{0.19\textwidth}
\centering
\scalebox{0.6}{
\begin{tikzpicture}
\begin{feynman}
	\node[dot] (f1) at (0,.75);
	\node[dot] (f2) at (0,-.75);
    \node[dot] (g1) at (.5, .75);
    \node[dot] (g2) at (1.5,1.5);
    \vertex[label = right: \(s\)] (h2) at(2.25,1.25);
    \vertex[label = right: \(s\)] (h1) at(2.25,1.75);
	\vertex(p3) at (1.5, .75);
	\vertex (p2) at (-1.5, .75);
	\vertex (p1) at (-1.5, -.75);
	\vertex (p4) at (1.5, -.75);
	\vertex [label= \(n\)] at (-1.7, .5);
	\vertex [label= \(\bar{n}\)] at (-1.7, -1);
    \vertex [label= \(n\)] at (1.7, .5);
	\vertex [label= \(\bar{n}\)] at (1.7, -1);
\diagram*{
	(f2)--[gluon, line width = 0.3mm	](f1),
	(p2)--[ line width = 0.3mm](f1)--[  line width = 0.3mm](p3),
	(p4)--[  line width = 0.3mm](f2)--[ line width = 0.3mm](p1),
    (h1)--[line width = 0.3 mm](g2),
    (h2)--[ line width = 0.3 mm](g2),
    (g2)--[gluon, line width = 0.3 mm](g1),
};
\end{feynman}
\end{tikzpicture}
}
\end{subfigure}
\begin{subfigure}[b]{0.19\textwidth}
\centering
\scalebox{0.6}{
\begin{tikzpicture}
\begin{feynman}
	\node[dot] (f1) at (0,.75);
	\node[dot] (f2) at (0,-.75);
    \node[dot] (g1) at (0, 0);
    \node[dot] (g2) at (1.25,0);
    \vertex[label = right: \(s\)] (h2) at(2,-.25);
    \vertex[label = right: \(s\)] (h1) at(2,.25);
	\vertex(p3) at (1.5, .75);
	\vertex (p2) at (-1.5, .75);
	\vertex (p1) at (-1.5, -.75);
	\vertex (p4) at (1.5, -.75);
	\vertex [label= \(n\)] at (-1.7, .5);
	\vertex [label= \(\bar{n}\)] at (-1.7, -1);
    \vertex [label= \(n\)] at (1.7, .5);
	\vertex [label= \(\bar{n}\)] at (1.7, -1);
\diagram*{
	(f2)--[gluon, line width = 0.3mm	](f1),
	(p2)--[ line width = 0.3mm](f1)--[  line width = 0.3mm](p3),
	(p4)--[  line width = 0.3mm](f2)--[ line width = 0.3mm](p1),
    (h1)--[ line width = 0.3 mm](g2),
    (h2)--[line width = 0.3 mm](g2),
    (g2)--[gluon, line width = 0.3 mm](g1),
};
\end{feynman}
\end{tikzpicture}
}
\end{subfigure}
\begin{subfigure}[b]{0.19\textwidth}
\centering
\scalebox{0.6}{
\begin{tikzpicture}
\begin{feynman}
	\node[dot] (f1) at (0,.75);
	\node[dot] (f2) at (0,-.75);
    \node[dot] (g1) at (-0.75, -.75);
    \node[dot] (g2) at (.25,-1.5);
    \vertex[label = right: \(s\)] (h1) at(1,-1.75);
    \vertex[label = right: \(s\)] (h2) at(1,-1.25);
    \vertex(p3) at (1.5, .75);
	\vertex (p2) at (-1.5, .75);
	\vertex (p1) at (-1.5, -.75);
	\vertex (p4) at (1.5, -.75);
	\vertex [label= \(n\)] at (-1.7, .5);
	\vertex [label= \(\bar{n}\)] at (-1.7, -1);
    \vertex [label= \(n\)] at (1.7, .5);
	\vertex [label= \(\bar{n}\)] at (1.7, -1);
\diagram*{
	(f2)--[gluon, line width = 0.3mm	](f1),
	(p2)--[ line width = 0.3mm](f1)--[  line width = 0.3mm](p3),
	(p4)--[  line width = 0.3mm](f2)--[ line width = 0.3mm](p1),
    (h1)--[line width = 0.3 mm](g2),
    (h2)--[ line width = 0.3 mm](g2),
    (g2)--[gluon, line width = 0.3 mm](g1),
};
\end{feynman}
\end{tikzpicture}
}
\end{subfigure}
\begin{subfigure}[b]{0.19\textwidth}
\centering
\scalebox{0.6}{
\begin{tikzpicture}
\begin{feynman}
	\node[dot] (f1) at (0,.75);
	\node[dot] (f2) at (0,-.75);
    \node[dot] (g1) at (0, -.75);
    \node[dot] (g2) at (1,-1.5);
    \vertex[label = right: \(s\)] (h2) at(1.75,-1.25);
    \vertex[label = right: \(s\)] (h1) at(1.75,-1.75);
	\vertex(p3) at (1.5, .75);
	\vertex (p2) at (-1.5, .75);
	\vertex (p1) at (-1.5, -.75);
	\vertex (p4) at (1.5, -.75);
	\vertex [label= \(n\)] at (-1.7, .5);
	\vertex [label= \(\bar{n}\)] at (-1.7, -1);
    \vertex [label= \(n\)] at (1.7, .5);
	\vertex [label= \(\bar{n}\)] at (1.7, -1);
\diagram*{
	(f2)--[gluon, line width = 0.3mm	](f1),
	(p2)--[ line width = 0.3mm](f1)--[  line width = 0.3mm](p3),
	(p4)--[  line width = 0.3mm](f2)--[ line width = 0.3mm](p1),
    (h1)--[ line width = 0.3 mm](g2),
    (h2)--[ line width = 0.3 mm](g2),
    (g2)--[gluon, line width = 0.3 mm](g1),
};
\end{feynman}
\end{tikzpicture}
}
\end{subfigure}
\begin{subfigure}[b]{0.19\textwidth}
\centering
\scalebox{0.6}{
\begin{tikzpicture}
\begin{feynman}
	\node[dot] (f1) at (0,.75);
	\node[dot] (f2) at (0,-.75);
    \node[dot] (g1) at (.5, -.75);
    \node[dot] (g2) at (1.5,-1.5);
    \vertex[label = right: \(s\)] (h2) at(2.25,-1.25);
    \vertex[label = right: \(s\)] (h1) at(2.25,-1.75);
	\vertex(p3) at (1.5, .75);
	\vertex (p2) at (-1.5, .75);
	\vertex (p1) at (-1.5, -.75);
	\vertex (p4) at (1.5, -.75);
	\vertex [label= \(n\)] at (-1.7, .5);
	\vertex [label= \(\bar{n}\)] at (-1.7, -1);
    \vertex [label= \(n\)] at (1.7, .5);
	\vertex [label= \(\bar{n}\)] at (1.7, -1);
\diagram*{
	(f2)--[gluon, line width = 0.3mm	](f1),
	(p2)--[ line width = 0.3mm](f1)--[  line width = 0.3mm](p3),
	(p4)--[  line width = 0.3mm](f2)--[ line width = 0.3mm](p1),
    (h1)--[ line width = 0.3 mm](g2),
    (h2)--[ line width = 0.3 mm](g2),
    (g2)--[gluon, line width = 0.3 mm](g1),
};
\end{feynman}
\end{tikzpicture}
}
\end{subfigure}
\begin{subfigure}[b]{0.19\textwidth}
\centering
\scalebox{0.6}{
\begin{tikzpicture}
\begin{feynman}
	\node[dot] (f1) at (0,.75);
	\node[dot] (f2) at (0,-.75);
    \node[dot] (g1) at (0, 0);
    \vertex[label = right: \(s\)] (h1) at(1,-.25);
    \node[dot] (g2) at (0,0);
    \vertex[label = right: \(s\)] (h2) at(1,.25);
	\vertex(p3) at (1.5, .75);
	\vertex (p2) at (-1.5, .75);
	\vertex (p1) at (-1.5, -.75);
	\vertex (p4) at (1.5, -.75);
	\vertex [label= \(n\)] at (-1.7, .5);
	\vertex [label= \(\bar{n}\)] at (-1.7, -1);
    \vertex [label= \(n\)] at (1.7, .5);
	\vertex [label= \(\bar{n}\)] at (1.7, -1);
\diagram*{
	(f2)--[gluon, line width = 0.3mm	](f1),
	(p2)--[ line width = 0.3mm](f1)--[  line width = 0.3mm](p3),
	(p4)--[  line width = 0.3mm](f2)--[ line width = 0.3mm](p1),
    (h1)--[ line width = 0.3 mm](g1),
    (h2)--[line width = 0.3 mm](g2),
};
\end{feynman}
\end{tikzpicture}
}
\end{subfigure}
\begin{subfigure}[b]{0.19\textwidth}
\centering
\scalebox{0.6}{
\begin{tikzpicture}
\begin{feynman}
	\node[dot] (f1) at (0,1);
	\node[dot] (f2) at (0,-1);
    \node[dot] (g1) at (0, .25);
    \vertex[label = right: \(s\)] (h1) at(1,.4);
    \node[dot] (g2) at (0,-.25);
    \vertex[label = right: \(s\)] (h2) at(1,-.4);
	\vertex(p3) at (1.5, 1);
	\vertex (p2) at (-1.5, 1);
	\vertex (p1) at (-1.5, -1);
	\vertex (p4) at (1.5, -1);
	\vertex [label= \(n\)] at (-1.7, .75);
	\vertex [label= \(\bar{n}\)] at (-1.7, -1.25);
    \vertex [label= \(n\)] at (1.7, .75);
	\vertex [label= \(\bar{n}\)] at (1.7, -1.25);
\diagram*{
	(f2)--[gluon, line width = 0.3mm	](g2),
    (g1)--[gluon, line width = 0.3 mm](f1),
	(p2)--[ line width = 0.3mm](f1)--[  line width = 0.3mm](p3),
	(p4)--[  line width = 0.3mm](f2)--[ line width = 0.3mm](p1),
    (h1)--[line width = 0.3mm](g1),
    (h2)--[ line width= 0.3mm ](g2),
    (g1)--[line width= 0.3mm ](g2),
};
\end{feynman}
\end{tikzpicture}
}
\end{subfigure}
\newline

\centering
\begin{subfigure}[b]{0.28\textwidth}
\subcaption{\qquad \qquad \qquad \qquad \qquad \qquad }
\centering
\begin{tikzpicture}
\begin{feynman}
	\vertex (f1) at (0,1);
	\vertex (f2) at (0,-1);
	\vertex(p3) at (1.5, 1);
	\vertex (p2) at (-1.5, 1);
	\vertex (p1) at (-1.5, -1);
	\vertex (p4) at (1.5, -1);
	\vertex [label= \(n\)] at (-1.7, .75);
	\vertex [label= \(\bar{n}\)] at (-1.7, -1.25);
    \vertex [label= \(n\)] at (1.7, .75);
	\vertex [label= \(\bar{n}\)] at (1.7, -1.25);
	\vertex (s1) at (0,0);
    \node[dot, color = SGreen](s2) at ( 1,0);
    \vertex[label = right: \(\textcolor{SGreen}{s}\)] (h2) at ( 1.5,.4);
    \vertex[label = right: \(\textcolor{SGreen}{s}\)] (h1) at ( 1.5,-.4);
\diagram*{
	(f1)--[scalar, red, line width = 0.3mm](s1),
	(f2)--[scalar, red, line width = 0.3mm](s1),
	(p2)--[scalar,  line width = 0.3mm](f1)--[ scalar,  line width = 0.3mm](p3),
	(p4)--[ scalar,  line width = 0.3mm](f2)--[ scalar,  line width = 0.3mm](p1),
    (s2)--[SGreen,gluon, line width = 0.3mm](s1),
    (h2)--[SGreen, line width = 0.3mm](s2),
    (h1)--[SGreen, line width = 0.3mm](s2),
};
\end{feynman}
	\filldraw[red] (0,1) ellipse (0.6mm and 1.2mm);
	\filldraw[red] (0, -1) ellipse (0.6mm and 1.2mm);
	\filldraw[red] (0,0) ellipse (0.6mm and 1.2mm);
\end{tikzpicture}
\end{subfigure}
\begin{subfigure}[b]{0.28\textwidth}
\centering
\begin{tikzpicture}
\begin{feynman}
	\vertex (f1) at (0,1);
	\vertex (f2) at (0,-1);
	\vertex(p3) at (1.5, 1);
	\vertex (p2) at (-1.5, 1);
	\vertex (p1) at (-1.5, -1);
	\vertex (p4) at (1.5, -1);
	\vertex [label= \(n\)] at (-1.7, .75);
	\vertex [label= \(\bar{n}\)] at (-1.7, -1.25);
    \vertex [label= \(n\)] at (1.7, .75);
	\vertex [label= \(\bar{n}\)] at (1.7, -1.25);
	\vertex (s1) at (0, .3333);
    \vertex(s2) at (0, -.3333);
    \vertex[label = right: \(\textcolor{SGreen}{s}\)] (h1) at ( 1,.5);
    \vertex[label = right: \(\textcolor{SGreen}{s}\)] (h2) at ( 1,-.5);
\diagram*{
	(f1)--[scalar, red, line width = 0.3mm](s1),
	(f2)--[scalar, red, line width = 0.3mm](s2),
	(p2)--[scalar,  line width = 0.3mm](f1)--[ scalar,  line width = 0.3mm](p3),
	(p4)--[ scalar,  line width = 0.3mm](f2)--[ scalar,  line width = 0.3mm](p1),
    (s2)--[SGreen, line width = 0.3mm](s1),
    (h2)--[SGreen, line width = 0.3mm](s2),
    (h1)--[SGreen, line width = 0.3mm](s1),
};
\end{feynman}
	\filldraw[red] (0,1) ellipse (0.6mm and 1.2mm);
	\filldraw[red] (0, -1) ellipse (0.6mm and 1.2mm);
	\filldraw[red] (0,.3333) ellipse (0.6mm and 1.2mm);
    \filldraw[red] (0,-.3333) ellipse (0.6mm and 1.2mm);
\end{tikzpicture}
\end{subfigure}
\begin{subfigure}[b]{0.28\textwidth}
\centering
\begin{tikzpicture}
\begin{feynman}
	\vertex (f1) at (0,1);
	\vertex (f2) at (0,-1);
	\vertex(p3) at (1.5, 1);
	\vertex (p2) at (-1.5, 1);
	\vertex (p1) at (-1.5, -1);
	\vertex (p4) at (1.5, -1);
	\vertex [label= \(n\)] at (-1.7, .75);
	\vertex [label= \(\bar{n}\)] at (-1.7, -1.25);
    \vertex [label= \(n\)] at (1.7, .75);
	\vertex [label= \(\bar{n}\)] at (1.7, -1.25);
	\vertex (s1) at (0,0);
    \vertex(s2) at ( 0,0);
    \vertex[label = right: \(\textcolor{SGreen}{s}\)] (h2) at ( 1,.5);
    \vertex[label = right: \(\textcolor{SGreen}{s}\)] (h1) at ( 1,-.5);
\diagram*{
	(f1)--[scalar, red, line width = 0.3mm](s1),
	(f2)--[scalar, red, line width = 0.3mm](s1),
	(p2)--[scalar,  line width = 0.3mm](f1)--[ scalar,  line width = 0.3mm](p3),
	(p4)--[ scalar,  line width = 0.3mm](f2)--[ scalar,  line width = 0.3mm](p1),
    (s2)--[SGreen, line width = 0.3mm](s1),
    (h2)--[SGreen,line width = 0.3mm](s2),
    (h1)--[SGreen, line width = 0.3mm](s2),
};
\end{feynman}
	\filldraw[red] (0,1) ellipse (0.6mm and 1.2mm);
	\filldraw[red] (0, -1) ellipse (0.6mm and 1.2mm);
	\filldraw[red] (0,0) ellipse (0.6mm and 1.2mm);
\end{tikzpicture}
\end{subfigure}
\caption{Matching for two soft scalar emissions.  In (a), we show the 9 full-theory diagrams.  In (b), we show the EFT contributions.  The first two involve time-ordered products of EFT operators, while the third is an insertion of the two scalar term of the soft operator.}
\end{figure}

We now match the soft scalar terms in the soft operator.  Here, in constructing the operator basis, we are aided by an additional symmetry of mass scalars, that is a symmetry of shifting by an additive constant, 
\begin{equation}
    \phi\rightarrow \phi + c.
\end{equation}
The EFT of course must also respect this symmetry.  This then requires that all scalars must come with a derivative in the combinations $\grad_\mu \phi$.  This then fixes the derivatives in the soft operator, leaving only the distinct ways indices can be contracted between the Wilson lines and the derivatives.  Thus, there are only two scalar operators one can write down:
\begin{align}
  O_1^\phi &= (S_{n}^T)_{--}^{\mu\nu}g_{\mu\rho}g_{\nu\sigma}g^{\alpha\beta}\frac{\kappa^2}{4}(\grad_\alpha \phi_S)(\grad_\beta \phi_S)(S_{\bar{n}})_{++}^{\rho\sigma},\nonumber\\
  O_2^\phi &= (S_{n}^T)_{--}^{\mu\nu}g_{\nu\sigma}\frac{\kappa^2}{4}(\grad_\mu \phi_S)(\grad_\rho \phi_S)(S_{\bar{n}})_{++}^{\rho\sigma}.
\end{align}
We can then match the coefficients of these operators by considering the scalar-scalar forward scattering with the emission of two additional soft scalars.  In the EFT there are three contributions, one which is a time-ordered product involving the gravitational Lipatov vertex, two from a time-ordered product of an $n$-$s$ and an $\bar{n}s$ scalar-scalar Glauber operator, and one from the soft scalar operator.  Meanwhile in the full theory there are 9 diagrams.  We are able to straightforwardly perform the calculations, and we find the Wilson coefficients to be
\begin{equation}
    \boxed{
    C_1^\phi = 0,\qquad  C_2^\phi = -2.
    }
\end{equation}

This completes the matching of the soft function for the specified matter fields.  In general, we can expect additional soft operator contributions for matter fields of different spins and different couplings to gravity.  Note that this does include a soft fermion emission operator.  In gravity, each additional matter field comes with a factor of $\kappa$, which reduce the mass dimension of the field by one; therefore two fermion fields come with a mass dimension of 1, and can satisfy the mass dimension constraint.  This with QCD, where a soft fermion bilinear has mass dimension 3 and is thus ruled out. 
Finally note that, as pointed out in the context of NRQCD \cite{Rothstein:2018dzq}, an advantage of building 
operators with   gauge invariant interpolating fields is that  we we do not need to consider operators with ghosts on external legs.

\bibliography{paper3} 
\end{document}